\documentclass[11pt]{article}
\usepackage{amsmath,amsfonts,amssymb,graphicx,epsfig,pdflscape,multirow}  
\usepackage[normalem]{ulem}
\usepackage{cite}
\usepackage[usenames]{color}
\usepackage{pstricks}
\usepackage{subfig}

\usepackage[nomessages]{fp}

\usepackage[backref=false]{hyperref}
\hypersetup{
colorlinks=true,
citecolor=red,
linkcolor=darkblue
}
\definecolor{darkblue}{rgb}{0,0,.8}
\definecolor{red}{rgb}{1,0,0}
\definecolor{myboxcolor}{rgb}{.95,.94,.99}

\numberwithin{equation}{section}

\graphicspath{{./eps/}}
\long\def\ignore#1{}
\definecolor{purple}{rgb}{1,0,1}
\definecolor{darkpurple}{rgb}{1,.2,1}
\definecolor{pink}{rgb}{1,.7,.7}

\definecolor{AliceBlue}{RGB}{240,248,255}
\definecolor{CornflowerBlue}{RGB}{100,149,237}
\definecolor{RoyalBlue}{RGB}{60,105,225}
\definecolor{RedViolet}{RGB}{199,21,133}
\definecolor{Blue}{rgb}{.01,.01,.6}

\newcommand{\nc}{\newcommand}
\nc\disp{\displaystyle}
\nc{\fh}{\hat{f}}
\nc{\muh}{\hat{\mu}}
\nc{\nuh}{\hat{\nu}}
\nc{\spos}[2]{\makebox(0,0)[#1]{$\sm{#2}$}}
\nc{\pos}[2]{\makebox(0,0)[#1]{\small ${#2}$}}
\nc{\sm}[1]{{\scriptstyle #1}}
\nc{\qbar}{\overline{q}}
\nc{\bib}{\bibitem}
\nc{\al}{\alpha}
\nc{\g}{\gamma}
\nc{\G}{\Gamma}
\nc{\D}{\Delta}
\nc{\eps}{\epsilon}
\nc{\la}{\lambda}
\nc{\La}{\Lambda}
\nc{\var}{\varphi}
\nc{\pa}{\partial}
\nc{\nn}{\nonumber \\ }
\nc{\hf}{\frac{1}{2}}
\nc{\dz}{\frac{dz}{2\pi i}}
\nc{\bin}[2]{\left(\!\!\!\begin{array}{c} {#1}\\ {#2} \end{array}\!\!\!\right)}
\nc{\be}{\begin{equation}}
\nc{\ee}{\end{equation}}
\nc{\bea}{\begin{eqnarray}}
\nc{\eea}{\end{eqnarray}}
\nc{\bra}[1]{\langle {#1}|}
\nc{\ket}[1]{|{#1}\rangle}
\nc{\ketw}[1]{({#1})^{\phantom{a}}_{{\cal W}}}
\nc{\chit}{\raisebox{0.25ex}{$\chi$}}
\nc{\chih}{\raisebox{0.25ex}{$\hat\chi$}}
\nc{\Db}{\mbox{\boldmath $D$}}
\nc{\Hb}{\mbox{\boldmath $H$}}
\nc{\calE}{{\cal E}}
\nc{\calH}{{\cal H}}
\nc{\calR}{{\cal R}}
\nc{\calL}{{\cal L}}
\nc{\calV}{{\cal V}}
\nc{\Hc}{{\cal H}}
\nc{\Rc}{{\cal R}}
\nc{\Lc}{{\cal L}}
\nc{\Vc}{{\cal V}}
\nc{\Ib}{\mbox{\boldmath $I$}}
\nc{\qb}{\bar{q}}
\nc{\Ac}{\mathcal{A}}
\nc{\Bc}{\mathcal{B}}
\nc{\Cc}{\mathcal{C}}
\nc{\Dc}{\mathcal{D}}
\nc{\Ec}{\mathcal{E}}
\nc{\Gc}{\mathcal{G}}
\nc{\Ic}{\mathcal{I}}
\nc{\Jc}{\mathcal{J}}
\nc{\Oc}{\mathcal{O}}
\nc{\Pc}{\mathcal{P}}
\nc{\Sc}{\mathcal{S}}
\nc{\Tc}{\mathcal{T}}
\nc{\Wc}{\mathcal{W}}
\nc{\Xc}{\mathcal{X}}
\nc{\Yc}{\mathcal{Y}}
\nc{\Zc}{\mathcal{Z}}
\nc{\fus}{\mbox{}\,\hat\otimes\,\mbox{}}
\nc{\Pch}{\hat{\Pc}}
\nc{\Rch}{\hat{\Rc}}
\nc{\Dh}{\hat{\Delta}}
\nc{\rh}{\hat{r}}
\nc{\sh}{\hat{s}}
\nc{\kb}{\bar{k}}
\nc{\taub}{\bar{\tau}}
\nc{\Jcb}{\Jc_{\mathrm{b}}}
\nc{\rtt}{\mathtt{r}}
\nc{\stt}{\mathtt{s}}
\nc{\cosR}{\cos\frac{\pi p'rr'}{p}}
\nc{\cosS}{\cos\frac{\pi pss'}{p'}}
\nc{\sinR}{\sin\frac{\pi p'rr'}{p}}
\nc{\sinS}{\sin\frac{\pi pss'}{p'}}
\def\vvdots{\mathinner{\mkern1mu\raise1pt\vbox{\kern7pt\hbox{.}}\mkern2mu
  \raise4pt\hbox{.}\mkern2mu\raise7pt\hbox{.}\mkern1mu}}
\def\gauss#1#2{\Big[\!\!\!\begin{array}{c} {#1}\\[-3pt] {#2} \end{array}\!\!\!\Big]}
\nc{\sbin}[2]{\left\{\!\!\!\begin{array}{c} {#1}\\ {#2} 
\end{array}\!\!\!\right\}}
\nc{\sbinlr}[2]{\Big\langle\!\!\begin{array}{c} {#1}\\ {#2} 
\end{array}\!\!\Big\rangle}
\nc{\bino}[2]{\left(\!\!\begin{array}{c} {#1}\\ {#2} \end{array}\!\!\right)}
\def\half {\mbox{$\textstyle \frac{1}{2}$} }
\def\vec#1{\mbox {\boldmath $#1$}}

\definecolor{lightblue}{rgb}{.61,.61,1}
\definecolor{lightblue}{rgb}{.61,.61,1}
\definecolor{midblue}{rgb}{.7,.7,1}
\definecolor{lightlightblue}{rgb}{.9,.9,1}
\definecolor{lightestblue}{rgb}{.96,.96,1}
\definecolor{lightpurple}{rgb}{1,.75,1}
\def\leftarc#1{\psarc[linecolor=blue,linewidth=1.5pt]#1{.5}{90}{270}}
\def\rightarc#1{\psarc[linecolor=blue,linewidth=1.5pt]#1{.5}{-90}{90}}
\def\loopa{
\psframe[linewidth=.25pt](0,0)(1,1)
\psarc[linewidth=1.5pt,linecolor=blue](1,0){.5}{90}{180}
\psarc[linewidth=1.5pt,linecolor=blue](0,1){.5}{-90}{0}
}
\def\loopb{
\psframe[linewidth=.25pt](0,0)(1,1)
\psarc[linewidth=1.5pt,linecolor=blue](0,0){.5}{0}{90}
\psarc[linewidth=1.5pt,linecolor=blue](1,1){.5}{180}{270}
}

\def\facegrid#1#2{
\psframe[fillstyle=solid,fillcolor=lightlightblue,linewidth=0pt]#1#2
\psgrid[gridlabels=0pt,subgriddiv=1]#1#2
}

\nc{\ch}{{\rm ch}}
\nc{\R}{{\cal R}}
\nc{\dkk}{\delta_{j,\{k,k'\}}^{(2)}}
\nc{\drr}{\delta_{j,\{r,r'\}}^{(2)}}
\nc{\ddkk}{\delta_{j,\{k,k'\}}^{(4)}}
\nc{\dddkk}{\delta_{j,\{k,k'\}}^{(8)}}
\nc{\dnn}{\delta_{j,\{n,n'\}}^{(2)}}
\nc{\ddnn}{\delta_{j,\{n,n'\}}^{(4)}}
\nc{\dddnn}{\delta_{j,\{n,n'\}}^{(8)}}

\nc{\chh}{\widehat{\mathrm{ch}}}
\nc{\ph}{\hat{p}}

\definecolor{pink}{rgb}{1,.65,.65}

\thicklines

\thicklines

\def \Monoid{
\rput(0,0){\MonoidCap}
\rput(0,1){\MonoidCup}
}

\def \MonoidCap{
\psbezier[linewidth=1pt](0,0)(0,.55)(1,.55)(1,0)
}

\def \MonoidCup{
\psbezier[linewidth=1pt](0,0)(0,-.55)(1,-.55)(1,0)
}

\def \cabledDefOne[#1]{
\psline[linewidth=1pt](-.4,0)(-.4,#1)
\psline[linewidth=1pt](.4,0)(.4,#1)
}

\def \squareProj[#1,#2,#3,#4,#5]{
\psline[linewidth=1pt](.4,0)(.4,#2)
\psline[linewidth=1pt](1.6,0)(1.6,#3)
\psline[linewidth=1pt](1.6,0)(1.6,-#4)
\psline[linewidth=1pt](.4,0)(.4,-#5)
\pspolygon[linewidth=1pt,fillstyle=solid,fillcolor=white](0,0)(1,1)(2,0)(1,-1)
\rput(1,0){$#1\lambda$}
}

\def \singleDef[#1]{
\psline(-.4,0)(-.4,#1)
}

\def \singleDefR[#1]{
\psline(.4,0)(.4,#1)
}

\def \proj[#1]{
\psellipse[linewidth=1pt](#1,0)(#1,.2)
}

\nc\drtm{{\vec D}}      		
\nc\face{\mathbb{X}} 		
\nc\faceK{\mathbb{K}} 		
\nc\genface[1]{\mathbb{X}^{(#1)}} 
\nc \ham{{\mathcal H}}			

\nc \nface{\mathbb{\hat X}}	

\nc\TLw{\omega} 			
\nc\TLb{\beta} 				

\nc\BMWw{\omega_2}		
\nc\BMWb{\beta_2}			

\nc\fTLw{\omega_2} 			
\nc\fTLb{\beta_2} 			

\nc{\genw}[1]{\omega_{#1}}  	
\nc{\Genw}[2]{\omega_{#1}^{(#2)}}
\nc{\genb}[1]{\beta_{#1}}       	

\def \trinomial[#1][#2][#3][#4]{\left[{#1\atop #2,#3,#4}\right]}

\def \superTrinomial[#1][#2]{
\left({#1 \atop #2} \right)_{\!2}
}

\def \qTrinomial[#1][#2][#3][#4]{\left[{#1\atop #2,#3,#4}\right]_{\!q}}

\def\Re{\mathop{\mbox{Re}}}

\def \bdyCoeff{h^{(\rho)}}

\nc{\XiShift}{\overline{\xi}_\rho}
\def\emptysquarea{
\facegrid{(0,0)}{(1,1)}
\psframe[linewidth=.15pt](0,0)(1,1)
}

\nc{\fws}{\small}

\def\rtri#1{
\pspolygon[fillstyle=solid,fillcolor=lightlightblue](0,1)(1,0)(1,2)
\rput(.6,1){\scriptsize $#1$}
}

\def\dddots{\mathinner{\mkern1mu\raise9pt\vbox{\kern7pt\hbox{.}}\mkern2mu
  \raise5pt\hbox{.}\mkern2mu\raise1pt\hbox{.}\mkern1mu}}

\nc{\smbin}[2]{\Big(\!\!\!\begin{array}{c} {#1}\\[-3pt] {#2} \end{array}\!\!\!\Big)}
  
\begin{document}

\topmargin -5mm
\oddsidemargin 5mm

\setcounter{page}{1}

\vspace{8mm}
\begin{center}
{\huge {\bf  Kac boundary conditions of the\\[4pt] logarithmic minimal models}}

\vspace{10mm}
{\Large Paul A. Pearce$^\ast$, Elena Tartaglia$^\ast$, Romain Couvreur$^\dagger$, }
\\[.4cm]
{\em {}$^\ast$Department of Mathematics and Statistics, University of Melbourne}\\
{\em Parkville, Victoria 3010, Australia}
\\[.4cm]
{\em {}$^\dagger${Department of Physics, ICFP,  \'{E}cole Normale Sup\'{e}rieure}}\\
{\em 24 rue Lhomond, 75005 Paris, France} 
\\[.4cm]
{\tt p.pearce\,@\,ms.unimelb.edu.au}
\qquad
{\tt elena.tartaglia\,@\,unimelb.edu.au}\\
\qquad
{\tt romain.couvreur\,@\,ens.fr}
\end{center}

\vspace{14pt}
\centerline{\textbf{\textit{This paper is dedicated to Jean-Bernard Zuber on the occasion of his retirement.}}}

\vspace{10mm}
\centerline{{\bf{Abstract}}}
\vskip.4cm
\noindent
We develop further the implementation and analysis of Kac boundary conditions in the general \mbox{logarithmic} \mbox{minimal} models ${\cal LM}(p,p')$ with $1\le p<p'$ and $p,p'$ coprime. 
Specifically, working in a strip geometry, we \mbox{consider} the $(r,s)$ Kac boundary conditions. These boundary conditions are organized into infinitely extended Kac tables labeled by the Kac labels $r,s=1,2,3,\ldots$.
They are conjugate to Virasoro Kac \mbox{representations} with conformal dimensions $\Delta_{r,s}$ given by the usual Kac formula. 
On a finite strip of width $N$, built from a square lattice, the associated integrable boundary conditions are constructed by acting on the \mbox{vacuum} $(1,1)$ boundary with an $s$-type seam of width $s-1$ columns and an $r$-type seam of width $\rho-1$ columns. The $r$-type seam contains an arbitrary boundary field $\xi$. While the usual fusion \mbox{construction} of the $r$-type seam relies on the existence of Wenzl-Jones projectors restricting its \mbox{application} to $r\le\rho<p'$, this limitation was recently removed by Pearce, Rasmussen and Villani who further conjectured that the conformal boundary conditions labeled by $r$ are realized, in particular, for $\rho=\rho(r)=\lfloor \frac{rp'}{p}\rfloor$. In this paper, we \mbox{confirm} this conjecture by performing extensive numerics on the commuting \mbox{double} row transfer \mbox{matrices} and their associated quantum Hamiltonian chains. 
Letting $[x]$ denote the \mbox{fractional} part, we fix the boundary field to the specialized values $\xi=\frac{\pi}{2}$ if $[\frac{\rho}{p'}]=0$ and $\xi=[\frac{\rho p}{p'}]\frac{\pi}{2}$ otherwise.
For these boundary conditions, we obtain the Kac conformal weights $\Delta_{r,s}$ by numerically \mbox{extrapolating} the finite-size corrections to the lowest eigenvalue of the quantum Hamiltonians out to sizes \mbox{$N\le 32-\rho-s$}.
\mbox{Additionally}, by solving local inversion relations, we obtain general analytic expressions for the \mbox{boundary} free energies allowing for more accurate estimates of the conformal data.

\newpage
\tableofcontents

\newpage
\hyphenpenalty=30000

\section{Introduction}

The universal critical behaviour of two dimensional lattice statistical systems (such as Ising and Potts models) with {\it local}\/ degrees of freedom are described, in the continuum scaling limit, by {\it rational}\/ Conformal Field Theories (CFTs)~\cite{MooreSeiberg}. Most notably, the minimal models ${\cal M}(m,m')$~\cite{BPZ}, with \mbox{$2\le m<m'$} and $m,m'$ coprime, are known to describe~\cite{Huse} the continuum scaling limit of the Forrester-Baxter Restricted-Solid-on-Solid (RSOS) models~\cite{FB}. Similarly, it is by now well established that the universal critical behaviour of two dimensional lattice statistical systems (such as polymers and percolation) with {\it nonlocal}\/ degrees of freedom are  described, in the continuum scaling limit, by {\it logarithmic}\/ CFTs. Logarithmic CFTs are characterized~\cite{Gurarie} by logarithmically growing or decaying correlation functions and the concomitant existence of reducible yet indecomposable repsentations of the Virasoro algebra. Logarithmic CFTs typically exhibit a very rich operator/representation content and their properties are profoundly different to rational CFTs. Nevertheless logarithmic CFTs offer a class of CFTs, beyond rational CFTs, that can potentially be understood in great depth. For these reasons, the theoretical study of logarithmic CFT has been steadily growing since the early nineties. 

Early studies~\cite{Tehran,FlohrIran,GabIran} of logarithmic CFTs focussed on an algebraic approach to a family of CFTs with central charges $c=c^{p,p'}$ given by (\ref{cKac}). This is the same formula that gives the central charges $c=c^{m,m'}$ of the minimal models ${\cal M}(m,m')$. Indeed this family of logarithmic CFTs was viewed informally as an extension of the minimal models. More recently, it has emerged that these theories can in fact be viewed as a {\it logarithmic limit}\/~\cite{RasmussenLogLimit} of the minimal models. More precisely, the logarithmic minimal models ${\cal LM}(p,p')$ are given symbolically by
\bea
{\cal LM}(p,p')=\lim_{m,m'\to\infty, \  \frac{m}{m'}\to \frac{p}{p'}} {\cal M}(m,m')
\label{logLimitEqn}
\eea
Perhaps most importantly, the logarithmic minimal models ${\cal LM}(p,p')$ inherit a coset structure from the minimal models ${\cal M}(m,m')$ enabling their CFTs to be constructed~\cite{PRcoset} as a GKO coset~\cite{GKO85}.
The family of logarithmic minimal models ${\cal LM}(p,p')$, with $1\le p<p'$ and $p,p'$ coprime, provides the simplest prototypical examples of logarithmic CFTs. 

In 2006 it was shown that, from a lattice perspective, the logarithmic minimal models ${\cal LM}(p,p')$ arise~\cite{PRZ} as the continuum scaling limit of a family of Yang-Baxter integrable~\cite{BaxBook} loop models on the square lattice. 
The first members of this family include critical dense polymers ${\cal LM}(1,2)$~\cite{PR2007} and critical bond percolation ${\cal LM}(2,3)$~\cite{percolation}. 
Specifically it has been shown~\cite{PRZ,PR2007,AMDSA} that, with suitable boundary conditions on the strip, the transfer matrices and associated quantum Hamiltonians exhibit rank 2 Jordan cells for finite systems. Remarkably, the patterns of Jordan cells robustly persist in the thermodynamic limit giving rise to the reducible yet indecomposable structures of the Virasoro dilatation operator $L_0$. These observations opened up a lattice approach that has reinvigorated the research activity on logarithmic CFT. A review of the current status of logarithmic CFT can be found in~\cite{SpecialIssue}.

The conformal properties of the logarithmic minimal models ${\cal LM}(p,p')$ have been well studied. These theories are nonunitary and non-rational. 
The central charges are
\be
c =c^{p,p'}= 1 - \frac{6(p'-p)^2}{pp'},\qquad 1\le p<p',\qquad \gcd(p,p')=1\label{cKac}
\ee
and the conformal weights are given by the infinitely extended Kac formula
\be
\Delta_{r,s} = \Delta^{p,p'}_{r,s}=\frac{(p'r-ps)^2-(p'-p)^2}{4pp'},\qquad r,s=1,2,3,\ldots\label{confWts}
\ee
The Kac characters of the associated $(r,s)$ Virasoro representations are 
\be
\chi_{r,s}(q)=\chi_{r,s}^{p,p'}(q)=\frac{q^{-\frac{c}{24}+\Delta_{r,s}^{p,p'}}}{(q)_\infty}\,(1-q^{rs}),\qquad (q)_\infty=\prod_{k=1}^\infty (1-q^k),\qquad r,s=1,2,3,\ldots
\label{KacChars}
\ee
where $q$ is the modular nome. In terms of conformal partition functions on a strip, these characters occur as the spectrum generating functions for the conjugate $(r,s)$ boundary conditions
\bea
Z_{(1,1)|(r,s)}^{p,p'}(q)=Z_{(1,s)|(r,1)}^{p,p'}(q)=\chi_{r,s}^{p,p'}(q)=q^{-\frac{c}{24}+\Delta_{r,s}^{p,p'}} \sum_E q^E\label{confZ}
\eea
where $(r,s)=(1,1)$ with $\Delta_{1,1}=0$ denotes the {\it vacuum} boundary condition. As an example, the infinitely extended Kac table of the logarithmic Yang-Lee model ${\cal LM}(2,5)$ is shown in Figure~\ref{YLKac}.

\begin{figure}[htb]
\psset{unit=.75cm}
\begin{center}
\begin{pspicture}(0,0)(7,11)
\psframe[linewidth=0pt,fillstyle=solid,fillcolor=lightestblue](0,0)(7,11)
\psframe[linewidth=1pt,fillstyle=solid,fillcolor=midblue](0,0)(1,4)
\psframe[linewidth=0pt,fillstyle=solid,fillcolor=lightlightblue](1,0)(2,11)
\psframe[linewidth=0pt,fillstyle=solid,fillcolor=lightlightblue](3,0)(4,11)
\psframe[linewidth=0pt,fillstyle=solid,fillcolor=lightlightblue](5,0)(6,11)
\psframe[linewidth=0pt,fillstyle=solid,fillcolor=lightlightblue](0,4)(7,5)
\psframe[linewidth=0pt,fillstyle=solid,fillcolor=lightlightblue](0,9)(7,10)
\multiput(0,0)(0,5){2}{\multiput(0,0)(2,0){3}{\psframe[linewidth=0pt,fillstyle=solid,fillcolor=lightblue](1,4)(2,5)}}
\multirput(2,1)(2,0){3}{\pswedge[fillstyle=solid,fillcolor=red,linecolor=red](0,0){.25}{180}{270}}
\multirput(2,2)(2,0){3}{\pswedge[fillstyle=solid,fillcolor=red,linecolor=red](0,0){.25}{180}{270}}
\multirput(2,3)(2,0){3}{\pswedge[fillstyle=solid,fillcolor=red,linecolor=red](0,0){.25}{180}{270}}
\multirput(2,4)(2,0){3}{\pswedge[fillstyle=solid,fillcolor=red,linecolor=red](0,0){.25}{180}{270}}
\multirput(2,5)(2,0){3}{\pswedge[fillstyle=solid,fillcolor=red,linecolor=red](0,0){.25}{180}{270}}
\multirput(1,5)(0,5){2}{\pswedge[fillstyle=solid,fillcolor=red,linecolor=red](0,0){.25}{180}{270}}
\multirput(2,5)(0,5){2}{\pswedge[fillstyle=solid,fillcolor=red,linecolor=red](0,0){.25}{180}{270}}
\psgrid[gridlabels=0pt,subgriddiv=1]
\rput(.5,10.65){$\vdots$}\rput(1.5,10.65){$\vdots$}\rput(2.5,10.65){$\vdots$}\rput(3.5,10.65){$\vdots$}\rput(4.5,10.65){$\vdots$}\rput(5.5,10.65){$\vdots$}\rput(6.5,10.5){$\vvdots$}
\rput(.5,9.5){$\frac {27}5$}\rput(1.5,9.5){$\frac{91}{40}$}\rput(2.5,9.5){$\frac 25$}\rput(3.5,9.5){$-\frac{9}{40}$}\rput(4.5,9.5){$\frac 25$}\rput(5.5,9.5){$\frac{91}{40}$}\rput(6.5,9.5){$\cdots$}
\rput(.5,8.5){$4$}\rput(1.5,8.5){$\frac{11}{8}$}\rput(2.5,8.5){$0$}\rput(3.5,8.5){$-\frac{1}{8}$}\rput(4.5,8.5){$1$}\rput(5.5,8.5){$\frac{27}{8}$}\rput(6.5,8.5){$\cdots$}
\rput(.5,7.5){$\frac {14}5$}\rput(1.5,7.5){$\frac {27}{40}$}\rput(2.5,7.5){$-\frac 15$}\rput(3.5,7.5){$\frac{7}{40}$}\rput(4.5,7.5){$\frac 95$}\rput(5.5,7.5){$\frac{187}{40}$}\rput(6.5,7.5){$\cdots$}
\rput(.5,6.5){$\frac 95$}\rput(1.5,6.5){$\frac {7}{40}$}\rput(2.5,6.5){$-\frac 15$}\rput(3.5,6.5){$\frac{27}{40}$}\rput(4.5,6.5){$\frac {14}5$}\rput(5.5,6.5){$\frac{247}{40}$}\rput(6.5,6.5){$\cdots$}
\rput(.5,5.5){$1$}\rput(1.5,5.5){$-\frac {1}{8}$}\rput(2.5,5.5){$0$}\rput(3.5,5.5){$\frac{11}{8}$}\rput(4.5,5.5){$4$}\rput(5.5,5.5){$\frac{63}{8}$}\rput(6.5,5.5){$\cdots$}
\rput(.5,4.5){$\frac 25$}\rput(1.5,4.5){$-\frac 9{40}$}\rput(2.5,4.5){$\frac 25$}\rput(3.5,4.5){$\frac{91}{40}$}\rput(4.5,4.5){$\frac {27}5$}\rput(5.5,4.5){$\frac{391}{40}$}\rput(6.5,4.5){$\cdots$}
\rput(.5,3.5){$0$}\rput(1.5,3.5){$-\frac 18$}\rput(2.5,3.5){$1$}\rput(3.5,3.5){$\frac{27}8$}\rput(4.5,3.5){$7$}\rput(5.5,3.5){$\frac{95}8$}\rput(6.5,3.5){$\cdots$}
\rput(.5,2.5){$-\frac 15$}\rput(1.5,2.5){$\frac 7{40}$}\rput(2.5,2.5){$\frac 95$}\rput(3.5,2.5){$\frac{187}{40}$}\rput(4.5,2.5){$\frac{44}5$}\rput(5.5,2.5){$\frac{567}{40}$}\rput(6.5,2.5){$\cdots$}
\rput(.5,1.5){$-\frac 15$}\rput(1.5,1.5){$\frac {27}{40}$}\rput(2.5,1.5){$\frac{14}5$}\rput(3.5,1.5){$\frac{247}{40}$}\rput(4.5,1.5){$\frac{54}5$}\rput(5.5,1.5){$\frac{667}{40}$}\rput(6.5,1.5){$\cdots$}
\rput(.5,.5){$0$}\rput(1.5,.5){$\frac {11}8$}\rput(2.5,.5){$4$}\rput(3.5,.5){$\frac{63}8$}\rput(4.5,.5){$13$}\rput(5.5,.5){$\frac{155}8$}\rput(6.5,.5){$\cdots$}
{\color{blue}
\rput(.5,-.5){$1$}
\rput(1.5,-.5){$2$}
\rput(2.5,-.5){$3$}
\rput(3.5,-.5){$4$}
\rput(4.5,-.5){$5$}
\rput(5.5,-.5){$6$}
\rput(6.5,-.5){$r$}
\rput(-.5,.5){$1$}
\rput(-.5,1.5){$2$}
\rput(-.5,2.5){$3$}
\rput(-.5,3.5){$4$}
\rput(-.5,4.5){$5$}
\rput(-.5,5.5){$6$}
\rput(-.5,6.5){$7$}
\rput(-.5,7.5){$8$}
\rput(-.5,8.5){$9$}
\rput(-.5,9.5){$10$}
\rput(-.5,10.5){$s$}}
\end{pspicture}
\bigskip
\caption{The infinitely extended Kac table of the logarithmic Yang-Lee model ${\cal LM}(2,5)$ with $c=-\frac{22}{5}$. The Kac table of the rational Yang-Lee model appears in the bottom left corner. The boxes indicated with red quadrants are associated to irreducible representations of the Virasoro algebra.\label{YLKac}}
\end{center}
\end{figure}

In this paper, we develop further the implementation and analysis~\cite{PRZ,PRV} of $(r,s)$ Kac boundary conditions in the general logarithmic minimal models. These boundary conditions are constructed~\cite{BP96,BPO96,BP2001} using fusion~\cite{KulReshSkly} and are conjugate to the $(r,s)$ Virasoro Kac representations. Following~\cite{PRV}, the $(r,s)$ boundaries are constructed in (\ref{bdyK}) using an $r$-type seam of width $\rho-1$ columns where
\bea
\rho= \rho(r) = \Big\lfloor \frac{rp'}{p} \Big\rfloor,\qquad r=1,2,3,\ldots\label{logGS}
\eea
In \cite{PRZ}, the fusion construction was implemented using Wenzl-Jones projectors~\cite{WenzlJones,Wenzl}. Requiring the existence of Wenzl-Jones projectors restricts this construction to $r\le \rho<p'$.
Here the fusion is implemented diagrammatically following \cite{PRV,PRT2014,MDPR14} so that $\rho,r=1,2,3,\ldots$ are unrestricted. 
Working in a strip geometry, we construct numerical transfer matrices with $(r,s)$ boundary conditions with Kac labels $r,s=1,2,3,\ldots$. 
The boundary weights involve a boundary field $\xi$ which we specialize appropriately. 
Extracting the finite-size spectra numerically and extrapolating, we confirm that the $(r,s)$ boundary conditions indeed lead to the conformal weights (\ref{confWts}) and conformal partition functions (\ref{confZ}). 

The layout of the paper is as follows. To explain the genesis of (\ref{logGS}) we consider, in Section~\ref{GSseq}, the relation of the logarithmic minimal models to the rational minimal models both at criticality and off-criticality. In Section~\ref{logMinSect}, we define the logarithmic minimal lattice models in terms of the linear and planar Temperley-Lieb algebras~\cite{TL,Jones}. We also define the planar link states on which the transfer matrices act and present the construction~\cite{PRZ,PRV} of the $(r,s)$ boundary conditions using diagrammatic fusion. 
This construction involves generalized Temperley-Lieb projectors whose properties are proved in Appendix~\ref{genTLproj}. 
In Section~\ref{TMHamSect}, the commuting double row transfer matrices and the associated quantum Hamiltonians for $(r,s)$ boundary conditions are explicitly constructed. We discuss the role of the boundary field $\xi$ and its specialization. 
The equivalence of two expressions for the specialization of $\xi$ is established in Appendix~\ref{equivxi}. The bulk and boundary free energies are also derived in this section by the solving the inversion relation derived in Appendix~\ref{InversionRelation}. The full details of this calculation are presented in Appendices~\ref{analyticity}, \ref{laplace:boundary} and \ref{analyticcontinuation}. In Section~\ref{numericalSection}, we present our numerical results for the central charges, conformal weights and conformal partition functions. We conclude with some final remarks in Section~\ref{Conclusion}.

\subsection{Ground state sequences}
\label{GSseq}

The logarithmic minimal models ${\cal LM}(p,p')$ are obtained as a {\it logarithmic limit} of the rational minimal models ${\cal M}(m,m')$. The nonunitary minimal models ${\cal M}(m,m')$ arise as the continuum scaling limit of the Forrester-Baxter Restricted-Solid-On-Solid (RSOS) models~\cite{FB}. Symbolically, we write this limit 
as in (\ref{logLimitEqn}) where the logarithmic limit (taken after the thermodynamic limit) is independent of the choice of sequence with $2\le m\le m'$ and $m,m'$ coprime. Explicitly, the limiting CFT data are given by
\begin{align}
 &\qquad\qquad  c^{m,m'}=1-{6(m'-m)^2\over mm'}\ {\to}\  1-{6(p'-p)^2\over pp'}=c^{p,p'}\\
&\Delta_{r,s}^{m,m'}={(rm'-sm)^2-(m'-m)^2\over 4mm'}\ {\to}\ {(rp'-sp)^2-(p'-p)^2\over 4pp'}=\Delta_{r,s}^{p,p'}\\
\mbox{ch}^{m,m'}_{r,s}\!(q)&={q^{-{c\over 24}+\Delta_{r,s}^{m,m'}}\over (q)_\infty}\!\!\! 
\sum_{k=-\infty}^\infty \!\!\!\big[q^{k(k m m'+rm'-sm)}\!-\!q^{(km+r)(km'+s)}\big] {\to}\,\disp q^{-{c\over 24}+\Delta_{r,s}^{p,p'}}\,
{(1\!-\!q^{rs})\over (q)_\infty}=\chi^{p,p'}_{r,s}\!(q)
\end{align}
The logarithmic limit can also be applied to off-critical ($\varphi_{1,3}$-perturbed) minimal models ${\cal M}(m,m';t)$ to obtain off-critical logarithmic minimal models ${\cal LM}(p,p';t)$~\cite{OffCrit}. Combining the logarithmic limit with the off-critical perturbation gives the commutative diagram
\bea
\begin{pspicture}[shift=-.9](0,-.3)(4,2)
\rput(0,0){${\cal LM}(p,p')$}
\rput(4,0){$\ \ {\cal LM}(p,p';t)$}
\rput(0,1.5){${\cal M}(m,m')$}
\rput(4,1.5){$\ \ {\cal M}(m,m';t)$}
\rput(1.8,.3){$\;t\sim \varphi_{1,3}$}
\rput(1.8,1.8){$\;t\sim \varphi_{1,3}$}
\rput(-.4,.9){\mbox{log}}
\rput(3.6,.9){\mbox{log}}
\psline[linewidth=1pt,arrowsize=7pt,linecolor=black]{->}(1,0)(3,0)
\psline[linewidth=1pt,arrowsize=7pt]{->}(1,1.5)(3,1.5)
\psline[linewidth=1pt,arrowsize=7pt]{->}(0,1.2)(0,.3)
\psline[linewidth=1pt,arrowsize=7pt]{->}(4,1.2)(4,.3)
\end{pspicture}
\eea
where the elliptic nome $t$ is the departure-from-criticality variable. 

\begin{figure}[htb]
\centering
\subfloat{
\psset{unit=0.7cm}
\begin{pspicture}(-3,-0.5)(6,11)
\pspolygon[fillstyle=solid,fillcolor=lightlightblue](0,0)(5,0)(5,10)(0,10)
\pspolygon[fillstyle=solid,fillcolor=lightblue](0,1)(5,1)(5,2)(0,2)
\pspolygon[fillstyle=solid,fillcolor=lightblue](0,4)(5,4)(5,5)(0,5)
\pspolygon[fillstyle=solid,fillcolor=lightblue](0,6)(5,6)(5,7)(0,7)
\pspolygon[fillstyle=solid,fillcolor=lightblue](0,9)(5,9)(5,10)(0,10)
\multirput(0,0)(0,5){2}{\psline[linewidth=2pt](0,5)(5,5)}
\psgrid[gridlabels=0pt,subgriddiv=1](0,0)(5,10)
\rput(-0.4,0){1}\rput(-0.4,1){2}\rput(-0.4,2){3}\rput(-0.4,3){4}\rput(-0.4,4){5}\rput(-0.4,5){6}\rput(-0.4,6){7}\rput(-0.4,7){8}\rput(-0.4,8){9}\rput(-0.4,9){10}\rput(-0.4,10){11}
\rput(-0.4,10.75){$\rho$}
\rput(5.4,1.5){$1$}\rput(5.4,4.5){$2$}\rput(5.4,6.5){$3$}\rput(5.4,9.5){$4$}
\rput(5.4,10.75){$r$}
\rput{90}(-0.9,2.5){$\overbrace{\hspace{3.5cm}}$}
\rput{90}(-0.9,7.5){$\overbrace{\hspace{3.5cm}}$}
\multirput(-1.9,2.5)(0,5){2}{$p'=5$}
\end{pspicture}
}
\subfloat{
\psset{unit=0.5cm}
\begin{pspicture}(-4.5,-0.75)(8,15.2)
\pspolygon[fillstyle=solid,fillcolor=lightlightblue](0,0)(7,0)(7,14)(0,14)
\pspolygon[fillstyle=solid,fillcolor=lightblue](0,0)(7,0)(7,1)(0,1)
\pspolygon[fillstyle=solid,fillcolor=lightblue](0,2)(7,2)(7,3)(0,3)
\pspolygon[fillstyle=solid,fillcolor=lightblue](0,4)(7,4)(7,5)(0,5)
\pspolygon[fillstyle=solid,fillcolor=lightblue](0,6)(7,6)(7,7)(0,7)
\pspolygon[fillstyle=solid,fillcolor=lightblue](0,7)(7,7)(7,8)(0,8)
\pspolygon[fillstyle=solid,fillcolor=lightblue](0,9)(7,9)(7,10)(0,10)
\pspolygon[fillstyle=solid,fillcolor=lightblue](0,11)(7,11)(7,12)(0,12)
\pspolygon[fillstyle=solid,fillcolor=lightblue](0,13)(7,13)(7,14)(0,14)
\multirput(0,0)(0,7){2}{\psline[linewidth=1.6pt](0,7)(7,7)}
\psgrid[gridlabels=0pt,subgriddiv=1](0,0)(7,14)
\rput(-0.6,0){1}\rput(-0.6,1){2}\rput(-0.6,2){3}\rput(-0.6,3){4}\rput(-0.6,4){5}\rput(-0.6,5){6}\rput(-0.6,6){7}\rput(-0.6,7){8}\rput(-0.6,8){9}\rput(-0.6,9){10}\rput(-0.6,10){11}\rput(-0.6,11){12}\rput(-0.6,12){13}\rput(-0.6,13){14}\rput(-0.6,14){15}
\rput(-0.6,15){$\rho$}
\rput(7.5,0.5){$1$}\rput(7.5,2.5){$2$}
\rput(7.5,4.5){$3$}\rput(7.5,6.5){$4$}\rput(7.5,7.5){$5$}\rput(7.5,9.5){$6$}\rput(7.5,11.5){$7$}\rput(7.5,13.5){$8$}
\rput(7.5,15){$r$}
\rput{90}(-1.3,3.5){$\overbrace{\hspace{3.5cm}}$}
\rput{90}(-1.3,10.5){$\overbrace{\hspace{3.5cm}}$}
\multirput(-2.7,3.5)(0,7){2}{$p'=7$}
\end{pspicture}
}
\caption{The shaded bands between heights $\rho$ and $\rho+1$ indicate the ``ground state" values of $\rho=\big\lfloor \frac{rp'}{p} \big\rfloor$ for ${\cal LM}(p,p')$ with $(p,p')=(2,5), (4,7)$ respectively. The patterns of shaded bands repeat periodically with a period $p'$.}
\label{fig:BandsofRho}
\end{figure}
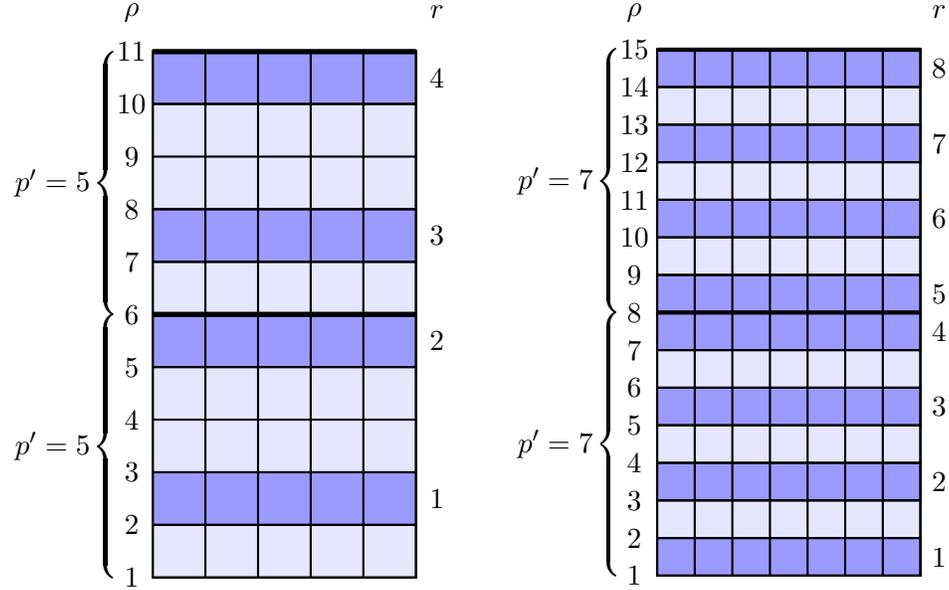

We recall that the $2(m-1)$ ground states of the off-critical RSOS model ${\cal M}(m,m',t)$ in Regime~III are given~\cite{FB,FodaW} by flat configurations on the square lattice where the heights on the two independent sublattices alternate between the values $\rho$ and $\rho+1$ with
\bea
\rho= \rho(r) = \Big\lfloor \frac{rm'}{m} \Big\rfloor,\qquad r=1,2,3,\ldots,m-1\label{minGS}
\eea
Taking the logarithmic limit of (\ref{minGS}) gives (\ref{logGS}) corresponding to the ground states of the off-critical logarithmic minimal models. In accord with the general {\it
correspondence principle} of the Kyoto school~\cite{Kyoto} relating off-critical one-dimensional configurational sums with finitized conformal characters, these ground state labels should also occur at criticality as the labels of the conformal boundary conditions. 
Two example sequences of $\rho$ values for prototypical logarithmic minimal models are
\begin{align}
{\cal LM}(2,5):\qquad& \rho(r)= \Big\lfloor \frac{5 r}{2} \Big\rfloor = 2,5,7,10,12,15,17,20,\ldots\\
{\cal LM}(4,7):\qquad& \rho(r)= \Big\lfloor \frac{7 r}{4} \Big\rfloor = 1, 3, 5, 7, 8, 10, 12, 14,\ldots
\end{align}
The sequences of consecutive differences $\rho(r+1)-\rho(r)$, and therefore the pattern of ground states, are periodic with a period $p'$ as shown by the shaded bands in Figure~\ref{fig:BandsofRho}. The value of $\rho$ fixes one possible choice for the number of columns in the $r$-type seam to obtain an $(r,s)$ boundary condition with Kac label $r$. 

The choice of $\rho$ to realize a particular value of the conformal label $r$, however, is not necessarily unique. More generally, the value of $r$ observed for given $\rho$ is given by the sequences
\bea
r= r(\rho) = \Big\lceil \frac{\rho p}{p'}\Big\rceil, \qquad \rho=1,2,3,\ldots
\eea
Two example sequences $r=r(\rho)$ for prototypical logarithmic minimal models are
\begin{align}
{\cal LM}(2,5):\qquad& r(\rho)= \Big\lceil \frac{2\rho}{5} \Big\rceil = 1,1,2,2,2,3,3,4,4,4,\ldots\\
{\cal LM}(4,7):\qquad& r(\rho)= \Big\lceil \frac{4\rho}{7} \Big\rceil = 1, 2, 2, 3, 3, 4, 4, 5, 6, 6, 7, 7, 8, 8, \ldots
\end{align}
In general, this means that multiple values of $\rho$ can correspond to the same value of $r$ so each conformal boundary condition can have multiple realizations on the lattice. 
We observe that the last occurrence of a given value of $r$ occurs at the position $\rho(r)$ corresponding to the ground state value. 
On the other hand, the first occurrence of a given value of $r$ occurs at positions given by the third sequence
\bea
\rho_-(r) = \Big\lfloor \frac{(r-1)p'}{p}\Big\rfloor + 1
\eea
Two example sequences $\rho_-=\rho_-(r)$ for prototypical logarithmic minimal models are
\begin{align}
{\cal LM}(2,5):\qquad& \rho_-(r)= \Big\lfloor \frac{5 (r-1)}{2}\Big\rfloor + 1 = 1,3,6,8,\ldots\\
{\cal LM}(4,7):\qquad& \rho_-(r)= \Big\lfloor \frac{7 (r-1)}{4}\Big\rfloor + 1 = 1, 2, 4, 6, 8, 9, 11, 13,\ldots
\end{align}
The relationship between these three sequences is illustrated in the following tables
\begin{align}
{\cal LM}(2,5):&\qquad
	  \mbox{\begin{tabular}{|c|c|c|c|c|c|c|c|c|c|c|}
    	\hline
	    ${\color{blue}\rho(r)},{\color{red}\rho_-(r)}$ &\color{red}\bf1&$\color{blue}\mathbf{2}	$&\color{red}\bf3&\textcolor{gray}4&$\color{blue}\mathbf{5}$&\color{red}\bf6&$\color{blue}\mathbf{7}$&\color{red}\bf8&\textcolor{gray}9 &$\color{blue}\mathbf{10}$ \\ 			\hline
	    $r$          &1&1&2&2&2&3&3&4&4&4 \\
    	\hline
	  \end{tabular}}\\[4pt]
{\cal LM}(4,7):&\qquad
	  \mbox{\begin{tabular}{|c|c|c|c|c|c|c|c|c|c|c|c|c|c|c|}
    	\hline
	    ${\color{blue}\rho(r)},{\color{red}\rho_-(r)}$ &{\color{blue}\bf1}&\color{red}\bf2&{\color{blue}\bf 3}&\color{red}\bf4&{\color{blue}\bf 5}&\color{red}\bf6&{\color{blue}\bf 7}&{\color{blue} \bf8}&\color{red}\bf9&{\color{blue}\bf 10}&\color{red}\bf11&{\color{blue}\bf 12}&\color{red}\bf13&{\color{blue}\bf 14}\\ 			\hline
	    $r$ &1&2&2&3&3&4&4&5&6&6&7&7&8&8          \\
    	\hline
	  \end{tabular}}
\end{align}
The bottom row in these tables give the sequence of $r(\rho)$ values. The top row in these tables gives the values of $\rho$ coloured to indicate the values in the subsequences 
$\rho(r)$ (blue) and $\rho_-(r)$ (red). Entries corresponding to singletons (which are members of both subsequences) are also coloured in blue. Values of $\rho$ corresponding to other repeated values of $r$ are shown in grey. 

\section{Logarithmic Minimal Lattice Models}
\label{logMinSect}

\subsection{Face operators and local relations}
The planar Temperley-Lieb (TL) algebra~\cite{TL,Jones}, is a diagrammatic algebra generated by the two tiles
\bea
\psset{unit=.95cm}
\begin{pspicture}[shift=-.45](1,1)
\facegrid{(0,0)}{(1,1)}
\rput[bl](0,0){\loopa}
\end{pspicture}\qquad\qquad
\begin{pspicture}[shift=-.42](1,1)
\facegrid{(0,0)}{(1,1)}
\rput[bl](0,0){\loopb}
\end{pspicture}
\eea
Within the planar algebra, these tiles are multiplied together (in arbitrary directions) by connecting the nodes at the midpoints of the edges of the faces thereby forming a planar web of connectivities. 

The face operators~\cite{PRZ} for the logarithmic minimal models are defined by
\bea
\psset{unit=.95cm}
\begin{pspicture}[shift=-.42](1,1)
\facegrid{(0,0)}{(1,1)}
\rput(.5,.5){$u$}
\psarc[linewidth=1pt,linecolor=red](0,0){.15}{0}{90}
\end{pspicture}\;=\;
s(\lambda-u)\;\begin{pspicture}[shift=-.45](1,1)
\facegrid{(0,0)}{(1,1)}
\rput[bl](0,0){\loopa}
\end{pspicture}\;+\,s(u)\;
\begin{pspicture}[shift=-.42](1,1)
\facegrid{(0,0)}{(1,1)}
\rput[bl](0,0){\loopb}
\end{pspicture}\;,\qquad s(u) = \frac{\sin u}{\sin\lambda}
 \label{eq:face1}
\eea
The face operators satisfy the Yang-Baxter equation \cite{BaxBook}
\be
\begin{array}{rcl}
\psset{unit=.36cm}
\begin{pspicture}[shift=-1.8](0,0)(7,3.5)
\pspolygon[linewidth=1pt,linecolor=black,fillstyle=solid,fillcolor=lightlightblue](0,2)(2,0)(5,0)(7,2)(5,4)(2,4)(0,2)
\pspolygon[linewidth=1pt,linecolor=black,fillstyle=solid,fillcolor=lightlightblue](0,2)(3,2)(5,0)(7,2)(5,4)(3,2)
\rput(2.5,1){$u$}
\rput(2.5,3){$v$}
\rput(5,2){$v\!-\!u$}
\psarc[linewidth=1pt,linecolor=red](2,0){.3}{0}{135}
\psarc[linewidth=1pt,linecolor=red](0,2){.5}{0}{45}
\psarc[linewidth=1pt,linecolor=red](3,2){.4}{-45}{45}
\end{pspicture}\ &=&
\psset{unit=.375cm}
\begin{pspicture}[shift=-1.8](0,0)(7,3.5)
\pspolygon[linewidth=1pt,linecolor=black,fillstyle=solid,fillcolor=lightlightblue](0,2)(2,0)(5,0)(7,2)(5,4)(2,4)(0,2)
\pspolygon[linewidth=1pt,linecolor=black,fillstyle=solid,fillcolor=lightlightblue](7,2)(4,2)(2,0)(0,2)(2,4)(4,2)
\rput(4.5,1){$v$}
\rput(4.5,3){$u$}
\rput(2,2){$v\!-\!u$}
\psarc[linewidth=1pt,linecolor=red](0,2){.4}{-45}{45}
\psarc[linewidth=1pt,linecolor=red](2,0){.5}{0}{45}
\psarc[linewidth=1pt,linecolor=red](4,2){.3}{0}{135}
\end{pspicture}
\end{array}
\label{YBE}
\ee
and the local inversion and crossing relations
\be
\psset{unit=0.65cm}
\begin{pspicture}[shift=-1.13](-.5,0.75)(4,3.25)
\pspolygon[fillstyle=solid,fillcolor=lightlightblue](0,2)(1,1)(2,2)(1,3)(0,2)
\pspolygon[fillstyle=solid,fillcolor=lightlightblue](2,2)(3,1)(4,2)(3,3)(2,2)
\psarc[linewidth=1pt,linecolor=red](0,2){.2}{-45}{45}
\psarc[linewidth=1pt,linecolor=red](2,2){.2}{-45}{45}
\psarc[linecolor=blue,linewidth=1.5pt](2,2){.7}{45}{135}
\psarc[linecolor=blue,linewidth=1.5pt](2,2){.7}{-135}{-45}
\rput(1,2){\small $u$}
\rput(3,2){\small $\!-u$}
\end{pspicture}
=s(\lambda-u)s(\lambda+u)\
\begin{pspicture}[shift=-1.13](1,0.75)(3.2,3.25)
\pspolygon[fillstyle=solid,fillcolor=lightlightblue](1,2)(2,1)(3,2)(2,3)(1,2)
\psarc[linecolor=blue,linewidth=1.5pt](2,1){.7}{45}{135}
\psarc[linecolor=blue,linewidth=1.5pt](2,3){.7}{-135}{-45}
\end{pspicture},\qquad \quad 
\psset{unit=.95cm}
\begin{pspicture}[shift=-.42](1,1)
\facegrid{(0,0)}{(1,1)}
\rput(.5,.5){$u$}
\psarc[linewidth=1pt,linecolor=red](0,0){.15}{0}{90}
\end{pspicture}\ =\ 
\begin{pspicture}[shift=-.42](1,1)
\facegrid{(0,0)}{(1,1)}
\rput(.5,.5){$\lambda\!-\!u$}
\psarc[linewidth=1pt,linecolor=red](1,0){.15}{90}{180}
\end{pspicture}
\label{Invrel}
\ee

\subsection{Linear Temperley-Lieb algebra}
Fixing the direction for multiplication in the planar algebra leads to the loop representation of the corresponding linear TL algebra.
The linear Temperley-Lieb algebra~\cite{TL}  ${\cal T\!L}(x;N)$ is a one-parameter algebra generated by the  identity $I$ and the monoids $e_j$, $j=1,\ldots,N$  subject to the relations
\bea
e_j^2&=&\TLb e_j \label{eq:TL1}\\
e_je_{j\pm 1}e_j&=&e_j \label{eq:TL2}\\
e_ie_j&=&e_je_i\qquad\quad |i-j|\ge 2 \label{eq:TL3}
\eea
The parameter $\beta=x+x^{-1}=2\cos\lambda$ is the {\em loop fugacity} where $x=e^{i\lambda}$ is a complex phase. 
For generic loop models, the crossing parameter $\lambda\in{\Bbb R}$ is an arbitrary real number. In contrast, for the logarithmic minimal models ${\cal LM}(p,p')$, $\lambda=\frac{(p'-p)\pi}{p'}$ is restricted to rational fractions of $\pi$ with $p,p'$ coprime. Since critical dense polymers ${\cal LM}(1,2)$ has been solved analytically~\cite{PRV} for all $(r,s)$ boundary conditions, we will assume throughout this paper that $\lambda\ne\frac{\pi}{2}$ so that $\beta\ne 0$.

A faithful representation of the (linear) TL algebra is given by the loop representation with generators 
\be
I = \;
\psset{unit=.7cm}
\begin{pspicture}[shift=-.75](0,-.45)(3.5,1)
\rput[B](0,-.5){1}\rput[B](1,-.5){2}\rput[B](2.5,-.5){$N\!-\!1$}\rput[B](3.5,-.5){$N$}
\rput(1.75,.5){\ldots}
\multirput(0,0)(2.5,0){2}{\multirput(0,0)(1,0){2}{\psline(0,0)(0,1)}}
\end{pspicture}\: , \qquad\qquad
e_j=\;
\begin{pspicture}[shift=-.75](0,-.45)(8,1)
\rput[B](0,-.5){1}\rput[B](1,-.5){2}\rput[B](3.5,-.5){$j$}\rput[B](4.5,-.5){$j\!+\!1$}\rput[B](7,-.5){$N\!-\!1$}\rput[B](8,-.5){$N$}
\multirput(1.75,.5)(4.5,0){2}{\ldots}
\multirput(0,0)(7,0){2}{\multirput(0,0)(1,0){2}{\psline(0,0)(0,1)}}
\multirput(2.5,0)(3,0){2}{\psline(0,0)(0,1)}
\rput(3.5,0){\Monoid}
\end{pspicture}
\ee
which act diagrammatically by vertical concatenation on a set of $N$ parallel strings. In diagrams, closed loops are removed and replaced with the scalar loop fugacity $\TLb$. For example, diagrammatically, relation \eqref{eq:TL1} becomes
\be
\psset{unit=.7cm}
\begin{pspicture}[shift=-1.25](0,-.45)(8,2)
\rput[B](0,-.5){1}\rput[B](1,-.5){2}\rput[B](3.5,-.5){$j$}\rput[B](4.5,-.5){$j\!+\!1$}\rput[B](7,-.5){$N\!-\!1$}\rput[B](8,-.5){$N$}
\multirput(1.75,1)(4.5,0){2}{\ldots}
\multirput(0,0)(0,1){2}{
\multirput(0,0)(7,0){2}{\multirput(0,0)(1,0){2}{\psline(0,0)(0,1)}}
\multirput(2.5,0)(3,0){2}{\psline(0,0)(0,1)}
\rput(3.5,0){\Monoid}
}
\end{pspicture}
\ =\  \TLb\ \;
\begin{pspicture}[shift=-.75](0,-.45)(8,1)
\rput[B](0,-.5){1}\rput[B](1,-.5){2}\rput[B](3.5,-.5){$j$}\rput[B](4.5,-.5){$j\!+\!1$}\rput[B](7,-.5){$N\!-\!1$}\rput[B](8,-.5){$N$}
\multirput(1.75,.5)(4.5,0){2}{\ldots}
\multirput(0,0)(7,0){2}{\multirput(0,0)(1,0){2}{\psline(0,0)(0,1)}}
\multirput(2.5,0)(3,0){2}{\psline(0,0)(0,1)}
\rput(3.5,0){\Monoid}
\end{pspicture}
\ee
The loop fugacity $\beta$ is a scalar weight assigned to closed loops
\bea
\psset{unit=.75cm}
\TLb=\begin{pspicture}[shift=-.4](0,0)(1,1)
\pscircle(.5,.5){.5}
\end{pspicture}
= \begin{pspicture}[shift=-.4](0,0)(1,1)
\psarc[arrowsize=6pt,arrowlength=1]{->}(.5,.5){.5}{10}{380}
\end{pspicture}
\;+\begin{pspicture}[shift=-.4](0,0)(1,1)
\psarc[arrowsize=6pt,arrowlength=1]{<-}(.5,.5){.5}{-25}{375}
\end{pspicture}\;
=x+x^{-1}=2\cos\lambda
\label{fugacities}
\eea
In the linear TL algebra the face operators, inversion relation and Yang-Baxter equation take the respective forms
\bea
\begin{array}{c}
X_j(u)=s(\lambda-u) I+s(u) e_j,\qquad X_j(u)X_j(-u)=s(\lambda-u)s(\lambda+u) I\\[10pt]
X_j(u)X_{j+1}(u+v)X_j(v)=X_{j+1}(v)X_j(u+v)X_{j+1}(u)
\end{array}
\eea

\subsection{Link states}
\label{LinkStates}
The TL link states are planar pairwise matchings of nodes. These form vector spaces by allowing arbitrary linear combinations. On a strip, the allowed link states depend on the choice of boundary conditions. We denote by $\mathcal{V}^{(N)}_{\rho,s}$ the vector space of link states with $N$ bulk sites in the presence of a boundary on the right side formed by an $r$-type seam of width $\rho-1$ columns and an $s$-type seam of width $s-1$ columns. The total width of the strip is then $N+\rho+s-2$. The labels $\rho$ and $s$ are quantum numbers labelling sectors of the theory but there may be more than one value of $\rho$ corresponding to a given value of $r$.
The simplest case is the {\em vacuum} $(r,s)=(1,1)$ sector with $\rho=r=1$. The 5 link states for bulk size $N=6$ are 
\bea
\psset{unit=0.25cm}
\begin{pspicture}(6,4)
\psarc[linecolor=black,linewidth=1.5pt](-1,0){1}{0}{180}
\psarc[linecolor=black,linewidth=1.5pt](3,0){1}{0}{180}
\psarc[linecolor=black,linewidth=1.5pt](7,0){1}{0}{180}
\rput(9,-.2){,}
\end{pspicture}\qquad\qquad
\begin{pspicture}(6,4)
\psarc[linecolor=black,linewidth=1.5pt](-1,0){1}{0}{180}
\psarc[linecolor=black,linewidth=1.5pt](5,0){3}{0}{180}
\psarc[linecolor=black,linewidth=1.5pt](5,0){1}{0}{180}
\rput(9,-.2){,}
\end{pspicture}\qquad\qquad
\begin{pspicture}(6,4)
\psarc[linecolor=black,linewidth=1.5pt](1,0){3}{0}{180}
\psarc[linecolor=black,linewidth=1.5pt](1,0){1}{0}{180}
\psarc[linecolor=black,linewidth=1.5pt](7,0){1}{0}{180}
\rput(9,-.2){,}
\end{pspicture}\qquad\qquad
\begin{pspicture}(6,4)
\psarc[linecolor=black,linewidth=1.5pt](3,0){5}{0}{180}
\psarc[linecolor=black,linewidth=1.5pt](1,0){1}{0}{180}
\psarc[linecolor=black,linewidth=1.5pt](5,0){1}{0}{180}
\rput(9,-.2){,}
\end{pspicture}\qquad\qquad
\begin{pspicture}(6,4)
\psarc[linecolor=black,linewidth=1.5pt](3,0){5}{0}{180}
\psarc[linecolor=black,linewidth=1.5pt](3,0){3}{0}{180}
\psarc[linecolor=black,linewidth=1.5pt](3,0){1}{0}{180}
\end{pspicture}
\eea
For even $N$, the number of such link states are given by Catalan numbers 
\be
\mbox{dim}\,\mathcal{V}^{(N)}_{1,1}={1\over n+1} \binom{2n}{n}, \qquad n=\half N
\ee

Moving to nontrivial $(r,s)$ boundary conditions means introducing $r$- and $s$-type seams on the right of the strip. This yields $N+\rho+s-2$ columns and thus $N+\rho+s-2$ nodes for the link states. As discussed in the next subsection, we thus need to impose the diagrammatic rule that there are no half-arcs closing within the $r$- or $s$-type seam. Half-arcs are allowed to close, however, between the $r$- and $s$-type seams. For $\rho=s=3$ and $N=4$, for example, the 6 allowed link states in $\mathcal{V}^{(4)}_{3,3}$ are
\bea
\begin{array}{c}
\psset{unit=.5cm}
\begin{pspicture}(7,4)
\psarc[linecolor=black,linewidth=1.5pt](.5,0){.5}{0}{180}
\psarc[linecolor=black,linewidth=1.5pt](3.5,0){.5}{0}{180}
\psarc[linecolor=black,linewidth=1.5pt](5.5,0){.5}{0}{180}
\psarc[linecolor=black,linewidth=1.5pt](4.5,0){2.5}{0}{180}
\psline[linewidth=1.pt,linecolor=red,linestyle=dashed](3.5,-.1)(3.5,3.8)
\psline[linewidth=1.pt,linecolor=red,linestyle=dashed](5.5,-.1)(5.5,3.8)
\end{pspicture}\qquad\qquad
\begin{pspicture}(7,4)
\psarc[linecolor=black,linewidth=1.5pt](1.5,0){.5}{0}{180}
\psarc[linecolor=black,linewidth=1.5pt](3.5,0){.5}{0}{180}
\psarc[linecolor=black,linewidth=1.5pt](5.5,0){.5}{0}{180}
\psarc[linecolor=black,linewidth=1.5pt](3.5,0){3.5}{0}{180}
\psline[linewidth=1.pt,linecolor=red,linestyle=dashed](3.5,-.1)(3.5,3.8)
\psline[linewidth=1.pt,linecolor=red,linestyle=dashed](5.5,-.1)(5.5,3.8)
\end{pspicture}\qquad\qquad
\begin{pspicture}(7,4)
\psarc[linecolor=black,linewidth=1.5pt](2.5,0){.5}{0}{180}
\psarc[linecolor=black,linewidth=1.5pt](2.5,0){1.5}{0}{180}
\psarc[linecolor=black,linewidth=1.5pt](5.5,0){.5}{0}{180}
\psarc[linecolor=black,linewidth=1.5pt](3.5,0){3.5}{0}{180}
\psline[linewidth=1.pt,linecolor=red,linestyle=dashed](3.5,-.1)(3.5,3.8)
\psline[linewidth=1.pt,linecolor=red,linestyle=dashed](5.5,-.1)(5.5,3.8)
\end{pspicture}
\\[0pt]
\psset{unit=.5cm}
\begin{pspicture}(7,4)
\psarc[linecolor=black,linewidth=1.5pt](.5,0){.5}{0}{180}
\psarc[linecolor=black,linewidth=1.5pt](2.5,0){.5}{0}{180}
\psarc[linecolor=black,linewidth=1.5pt](5.5,0){.5}{0}{180}
\psarc[linecolor=black,linewidth=1.5pt](5.5,0){1.5}{0}{180}
\psline[linewidth=1.pt,linecolor=red,linestyle=dashed](3.5,-.1)(3.5,3.8)
\psline[linewidth=1.pt,linecolor=red,linestyle=dashed](5.5,-.1)(5.5,3.8)
\end{pspicture}\qquad\qquad
\begin{pspicture}(7,4)
\psarc[linecolor=black,linewidth=1.5pt](1.5,0){.5}{0}{180}
\psarc[linecolor=black,linewidth=1.5pt](1.5,0){1.5}{0}{180}
\psarc[linecolor=black,linewidth=1.5pt](5.5,0){.5}{0}{180}
\psarc[linecolor=black,linewidth=1.5pt](5.5,0){1.5}{0}{180}
\psline[linewidth=1.pt,linecolor=red,linestyle=dashed](3.5,-.1)(3.5,3.8)
\psline[linewidth=1.pt,linecolor=red,linestyle=dashed](5.5,-.1)(5.5,3.8)
\end{pspicture}\qquad\qquad
\begin{pspicture}(7,4)
\psarc[linecolor=black,linewidth=1.5pt](3.5,0){.5}{0}{180}
\psarc[linecolor=black,linewidth=1.5pt](3.5,0){1.5}{0}{180}
\psarc[linecolor=black,linewidth=1.5pt](3.5,0){2.5}{0}{180}
\psarc[linecolor=black,linewidth=1.5pt](3.5,0){3.5}{0}{180}
\psline[linewidth=1.pt,linecolor=red,linestyle=dashed](3.5,-.1)(3.5,3.8)
\psline[linewidth=1.pt,linecolor=red,linestyle=dashed](5.5,-.1)(5.5,3.8)
\end{pspicture}
\end{array}
\label{linkstates}
\eea
where the red dashed lines delimit the boundary seams on the right. 
In general, the counting of link states is given by the difference of two binomial coefficients
\be
\dim \mathcal{V}^{(N)}_{\rho,s} = \smbin{N}{\rule{0pt}{12pt}{N-\rho+s\over 2}}-\smbin{N}{\rule{0pt}{12pt}{N-\rho-s\over 2}}
\label{countstates}
\ee
where the total number of nodes $N+\rho+s-2$ is even.

\subsection{Boundary operator construction using diagrammatic fusion}
\label{sec:bdyK}
In this section, we recall the construction~\cite{PRZ,PRV} of the boundary operator $ K^{(\rho,s)}(u,\xi)$
\bea
\psset{unit=1.185cm}
\setlength{\unitlength}{1.185cm}
{\begin{pspicture}[shift=-1.5](9.5,0)(5.,3.2)
\put(1,.5){\color{lightlightblue}\rule{5\unitlength}{2\unitlength}}
\put(0.45,1.5){\makebox(0,0)[]{$=$}}
\put(.45,2.83){\pos{bc}{\color{blue}=}}
\put(-.1,2.8){\pos{bc}{\color{blue}(r,s)}}
\put(3.9,2.8){\pos{bc}{\color{blue}(r,1)}}
\put(6,2.86){\pos{bc}{\color{blue}\otimes}}
\put(-0.1,0.5){\line(0,1){2}}
\multiput(1,0.5)(1,0){4}{\line(0,1){2}}
\multiput(5,0.5)(1,0){2}{\line(0,1){2}}
\multiput(1,0.5)(0,1){3}{\line(1,0){5}}
\put(-0.6,1.5){\line(1,2){0.5}}
\put(-0.6,1.5){\line(1,-2){0.5}}
\pspolygon[linewidth=1pt,fillstyle=solid, fillcolor=lightlightblue](-0.6,1.5)(-.1,2.5)(-.1,.5)
\psarc[linewidth=1pt,linecolor=red](5,.5){.15}{0}{90}
\psarc[linewidth=1pt,linecolor=red](5,1.5){.15}{0}{90}
\psarc[linewidth=1pt,linecolor=red](1,.5){.15}{0}{90}
\psarc[linewidth=1pt,linecolor=red](1,1.5){.15}{0}{90}
\psarc[linewidth=1pt,linecolor=red](2,.5){.15}{0}{90}
\psarc[linewidth=1pt,linecolor=red](2,1.5){.15}{0}{90}
\put(1.5,1){\pos{}{u\!-\!\xi_{\rho\!-\!1}}}
\put(2.5,1){\pos{}{u\!-\!\xi_{\rho\!-\!2}}}
\put(5.5,1){\pos{}{u\!-\!\xi_1}}
\put(1.5,2){\pos{}{-\!u\!\!-\!\!\xi_{\rho\!-\!2}}}
\put(2.5,2){\pos{}{-\!u\!\!-\!\!\xi_{\rho\!-\!3}}}
\multirput(0,0)(0,1){2}{\multirput(0,0)(1,0){2}{\rput(3.5,1){$\cdots$}}}
\put(5.5,2){\pos{}{-\!u\!-\!\xi_0}}
\put(-0.31,1.5){\spos{}{u,\xi}}
\multiput(0,0)(1,0){5}{\psarc[linewidth=2pt,linecolor=blue](.5,2.5){1}{0}{40}}
\multiput(0,0)(1,0){5}{\psline[linewidth=2pt,linecolor=blue](1.5,.3)(1.5,.5)}
\psline[linewidth=2pt,linecolor=blue](.8,1)(1,1)
\psline[linewidth=2pt,linecolor=blue](.8,2)(1,2)
\psline[linewidth=2pt,linecolor=blue](-.55,1)(-.35,1)
\psline[linewidth=2pt,linecolor=blue](-.55,2)(-.35,2)
\psline[linewidth=3pt,linecolor=Blue](1,.5)(6,.5)
\end{pspicture}}
{\begin{pspicture}[shift=-1.5](6,0)(4.5,3.2)
\put(1,.5){\color{lightpurple}\rule{5\unitlength}{2\unitlength}}
\put(2.9,2.8){\pos{bc}{\color{blue}(1,s)}}
\put(6.5,2.8){\pos{bc}{\color{blue}(1,1)}}
\put(6,2.86){\pos{bc}{\color{blue}\otimes}}
\put(6.5,0.5){\line(0,1){2}}
\multiput(1,0.5)(1,0){4}{\line(0,1){2}}
\multiput(5,0.5)(1,0){2}{\line(0,1){2}}
\multiput(1,0.5)(0,1){3}{\line(1,0){5}}
\put(6,1.5){\line(1,2){0.5}}
\put(6,1.5){\line(1,-2){0.5}}
\pspolygon[linewidth=1pt,fillstyle=solid, fillcolor=lightlightblue](6,1.5)(6.5,2.5)(6.5,.5)
\multiput(0,0)(1,0){5}{\psarc[linewidth=2pt,linecolor=blue](.5,2.5){1}{0}{40}}
\multiput(0,0)(1,0){5}{\psline[linewidth=2pt,linecolor=blue](1.5,.3)(1.5,.5)}
\multiput(6,0.5)(0,2){2}{\makebox(0.5,0){\dotfill}}
\psarc[linewidth=1.5pt,linecolor=blue](5.7,1.5){.58}{-61}{61}
\rput[bl](1,.5){\loopa}
\rput[bl](2,.5){\loopa}
\rput[bl](3,.5){\loopa}
\rput[bl](4,.5){\loopa}
\rput[bl](5,.5){\loopa}
\rput[bl](1,1.5){\loopb}
\rput[bl](2,1.5){\loopb}
\rput[bl](3,1.5){\loopb}
\rput[bl](4,1.5){\loopb}
\rput[bl](5,1.5){\loopb}
\psline[linewidth=2pt,linecolor=red,linestyle=dashed](1,-.2)(1,2.7)
\psline[linewidth=3pt,linecolor=Blue](1,.5)(6,.5)
\end{pspicture}}\\[-14pt]
\mbox{}
\underbrace{\hspace{5.9cm}}_{\mbox{\small $\rho-1$}}\underbrace{\hspace{5.9cm}}_{\mbox{\small $s-1$}}\nonumber
\label{dressedVac}
\eea
where $\xi_n=\xi+n\lambda$ are column inhomogeneities. Following the usual construction~\cite{BP96,BPO96,BP2001}, these boundary triangle weights satisfy the Boundary Yang-Baxter Equation (BYBE) so, in this sense, they implement integrable $(r,s)$ boundary conditions. Usually, fusion is implemented, along the lower edge of each seam, using Wenzl-Jones projectors \cite{WenzlJones,Wenzl}. 
These act to project out configurations with closed half-arcs on the lower edges of the seams. 
These projectors, however, do not exist for $\rho\ge p'$. To apply fusion for $\rho\ge p'$, we instead use a diagrammatic implementation of fusion.
On the lower edge of the boundary, we impose a diagrammatic rule that forbids closed half-arcs within either the $r$-type or $s$-type seam. There can, however, be closed half-arcs linking the $r$- to the $s$-type boundary seam. The diagrammatic fusion rule can be pushed through from the bottom row to the top row of each seam where it acts to restrict the allowed link states as in Section~\ref{LinkStates}.

To fully describe diagrammatic fusion, let us next recall the push-through property. If there is a closed half arc anywhere in a TL link state, then there must be a small closed half arc between neighbouring nodes somewhere in the link state. Such small external closed half-arcs have a drop-down property. Specifically, if there is a small half-arc in an in-link state at the top acted upon by a seam then, by a simple trigonometric identity, there must also be a small closed half-arc in the out-link state at the bottom of the seam
\psset{unit=.8cm}
\bea
&&\begin{pspicture}[shift=-.67](-.2,-.3)(2.2,1.8)
\psarc[linewidth=1.5pt,linecolor=blue](1,1){.5}{0}{180}
\psline[linewidth=1.5pt,linecolor=blue](-.3,.5)(2.3,.5)
\psline[linewidth=1.5pt,linecolor=blue](.5,-.3)(.5,.3)
\psline[linewidth=1.5pt,linecolor=blue](1.5,-.3)(1.5,.3)
\facegrid{(0,0)}{(2,1)}
\rput[B](.5,.4){\small$u\!\!-\!\!\lambda$}
\rput[B](1.5,.4){\small $u$}
\psarc[linewidth=1pt,linecolor=red](0,0){.15}{0}{90}
\psarc[linewidth=1pt,linecolor=red](1,0){.15}{0}{90}
\end{pspicture}\ \;=\;
s(2\lambda-u)s(u)\ \ 
\begin{pspicture}[shift=-.67](-.2,-.3)(2.2,1.8)
\psarc[linewidth=1.5pt,linecolor=blue](1,1){.5}{0}{180}
\psline[linewidth=1.5pt,linecolor=blue](-.3,.5)(2.3,.5)
\psline[linewidth=1.5pt,linecolor=blue](.5,-.3)(.5,.3)
\psline[linewidth=1.5pt,linecolor=blue](1.5,-.3)(1.5,.3)
\facegrid{(0,0)}{(2,1)}
\psarc[linewidth=1.5pt,linecolor=blue](1,0){.5}{0}{180}
\psarc[linewidth=1.5pt,linecolor=blue](0,1){.5}{-90}{0}
\psarc[linewidth=1.5pt,linecolor=blue](2,1){.5}{180}{270}
\end{pspicture}\ +
s(2\lambda-u)s(\lambda-u)\ \ 
\begin{pspicture}[shift=-.67](-.2,-.3)(2.2,1.8)
\psarc[linewidth=1.5pt,linecolor=blue](1,1){.5}{0}{180}
\psline[linewidth=1.5pt,linecolor=blue](-.3,.5)(2.3,.5)
\psline[linewidth=1.5pt,linecolor=blue](.5,-.3)(.5,.3)
\psline[linewidth=1.5pt,linecolor=blue](1.5,-.3)(1.5,.3)
\facegrid{(0,0)}{(2,1)}
\psarc[linewidth=1.5pt,linecolor=blue](1,0){.5}{90}{180}
\psarc[linewidth=1.5pt,linecolor=blue](2,0){.5}{90}{180}
\psarc[linewidth=1.5pt,linecolor=blue](0,1){.5}{270}{0}
\psarc[linewidth=1.5pt,linecolor=blue](1,1){.5}{270}{0}
\end{pspicture}\qquad\\
&\!\!+\!\!&
s(u-\lambda)s(\lambda-u)\ \ 
\begin{pspicture}[shift=-.67](-.2,-.3)(2.2,1.8)
\psarc[linewidth=1.5pt,linecolor=blue](1,1){.5}{0}{180}
\psline[linewidth=1.5pt,linecolor=blue](-.3,.5)(2.3,.5)
\psline[linewidth=1.5pt,linecolor=blue](.5,-.3)(.5,.3)
\psline[linewidth=1.5pt,linecolor=blue](1.5,-.3)(1.5,.3)
\facegrid{(0,0)}{(2,1)}
\psarc[linewidth=1.5pt,linecolor=blue](0,0){.5}{0}{90}
\psarc[linewidth=1.5pt,linecolor=blue](2,0){.5}{90}{180}
\psarc[linewidth=1.5pt,linecolor=blue](1,1){.5}{180}{360}
\end{pspicture}\ +
s(u-\lambda)s(u)\ \ 
\begin{pspicture}[shift=-.67](-.2,-.3)(2.2,1.8)
\psarc[linewidth=1.5pt,linecolor=blue](1,1){.5}{0}{180}
\psline[linewidth=1.5pt,linecolor=blue](-.3,.5)(2.3,.5)
\psline[linewidth=1.5pt,linecolor=blue](.5,-.3)(.5,.3)
\psline[linewidth=1.5pt,linecolor=blue](1.5,-.3)(1.5,.3)
\facegrid{(0,0)}{(2,1)}
\psarc[linewidth=1.5pt,linecolor=blue](0,0){.5}{0}{90}
\psarc[linewidth=1.5pt,linecolor=blue](1,1){.5}{180}{270}
\psarc[linewidth=1.5pt,linecolor=blue](1,0){.5}{0}{90}
\psarc[linewidth=1.5pt,linecolor=blue](2,1){.5}{180}{270}
\end{pspicture}\ =s(2\lambda-u)s(u)\ \ 
\begin{pspicture}[shift=-.67](-.2,-.3)(2.2,1.8)
\psarc[linewidth=1.5pt,linecolor=blue](1,1){.5}{0}{180}
\psline[linewidth=1.5pt,linecolor=blue](-.3,.5)(2.3,.5)
\psline[linewidth=1.5pt,linecolor=blue](.5,-.3)(.5,.3)
\psline[linewidth=1.5pt,linecolor=blue](1.5,-.3)(1.5,.3)
\facegrid{(0,0)}{(2,1)}
\psarc[linewidth=1.5pt,linecolor=blue](1,0){.5}{0}{180}
\psarc[linewidth=1.5pt,linecolor=blue](0,1){.5}{-90}{0}
\psarc[linewidth=1.5pt,linecolor=blue](2,1){.5}{180}{270}
\end{pspicture}\nonumber\smallskip
\label{pushthrough}
\eea
This drop-down property means that, if there are no internal closed half loops on the bottom edge,
then there must be no closed half loops on the in-link state at the top. 
We denote this diagrammatic rule by a solid line along the bottom of a row
\be
\psset{unit=.75cm}
\begin{pspicture}[shift=-0.35](0,0)(3,1)
\facegrid{(0,0)}{(3,1)}
\psline[linewidth=3pt,linecolor=Blue](0,0)(3,0)
\psarc[linewidth=1.5pt,linecolor=blue](1,0){.5}{0}{180}
\end{pspicture}
\; =\, 0
\ee
The diagrammatic rule
can be pushed through from the bottom to the top so the diagrammatic rule of no closed half-arcs applies to all intermediate rows
\be
\psset{unit=.75cm}
\begin{pspicture}[shift=-1](0,0)(3,2.2)
\multirput(0,0)(1,0){4}{\psline(0,0)(0,2)}
\facegrid{(0,0)}{(3,2)}
\psline[linewidth=3pt,linecolor=Blue](0,0)(3,0)
\end{pspicture}
=
\begin{pspicture}[shift=-1](0,0)(3,2.2)
\multirput(0,0)(1,0){4}{\psline(0,0)(0,2)}
\facegrid{(0,0)}{(3,1)}
\psline[linewidth=3pt,linecolor=Blue](0,0)(3,0)
\rput(0,1.2){\facegrid{(0,0)}{(3,1)}}
\psline[linewidth=3pt,linecolor=Blue](0,1.2)(3,1.2)
\end{pspicture}
 \ee
If the Wenzl-Jones projector exists, then the action of the diagrammatic fusion described above agrees with the action of the Wenzl-Jones projector. 
However, diagrammatic fusion also makes sense when the Wenzl-Jones projector fails to exist.
We note that this diagrammatic rule is directional in the sense that it only kills upward and not downward closed half-arcs
\be
\psset{unit=.75cm}
\begin{pspicture}[shift=-1](0,0)(3,2.2)
\multirput(0,0)(1,0){4}{\psline(0,0)(0,2)}
\facegrid{(0,0)}{(3,1)}
\psline[linewidth=3pt,linecolor=Blue](0,0)(3,0)
%
\rput(0,1.2){\facegrid{(0,0)}{(3,1)}}
\psline[linewidth=3pt,linecolor=Blue](0,1.2)(3,1.2)
\psarc[linewidth=1.5pt,linecolor=blue](1,1.2){.5}{0}{180}
\end{pspicture}
 \;=\,0 \qquad
 \begin{pspicture}[shift=-1](0,0)(3,2.2)
\multirput(0,0)(1,0){4}{\psline(0,0)(0,2)}
\facegrid{(0,0)}{(3,1)}
\psline[linewidth=3pt,linecolor=Blue](0,0)(3,0)
\psarc[linewidth=1.5pt,linecolor=blue](1,1){.5}{180}{0}
\rput(0,1.2){\facegrid{(0,0)}{(3,1)}}
\psline[linewidth=3pt,linecolor=Blue](0,1.2)(3,1.2)
\end{pspicture}
\; \neq \,0
\ee

Since $s$-type seams are obtained from $r$-type seams in the the braid limit $u\to\pm i\infty$, let us focus on $r$-type seams of width $\rho-1$
\begin{align}
 K^{(\rho)}(u,\xi)&=K^{(\rho,1)}(u,\xi)={1\over \eta^{(\rho)}(u,\xi)}
\quad\
\psset{unit=1.25cm}
\begin{pspicture}[shift=-1.4](0,-.5)(5.5,2.2)
\multirput(0,0)(1,0){4}{\psline[linewidth=1.5pt,linecolor=blue](.5,-.2)(.5,2.2)}
\facegrid{(0,0)}{(4,2)}
\rput(.5,.5){\fws $u\!-\!\xi_{\rho\!-\!1}$}
\rput(.5,1.5){\fws $-\!u\!\!-\!\!\xi_{\rho\!-\!2}$}
\rput(1.55,.5){\fws $\cdots$}
\rput(1.55,1.5){\fws $\cdots$}
\rput(2.5,.5){\fws $u\!-\!\xi_2$}
\rput(2.5,1.5){\fws $-u\!-\!\xi_1$}
\rput(3.5,.5){\fws $u\!-\!\xi_1$}
\rput(3.5,1.5){\fws $-u\!-\!\xi_0$}
\psarc[linewidth=.5pt,linecolor=red](0,0){.15}{0}{90}
\psarc[linewidth=.5pt,linecolor=red](0,1){.15}{0}{90}
\psarc[linewidth=.5pt,linecolor=red](2,0){.15}{0}{90}
\psarc[linewidth=.5pt,linecolor=red](2,1){.15}{0}{90}
\psarc[linewidth=.5pt,linecolor=red](3,0){.15}{0}{90}
\psarc[linewidth=.5pt,linecolor=red](3,1){.15}{0}{90}
\psline[linewidth=1.5pt,linecolor=blue](-0.3,0.5)(0,0.5)
\psline[linewidth=1.5pt,linecolor=blue](-0.3,1.5)(0,1.5)
\rput(4,0){\rtri{}}
\rightarc{(4,1)}
\rput(2,-0.4){$\underbrace{\hspace{4.9cm}}$}
\rput(2,-0.7){$\rho-1$}
%
\psline[linewidth=3pt,linecolor=Blue](0,0)(4,0)
\end{pspicture}
\label{bdyK}
\end{align}
where
\begin{align}
\eta^{(\rho)}(u,\xi) &= \prod_{j=1}^{\rho-1} s(u+\xi_{j-1}) s(u-\xi_{j+1})=\frac{s(u+\xi)s(\xi_\rho-u)}{\beta s(\xi)s(\xi_\rho)}\,{\cal N}^{(\rho)}(u,\xi)\\
&=s(u+\xi)s(u-\xi_\rho)\prod_{j=1}^{\rho-2} s(u+\xi_j)s(u-\xi_{j+1})
\end{align}
and ${\cal N}^{(\rho)}(u,\xi)$ is the overall normalization (\ref{normalization}) of the double row transfer matrix. 
The last product removes the $2(\rho-2)$ common factors produced by the fusion process. 
The diagrammatic fusion rule can be pushed through so that the no closed half-arc rule also applies on the intermediate row. 
As shown by Pearce, Rasmussen and Villani~\cite{PRV}, acting on ${\cal V}^{(N)}_{\rho,1}$, the boundary operators simplify to
\be
K^{(\rho)}_N(u,\xi) = I + \frac{s(2u)}{s(u+\xi)s(u-\xi_\rho)}P_N^{(\rho)}\label{simpK}
\ee
These operators act as the identity in the bulk for $j<N$ but act non-trivially on the boundary sites with $j\ge N$. The {\em generalized Temperley-Lieb projectors} are
\begin{align}
P_N^{(\rho)} &= \sum_{k=0}^{\rho-2} (-1)^k s((\rho-k-1)\lambda)e_N^{(k)},
 \qquad e_N^{(k)}=\prod_{j=0}^k e_{N+j}=e_Ne_{N+1}\cdots e_{N+k}
\label{eq:genproj}
\end{align}
where the product is ordered and the functions $s(m\lambda) = U_{m-1}(\beta/2)$ are Chebyshev polynomials of the second kind of order $m-1$. 

The first few generalized TL projectors are
\be
P_N^{(2)}=e_N, \quad P_N^{(3)} = \beta  e_N-e_N e_{N+1}, \quad P_N^{(4)} = (\beta ^2-1) e_N-\beta  e_N e_{N+1} +e_N e_{N+1} e_{N+2} 
\ee
As shown in Appendix~\ref{genTLproj}, the generalized TL projectors satisfy the properties
\be
P^{(\rho')}_j P^{(\rho)}_j = U_{\rho'-1}(\beta/2) P^{(\rho)}_j, \qquad N \leq j \leq N+\rho - 1
\label{eq:genProjProp1}
\ee
\be
P^{(\rho)}_{N+1} e_N = 0, \qquad e_N P^{(\rho)}_{N+1} = U_{\rho-1}(\beta/2) e_N - P^{(\rho+1)}_N, \qquad e_N P^{(\rho)}_{N+1} e_N = U_{\rho-2}(\beta/2)e_N
\label{eq:genProjProp2}
\ee
These properties only hold when acting from ${\cal V}^{(N)}_{\rho,1}$ back onto itself, that is, from restricted states (with no closed half-arcs) back to restricted states. 
This action kills any TL words not of the form $I$ or $e_N^{(k)}$. We note that, even though these operators are called generalized projectors, they can only be normalized to give proper projectors when $U_{\rho-1}(\beta/2)\neq0$, that is, if $\rho$ is not a multiple of $p'$.

\medskip
For completeness, we recall the proof of (\ref{simpK}) from \cite{PRV}:

\medskip\noindent
{\bf Proof:}\quad The double row in (\ref{simpK}) is opened up by rotating the lower row by 45 degrees anticlockwise and the upper row by 45 degrees clockwise. 
This allows the manipulations to be carried out in the linear TL algebra. 
For one column, that is $\rho=2$, the result follows immediately from the identity
\bea
 X_N(u-v)X_N(u+v)=s(u-v-\lambda)s(u+v-\lambda)I+s(2u)e_N
\eea
with $v=\xi_1$. For $\rho>2$, we proceed by induction in $\rho$ using
\bea
 K_N^{(\rho+1)}(u,\xi)=\frac{\eta^{(\rho)}(u,\xi)}{\eta^{(\rho+1)}(u,\xi)}\,
   X_N(u-\xi_{\rho})K_{N+1}^{(\rho)}(u,\xi)X_N(u+\xi_{\rho})
\eea
with
\begin{align}
{\eta^{(\rho+1)}(u,\xi)\over \eta^{(\rho)}(u,\xi)}&=s(u+\xi_{\rho-1})s(u-\xi_{\rho+1})\\
 K_{N+1}^{(\rho)}(u,\xi)= I+\frac{s(2u)}{s(u+\xi)s(u-\xi_{\rho})}\,P_{N+1}^{(\rho)},\quad&\quad
 P_{N+1}^{(\rho)}=\sum_{k=1}^{\rho-1}(-1)^{k-1}s\big((\rho-k)\lambda\big)
  e_{N+1}\ldots e_{N+k} 
\end{align}
Explicitly, using (\ref{eq:genProjProp2}), it follows that
\begin{align}
&\phantom{\;=\;}s(u\!+\!\xi)s(u\!-\!\xi_\rho)\,{\eta^{(\rho+1)}(u,\xi)\over \eta^{(\rho)}(u,\xi)}\,K_N^{(\rho+1)}(u,\xi)\nonumber\\
&=s(u\!+\!\xi)s(u\!-\!\xi_\rho) X_N(u\!-\!\xi_\rho)K_{N+1}^{(\rho)}(u,\xi)X_N(u\!+\!\xi_\rho)\nonumber\\
&=s(u\!+\!\xi)s(u\!-\!\xi_\rho)X_N(u\!-\!\xi_\rho)X_N(u\!+\!\xi_\rho)
+s(2u)X_N(u\!-\!\xi_\rho)P_{N+1}^{(\rho)}X_N(u\!+\!\xi_\rho)\qquad\nonumber\\[4pt]
&=s(u\!+\!\xi)s(u\!-\!\xi_\rho)\big[s(u\!+\!\xi_{\rho-1})s(u\!-\!\xi_{\rho+1})I+s(2u)e_N\big]\\
&\mbox{}\qquad+s(2u)s(u\!-\!\xi_\rho)s(u\!+\!\xi_{\rho-1})[P_N^{(\rho+1)}\!-\!U_{\rho-1}(\beta/2)e_N]
+s(2u)s(u\!-\!\xi_\rho)s(u\!+\!\xi_\rho)U_{\rho-2}(\beta/2)e_0\qquad\nonumber\\
&=s(u+\xi_{\rho-1})s(u-\xi_\rho) [s(u\!+\!\xi)s(u\!-\!\xi_{\rho+1})I\!+\!s(2u) P_N^{(\rho+1)}]\nonumber
\end{align}
where all terms involving $e_N$ cancel out. The term proportional to $P_{N+1}^{(\rho)}$ is discarded since it is killed by the restriction on link states.

\section{Transfer Matrices and Hamiltonians}
\label{TMHamSect}
\subsection{Commuting double row transfer matrices}
\label{commutingDTM}
In this section, we explicitly construct the commuting double row transfer matrices $\vec D(u)$ of the logarithmic minimal models  ${\cal LM}(p,p')$ on a strip with a vacuum $(r,s)=(1,1)$ boundary condition applied on the left edge and an $(r,s)$ boundary conditions applied on the right edge. These transfer matrices act diagrammatically on the vector space of link states ${\cal V}_{\rho,s}^{(N)}$ of Section~\ref{LinkStates}. The double row transfer matrix is defined~\cite{PRZ,PRV} diagrammatically by 
\be
\drtm(u)=\drtm(u,\xi)=
\psset{unit=1.1cm}
\begin{pspicture}[shift=-1.5](-1,-.6)(6,2.3)
\rput(-1,0){\pspolygon[linewidth=0.8pt,linecolor=black,fillstyle=solid,fillcolor=lightlightblue](0,0)(1,1)(0,2)}
\leftarc{(0,1)}
\rightarc{(5,1)}
\rput(5,0){\pspolygon[linewidth=0.8pt,linecolor=black,fillstyle=solid,fillcolor=lightlightblue](0,1)(1,0)(1,2)}
\facegrid{(0,0)}{(5,2)}
\psline[linewidth=1.5pt,linecolor=red,linestyle=dashed](5,-.3)(5,2.3)
\multiput(0,0)(3,0){2}{
	\multirput(0,0)(1,0){2}{
		\multirput(0,0)(0,1){2}{
			\psarc[linewidth=.5pt,linecolor=red](0,0){.15}{0}{90}}}}
\rput(0.5,.5){\fws $u$}
\rput(0.5,1.5){\fws $\lambda-u$}
\rput(1.5,.5){\fws $u$}
\rput(1.5,1.5){\fws $\lambda-u$}
\rput(2.55,.5){\fws $\cdots$}
\rput(2.55,1.5){\fws $\cdots$}
\rput(3.5,.5){\fws $u$}
\rput(3.5,1.5){\fws $\lambda-u$}
\rput(4.5,.5){\fws $u$}
\rput(4.5,1.5){\fws $\lambda-u$}
\rput(5.6,1){$u,\xi$}
\rput(2.5,-.2){$\underbrace{\rule{5.5cm}{0pt}}$}
\rput(2.5,-.45){$N$}
\end{pspicture}
\ee
We usually suppress the dependence on the boundary field $\xi$ which is typically fixed to a specialized value. 
Since we are interested in the properties of the $r$-type seam and it is easy to add an $s$-type seam, we 
focus on the case $(r,s)=(r,1)$. In this case, the double row transfer matrix takes the explicit form
\be
\drtm(u)=
\psset{unit=1.1cm}
 \begin{pspicture}[shift=-1.5](-1,-.6)(7,2.3)
\rput(-1,0){\pspolygon[linewidth=0.8pt,linecolor=black,fillstyle=solid,fillcolor=lightlightblue](0,0)(1,1)(0,2)}
\rput(6,0){\pspolygon[linewidth=0.8pt,linecolor=black,fillstyle=solid,fillcolor=lightlightblue](0,1)(1,0)(1,2)}
\facegrid{(0,0)}{(6,2)}
\psline[linewidth=1.5pt,linecolor=red,linestyle=dashed](3,-.3)(3,2.3)
\leftarc{(0,1)}
\rightarc{(6,1)}
\rput(0.5,.5){\fws $u$}
\rput(0.5,1.5){\fws $\lambda-u$}
\rput(1.55,.5){\fws $\cdots$}
\rput(1.55,1.5){\fws $\cdots$}
\rput(2.5,.5){\fws $u$}
\rput(2.5,1.5){\fws $\lambda-u$}
\rput(3.5,.5){\fws $u\!-\!\xi_{\rho\!-\!1}$}
\rput(3.5,1.5){\fws $-\!u\!\!-\!\!\xi_{\rho\!-\!2}$}
\rput(4.55,.5){\fws $\cdots$}
\rput(4.55,1.5){\fws $\cdots$}
\rput(5.5,.5){\fws $u-\xi_1$}
\rput(5.5,1.5){\fws $-u\!-\!\xi_0$}
\rput(1.5,-.2){$\underbrace{\rule{3.3cm}{0pt}}$}
\rput(1.5,-.45){$N$}
\rput(4.5,-.2){$\underbrace{\rule{3.3cm}{0pt}}$}
\rput(4.5,-.45){$\rho-1$}
\multirput(0,0)(1,0){6}{\psarc[linecolor=red](0,0){.15}{0}{90}}
\multirput(0,1)(1,0){6}{\psarc[linecolor=red](0,0){.15}{0}{90}}
\end{pspicture}
\label{DTM}
\ee
where there are $N$ bulk columns, the $r$-type seam has width $\rho-1$ columns and the total number of columns is $N+\rho-1$.

Following the methods of \cite{BPO96}, the double row transfer matrices form a one-parameter commuting family and satisfy crossing symmetry
\be
 [\drtm(u),\drtm(v)]=0, \qquad 
   \drtm \big(\lambda-u\big)=\drtm(u)
\label{cross}
\ee
The double row transfer matrix is normalized by the crossing symmetric factor 
\be 
{\cal N}^{(\rho)}(u,\xi)=(-1)^{\rho-1}\beta s(\xi)s(\xi_\rho)\prod_{j=1}^{\rho-2}s(u+\xi_j)s(\xi_{j+1}-u) 
\label{normalization}
\ee
so that, in addition, it satisfies $\vec D(0)=\vec D(\lambda)=I$. 
Although there is no proof, we observe numerically that the double row transfer matrices and Hamiltonians are diagonalizable with real eigenvalues for all $(r,s)$ boundary conditions. The ground state eigenvalue is the highest eigenvalue of the double row transfer matrix in each $(r,s)$ sector. The ground state of the associated Hamiltonian is the lowest eigenvalue in each $(r,s)$ sector.

\subsection{Quantum Hamiltonians}

\subsubsection{Hamiltonian limit}
\psset{unit=0.7cm}
\setlength{\unitlength}{0.9cm}
In this section, we calculate explicitly the quantum Hamiltonians associated with the double row transfer matrices (\ref{DTM}) with an $r$-type seam of width $\rho-1$. We do this by taking the Hamiltonian limit
\bea
{\cal H}=-\half \sin\lambda \lim_{u\to 0} \frac{\partial}{\partial u} \log \vec D(u),\qquad \vec D(u)=\vec D(0)\,e^{-2u{\cal H}/\sin\lambda+\text{O}(u^2)}
\label{HamLimit}
\eea
where it is often convenient to shift the zero of energy for ${\cal H}$ by adding a multiple of the identity $I$.

Let us represent the generalized projector \eqref{eq:genproj} diagrammatically by
\bea
&&\hspace{-2.5cm}
\psset{unit=.6cm}
\begin{pspicture}[shift=-.9](0,0)(2,2.1)
\pspolygon[linewidth=1pt,linecolor=black,fillstyle=solid,fillcolor=lightpurple](0,0)(2,0)(2,2)(0,2)
\rput(1,1){\large $P_N^{(\rho)}$}
\end{pspicture}
\ \ =\ 
\, a_0^{(\rho)}
\quad
\begin{pspicture}[shift=-.9](0,0)(4,2.1)
\facegrid{(0,0)}{(4,2)}
\psline[linewidth=1.5pt,linecolor=blue](-0.3,0.5)(0,0.5)
\psline[linewidth=1.5pt,linecolor=blue](-0.3,1.5)(0,1.5)
\psarc[linewidth=1.5pt,linecolor=blue](0,2){.5}{-90}{0}
\psarc[linewidth=1.5pt,linecolor=blue](0,0){.5}{0}{90}
\psline[linewidth=1.5pt,linecolor=blue](1.5,0)(1.5,2)
\psline[linewidth=1.5pt,linecolor=blue](2.5,0)(2.5,2)
\psline[linewidth=1.5pt,linecolor=blue](3.5,0)(3.5,2)
\end{pspicture}
\ +a_1^{(\rho)}
\quad
 \begin{pspicture}[shift=-.9](0,0)(4,2.3)
\facegrid{(0,0)}{(4,2)}
\psline[linewidth=1.5pt,linecolor=blue](-0.3,0.5)(0,0.5)
\psline[linewidth=1.5pt,linecolor=blue](-0.3,1.5)(0,1.5)
\psline[linewidth=1.5pt,linecolor=blue](2.5,0)(2.5,2)
\psline[linewidth=1.5pt,linecolor=blue](3.5,0)(3.5,2)
\rput(0,0){\loopb}
\rput(0,1){\loopb}
\psarc[linewidth=1.5pt,linecolor=blue](1,0){.5}{0}{90}
\psarc[linewidth=1.5pt,linecolor=blue](1,2){.5}{-90}{0}
\end{pspicture}
\label{Bexp}
\\
%
%
&&\hspace{-.65cm}
\psset{unit=.6cm}
\ +\ a_2^{(\rho)}
\quad
 \begin{pspicture}[shift=-.9](0,0)(4,2.3)
\facegrid{(0,0)}{(4,2)}
\psline[linewidth=1.5pt,linecolor=blue](-0.3,0.5)(0,0.5)
\psline[linewidth=1.5pt,linecolor=blue](-0.3,1.5)(0,1.5)
\psline[linewidth=1.5pt,linecolor=blue](3.5,0)(3.5,2)
\rput(0,0){\loopb}
\rput(0,1){\loopb}
\rput(1,0){\loopb}
\rput(1,1){\loopb}
\psarc[linewidth=1.5pt,linecolor=blue](2,0){.5}{0}{90}
\psarc[linewidth=1.5pt,linecolor=blue](2,2){.5}{-90}{0}
\end{pspicture}
\ +\ldots +a_{\rho-1}^{(\rho)}
\quad
 \begin{pspicture}[shift=-.9](0,0)(4,2.3)
\facegrid{(0,0)}{(4,2)}
\psline[linewidth=1.5pt,linecolor=blue](-0.3,0.5)(0,0.5)
\psline[linewidth=1.5pt,linecolor=blue](-0.3,1.5)(0,1.5)
\rput(0,0){\loopb}
\rput(0,1){\loopb}
\rput(1,0){\loopb}
\rput(1,1){\loopb}
\rput(2,0){\loopb}
\rput(2,1){\loopb}
\psarc[linewidth=1.5pt,linecolor=blue](3,0){.5}{0}{90}
\psarc[linewidth=1.5pt,linecolor=blue](3,2){.5}{-90}{0}
\end{pspicture}
\eea
\psset{unit=0.6cm}
Diagrammatically the first order expansion of the normalized double row transfer matrix is then 
\begin{align}
&\! \beta s(\xi)s(\xi_\rho)\vec D(u)\;=s(\xi\!+\!u)s(\xi_\rho\!-\!u)\;
\begin{pspicture}[shift=-0.85](-.5,0)(7,2)
\rput[bl](0,0){\emptysquarea}
\rput[bl](0,1){\emptysquarea}
\rput[bl](1,0){\emptysquarea}
\rput[bl](1,1){\emptysquarea}
\rput[bl](2,0){\emptysquarea}
\rput[bl](2,1){\emptysquarea}
\rput[bl](3,0){\emptysquarea}
\rput[bl](3,1){\emptysquarea}
\rput(.5,.5){\tiny $u$}
\rput(.5,1.5){\tiny $\lambda\!-\!u$}
\rput(3.5,.5){\tiny $u$}
\rput(3.5,1.5){\tiny $\lambda\!-\!u$}
\rput[bl](4,0){\emptysquarea}
\rput[bl](4,1){\emptysquarea}
\rput[bl](5,0){\emptysquarea}
\rput[bl](5,1){\emptysquarea}
\rput[bl](6,0){\emptysquarea}
\rput[bl](6,1){\emptysquarea}
\rput(4,0){
\psline[linewidth=1.5pt,linecolor=blue](0.5,0)(0.5,2)
\psline[linewidth=1.5pt,linecolor=blue](1.5,0)(1.5,2)
\psline[linewidth=1.5pt,linecolor=blue](2.5,0)(2.5,2)
\psarc[linewidth=1.5pt,linecolor=blue](-.33,1){.61}{-60}{60}
}
\psline[linewidth=1pt,linecolor=red,linestyle=dashed](4,-0.4)(4,2.4)
\multirput(0,0)(1,0){4}{\psarc[linecolor=red](0,0){.15}{0}{90}}
\multirput(0,1)(1,0){4}{\psarc[linecolor=red](0,0){.15}{0}{90}}
\end{pspicture}
\;-\;s(2u)\;
\begin{pspicture}[shift=-0.85](-.5,0)(6,2)\
\rput[bl](0,0){\emptysquarea}
\rput[bl](0,1){\emptysquarea}
\rput[bl](1,0){\emptysquarea}
\rput[bl](1,1){\emptysquarea}
\rput[bl](2,0){\emptysquarea}
\rput[bl](2,1){\emptysquarea}
\rput[bl](3,0){\emptysquarea}
\rput[bl](3,1){\emptysquarea}
\rput(.5,.5){\tiny $u$}
\rput(.5,1.5){\tiny $\lambda\!-\!u$}
\rput(3.5,.5){\tiny $u$}
\rput(3.5,1.5){\tiny $\lambda\!-\!u$}
\psarc[linewidth=1.5pt,linecolor=blue](0,1){.5}{90}{270}
\pspolygon[linewidth=1pt,linecolor=black,fillstyle=solid,fillcolor=lightpurple](4,0)(6,0)(6,2)(4,2)
\rput(5,1){\large $P_N^{(\rho)}$}
\multirput(0,0)(1,0){4}{\psarc[linecolor=red](0,0){.15}{0}{90}}
\multirput(0,1)(1,0){4}{\psarc[linecolor=red](0,0){.15}{0}{90}}
\end{pspicture}\qquad\mbox{}\nonumber\\[16pt]
&=
\beta s(\lambda\!-\!u)^{2N}\Bigg(s(\xi\!+\!u)s(\xi_\rho\!-\!u) \ 
\begin{pspicture}[shift=-0.85](-.5,0)(7,2)\
\rput[bl](0,0){\emptysquarea}
\rput[bl](0,1){\emptysquarea}
\rput[bl](1,0){\emptysquarea}
\rput[bl](1,1){\emptysquarea}
\rput[bl](2,0){\emptysquarea}
\rput[bl](2,1){\emptysquarea}
\rput[bl](3,0){\emptysquarea}
\rput[bl](3,1){\emptysquarea}
\rput[bl](0,0){\loopa}
\rput[bl](0,1){\loopb}
\rput[bl](1,0){\loopa}
\rput[bl](1,1){\loopb}
\rput[bl](2,0){\loopa}
\rput[bl](2,1){\loopb}
\rput[bl](3,0){\loopa}
\rput[bl](3,1){\loopb}
\psarc[linewidth=1.5pt,linecolor=blue](0,1){.5}{90}{270}
\rput[bl](4,0){\emptysquarea}
\rput[bl](4,1){\emptysquarea}
\rput[bl](5,0){\emptysquarea}
\rput[bl](5,1){\emptysquarea}
\rput[bl](6,0){\emptysquarea}
\rput[bl](6,1){\emptysquarea}
\rput(4,0){
\psline[linewidth=1.5pt,linecolor=blue](0.5,0)(0.5,2)
\psline[linewidth=1.5pt,linecolor=blue](1.5,0)(1.5,2)
\psline[linewidth=1.5pt,linecolor=blue](2.5,0)(2.5,2)
\psarc[linewidth=1.5pt,linecolor=blue](-.33,1){.61}{-60}{60}
}
\psline[linewidth=1pt,linecolor=red,linestyle=dashed](4,-0.4)(4,2.4)
\end{pspicture}
\;-\;
{s(2u)}\;
\begin{pspicture}[shift=-0.85](-.5,0)(6,2)\
\rput[bl](0,0){\emptysquarea}
\rput[bl](0,1){\emptysquarea}
\rput[bl](1,0){\emptysquarea}
\rput[bl](1,1){\emptysquarea}
\rput[bl](2,0){\emptysquarea}
\rput[bl](2,1){\emptysquarea}
\rput[bl](3,0){\emptysquarea}
\rput[bl](3,1){\emptysquarea}
\rput[bl](0,0){\loopa}
\rput[bl](0,1){\loopb}
\rput[bl](1,0){\loopa}
\rput[bl](1,1){\loopb}
\rput[bl](2,0){\loopa}
\rput[bl](2,1){\loopb}
\rput[bl](3,0){\loopa}
\rput[bl](3,1){\loopb}
\psarc[linewidth=1.5pt,linecolor=blue](0,1){.5}{90}{270}
\pspolygon[linewidth=1pt,linecolor=black,fillstyle=solid,fillcolor=lightpurple](4,0)(6,0)(6,2)(4,2)
\rput(5,1){\large $P_N^{(\rho)}$}
\end{pspicture}\ \Bigg)
\qquad\mbox{}\nonumber\\[16pt]
&+
\beta s(\xi\!+\!u)s(\xi_\rho\!-\!u)s(u)s(\lambda\!-\!u)^{2N-1}\Bigg(
\begin{pspicture}[shift=-0.85](-.5,0)(7,2)\
\rput[bl](0,0){\emptysquarea}
\rput[bl](0,1){\emptysquarea}
\rput[bl](1,0){\emptysquarea}
\rput[bl](1,1){\emptysquarea}
\rput[bl](2,0){\emptysquarea}
\rput[bl](2,1){\emptysquarea}
\rput[bl](3,0){\emptysquarea}
\rput[bl](3,1){\emptysquarea}
\rput[bl](0,0){\loopa}
\rput[bl](0,1){\loopb}
\rput[bl](1,0){\loopb}
\rput[bl](1,1){\loopb}
\rput[bl](2,0){\loopa}
\rput[bl](2,1){\loopb}
\rput[bl](3,0){\loopa}
\rput[bl](3,1){\loopb}
\psarc[linewidth=1.5pt,linecolor=blue](0,1){.5}{90}{270}
\rput[bl](4,0){\emptysquarea}
\rput[bl](4,1){\emptysquarea}
\rput[bl](5,0){\emptysquarea}
\rput[bl](5,1){\emptysquarea}
\rput[bl](6,0){\emptysquarea}
\rput[bl](6,1){\emptysquarea}
\rput(4,0){
\psline[linewidth=1.5pt,linecolor=blue](0.5,0)(0.5,2)
\psline[linewidth=1.5pt,linecolor=blue](1.5,0)(1.5,2)
\psline[linewidth=1.5pt,linecolor=blue](2.5,0)(2.5,2)
\psarc[linewidth=1.5pt,linecolor=blue](-.33,1){.61}{-60}{60}
}
\psline[linewidth=1pt,linecolor=red,linestyle=dashed](4,-0.4)(4,2.4)
\end{pspicture}
\;+\;\cdots\;+\;
\begin{pspicture}[shift=-0.85](-.5,0)(7,2)\
\rput[bl](0,0){\emptysquarea}
\rput[bl](0,1){\emptysquarea}
\rput[bl](1,0){\emptysquarea}
\rput[bl](1,1){\emptysquarea}
\rput[bl](2,0){\emptysquarea}
\rput[bl](2,1){\emptysquarea}
\rput[bl](3,0){\emptysquarea}
\rput[bl](3,1){\emptysquarea}
\rput[bl](0,0){\loopa}
\rput[bl](0,1){\loopb}
\rput[bl](1,0){\loopa}
\rput[bl](1,1){\loopb}
\rput[bl](2,0){\loopa}
\rput[bl](2,1){\loopb}
\rput[bl](3,0){\loopb}
\rput[bl](3,1){\loopb}
\psarc[linewidth=1.5pt,linecolor=blue](0,1){.5}{90}{270}
\rput[bl](4,0){\emptysquarea}
\rput[bl](4,1){\emptysquarea}
\rput[bl](5,0){\emptysquarea}
\rput[bl](5,1){\emptysquarea}
\rput[bl](6,0){\emptysquarea}
\rput[bl](6,1){\emptysquarea}
\rput(4,0){
\psline[linewidth=1.5pt,linecolor=blue](0.5,0)(0.5,2)
\psline[linewidth=1.5pt,linecolor=blue](1.5,0)(1.5,2)
\psline[linewidth=1.5pt,linecolor=blue](2.5,0)(2.5,2)
\psarc[linewidth=1.5pt,linecolor=blue](-.33,1){.61}{-60}{60}
}
\psline[linewidth=1pt,linecolor=red,linestyle=dashed](4,-0.4)(4,2.4)
\end{pspicture}\ \Bigg)\nonumber\\[16pt]
&+
\beta s(\xi\!+\!u)s(\xi_\rho\!-\!u)s(u)s(\lambda-u)^{2N-1}\Bigg(
\begin{pspicture}[shift=-0.85](-.5,0)(7,2)\
\rput[bl](0,0){\emptysquarea}
\rput[bl](0,1){\emptysquarea}
\rput[bl](1,0){\emptysquarea}
\rput[bl](1,1){\emptysquarea}
\rput[bl](2,0){\emptysquarea}
\rput[bl](2,1){\emptysquarea}
\rput[bl](3,0){\emptysquarea}
\rput[bl](3,1){\emptysquarea}
\rput[bl](0,0){\loopa}
\rput[bl](0,1){\loopb}
\rput[bl](1,0){\loopa}
\rput[bl](1,1){\loopa}
\rput[bl](2,0){\loopa}
\rput[bl](2,1){\loopb}
\rput[bl](3,0){\loopa}
\rput[bl](3,1){\loopb}
\psarc[linewidth=1.5pt,linecolor=blue](0,1){.5}{90}{270}
\rput[bl](4,0){\emptysquarea}
\rput[bl](4,1){\emptysquarea}
\rput[bl](5,0){\emptysquarea}
\rput[bl](5,1){\emptysquarea}
\rput[bl](6,0){\emptysquarea}
\rput[bl](6,1){\emptysquarea}
\rput(4,0){
\psline[linewidth=1.5pt,linecolor=blue](0.5,0)(0.5,2)
\psline[linewidth=1.5pt,linecolor=blue](1.5,0)(1.5,2)
\psline[linewidth=1.5pt,linecolor=blue](2.5,0)(2.5,2)
\psarc[linewidth=1.5pt,linecolor=blue](-.33,1){.61}{-60}{60}
}
\psline[linewidth=1pt,linecolor=red,linestyle=dashed](4,-0.4)(4,2.4)
\end{pspicture}\;+\;\cdots\;+\;
\begin{pspicture}[shift=-0.85](-.5,0)(7,2)\
\rput[bl](0,0){\emptysquarea}
\rput[bl](0,1){\emptysquarea}
\rput[bl](1,0){\emptysquarea}
\rput[bl](1,1){\emptysquarea}
\rput[bl](2,0){\emptysquarea}
\rput[bl](2,1){\emptysquarea}
\rput[bl](3,0){\emptysquarea}
\rput[bl](3,1){\emptysquarea}
\rput[bl](0,0){\loopa}
\rput[bl](0,1){\loopb}
\rput[bl](1,0){\loopa}
\rput[bl](1,1){\loopb}
\rput[bl](2,0){\loopa}
\rput[bl](2,1){\loopb}
\rput[bl](3,0){\loopa}
\rput[bl](3,1){\loopa}
\psarc[linewidth=1.5pt,linecolor=blue](0,1){.5}{90}{270}
\rput[bl](4,0){\emptysquarea}
\rput[bl](4,1){\emptysquarea}
\rput[bl](5,0){\emptysquarea}
\rput[bl](5,1){\emptysquarea}
\rput[bl](6,0){\emptysquarea}
\rput[bl](6,1){\emptysquarea}
\rput(4,0){
\psline[linewidth=1.5pt,linecolor=blue](0.5,0)(0.5,2)
\psline[linewidth=1.5pt,linecolor=blue](1.5,0)(1.5,2)
\psline[linewidth=1.5pt,linecolor=blue](2.5,0)(2.5,2)
\psarc[linewidth=1.5pt,linecolor=blue](-.33,1){.61}{-60}{60}
}
\psline[linewidth=1pt,linecolor=red,linestyle=dashed](4,-0.4)(4,2.4)
\end{pspicture}\ \Bigg)
\;\mbox{}\nonumber\\[16pt]
&+
s(\xi\!+\!u)s(\xi_\rho\!-\!u)s(u)s(\lambda\!-\!u)^{2N-1}\Bigg(
\begin{pspicture}[shift=-0.85](-.5,0)(7,2)\
\rput[bl](0,0){\emptysquarea}
\rput[bl](0,1){\emptysquarea}
\rput[bl](1,0){\emptysquarea}
\rput[bl](1,1){\emptysquarea}
\rput[bl](2,0){\emptysquarea}
\rput[bl](2,1){\emptysquarea}
\rput[bl](3,0){\emptysquarea}
\rput[bl](3,1){\emptysquarea}
\rput[bl](0,0){\loopa}
\rput[bl](0,1){\loopa}
\rput[bl](1,0){\loopa}
\rput[bl](1,1){\loopb}
\rput[bl](2,0){\loopa}
\rput[bl](2,1){\loopb}
\rput[bl](3,0){\loopa}
\rput[bl](3,1){\loopb}
\psarc[linewidth=1.5pt,linecolor=blue](0,1){.5}{90}{270}
\rput[bl](4,0){\emptysquarea}
\rput[bl](4,1){\emptysquarea}
\rput[bl](5,0){\emptysquarea}
\rput[bl](5,1){\emptysquarea}
\rput[bl](6,0){\emptysquarea}
\rput[bl](6,1){\emptysquarea}
\rput(4,0){
\psline[linewidth=1.5pt,linecolor=blue](0.5,0)(0.5,2)
\psline[linewidth=1.5pt,linecolor=blue](1.5,0)(1.5,2)
\psline[linewidth=1.5pt,linecolor=blue](2.5,0)(2.5,2)
\psarc[linewidth=1.5pt,linecolor=blue](-.33,1){.61}{-60}{60}
}
\psline[linewidth=1pt,linecolor=red,linestyle=dashed](4,-0.4)(4,2.4)
\end{pspicture}
\;+\;
\begin{pspicture}[shift=-0.85](-.5,0)(7,2)\
\rput[bl](0,0){\emptysquarea}
\rput[bl](0,1){\emptysquarea}
\rput[bl](1,0){\emptysquarea}
\rput[bl](1,1){\emptysquarea}
\rput[bl](2,0){\emptysquarea}
\rput[bl](2,1){\emptysquarea}
\rput[bl](3,0){\emptysquarea}
\rput[bl](3,1){\emptysquarea}
\rput[bl](0,0){\loopb}
\rput[bl](0,1){\loopb}
\rput[bl](1,0){\loopa}
\rput[bl](1,1){\loopb}
\rput[bl](2,0){\loopa}
\rput[bl](2,1){\loopb}
\rput[bl](3,0){\loopa}
\rput[bl](3,1){\loopb}
\psarc[linewidth=1.5pt,linecolor=blue](0,1){.5}{90}{270}
\rput[bl](4,0){\emptysquarea}
\rput[bl](4,1){\emptysquarea}
\rput[bl](5,0){\emptysquarea}
\rput[bl](5,1){\emptysquarea}
\rput[bl](6,0){\emptysquarea}
\rput[bl](6,1){\emptysquarea}
\rput(4,0){
\psline[linewidth=1.5pt,linecolor=blue](0.5,0)(0.5,2)
\psline[linewidth=1.5pt,linecolor=blue](1.5,0)(1.5,2)
\psline[linewidth=1.5pt,linecolor=blue](2.5,0)(2.5,2)
\psarc[linewidth=1.5pt,linecolor=blue](-.33,1){.61}{-60}{60}
}
\psline[linewidth=1pt,linecolor=red,linestyle=dashed](4,-0.4)(4,2.4)
\end{pspicture}\ \Bigg)
+\mbox{O}(u^2)\mbox{}\nonumber
\end{align}

Collecting connectivity classes together gives
\bea
\vec D(u)
&\!\!=\!\!&
\left[1+u\Big(\!\cot\xi-\cot\xi_\rho-2N\cot\lambda+\frac{2\beta^{-1}}{\sin\lambda}\Big) \right]\;
\begin{pspicture}[shift=-0.8](0,0)(7,2)
\rput[bl](0,0){\emptysquarea}
\rput[bl](0,1){\emptysquarea}
\rput[bl](1,0){\emptysquarea}
\rput[bl](1,1){\emptysquarea}
\rput[bl](2,0){\emptysquarea}
\rput[bl](2,1){\emptysquarea}
\rput[bl](3,0){\emptysquarea}
\rput[bl](3,1){\emptysquarea}
\rput[bl](4,0){\emptysquarea}
\rput[bl](4,1){\emptysquarea}
\rput[bl](5,0){\emptysquarea}
\rput[bl](5,1){\emptysquarea}
\rput[bl](6,0){\emptysquarea}
\rput[bl](6,1){\emptysquarea}
\psline[linewidth=1.5pt,linecolor=blue](.5,0)(.5,2)
\psline[linewidth=1.5pt,linecolor=blue](1.5,0)(1.5,2)
\psline[linewidth=1.5pt,linecolor=blue](2.5,0)(2.5,2)
\psline[linewidth=1.5pt,linecolor=blue](3.5,0)(3.5,2)
\psline[linewidth=1.5pt,linecolor=blue](4.5,0)(4.5,2)
\psline[linewidth=1.5pt,linecolor=blue](5.5,0)(5.5,2)
\psline[linewidth=1.5pt,linecolor=blue](6.5,0)(6.5,2)
\psline[linewidth=1pt,linecolor=red,linestyle=dashed](4,-0.4)(4,2.4)
\end{pspicture}\nonumber\\[16pt]
&\!\!+\!\!&
{2u\over\sin\lambda}\Bigg(\;
\begin{pspicture}[shift=-.8](0,0)(7,2)
\rput[bl](0,0){\emptysquarea}
\rput[bl](0,1){\emptysquarea}
\rput[bl](1,0){\emptysquarea}
\rput[bl](1,1){\emptysquarea}
\rput[bl](2,0){\emptysquarea}
\rput[bl](2,1){\emptysquarea}
\rput[bl](3,0){\emptysquarea}
\rput[bl](3,1){\emptysquarea}
\rput[bl](4,0){\emptysquarea}
\rput[bl](4,1){\emptysquarea}
\rput[bl](5,0){\emptysquarea}
\rput[bl](5,1){\emptysquarea}
\rput[bl](6,0){\emptysquarea}
\rput[bl](6,1){\emptysquarea}
\psline[linewidth=1.5pt,linecolor=blue](2.5,0)(2.5,2)
\psline[linewidth=1.5pt,linecolor=blue](3.5,0)(3.5,2)
\psline[linewidth=1.5pt,linecolor=blue](4.5,0)(4.5,2)
\psline[linewidth=1.5pt,linecolor=blue](5.5,0)(5.5,2)
\psline[linewidth=1.5pt,linecolor=blue](6.5,0)(6.5,2)
\psarc[linewidth=1.5pt,linecolor=blue](1,0){.5}{0}{180}
\psarc[linewidth=1.5pt,linecolor=blue](1,2){.5}{-180}{0}
\psline[linewidth=1pt,linecolor=red,linestyle=dashed](4,-0.4)(4,2.4)
\end{pspicture}
\;+\;\cdots\;+\;
\begin{pspicture}[shift=-.8](0,0)(7,2)
\rput[bl](0,0){\emptysquarea}
\rput[bl](0,1){\emptysquarea}
\rput[bl](1,0){\emptysquarea}
\rput[bl](1,1){\emptysquarea}
\rput[bl](2,0){\emptysquarea}
\rput[bl](2,1){\emptysquarea}
\rput[bl](3,0){\emptysquarea}
\rput[bl](3,1){\emptysquarea}
\rput[bl](4,0){\emptysquarea}
\rput[bl](4,1){\emptysquarea}
\rput[bl](5,0){\emptysquarea}
\rput[bl](5,1){\emptysquarea}
\rput[bl](6,0){\emptysquarea}
\rput[bl](6,1){\emptysquarea}
\psline[linewidth=1.5pt,linecolor=blue](.5,0)(.5,2)
\psline[linewidth=1.5pt,linecolor=blue](1.5,0)(1.5,2)
\psline[linewidth=1.5pt,linecolor=blue](4.5,0)(4.5,2)
\psline[linewidth=1.5pt,linecolor=blue](5.5,0)(5.5,2)
\psline[linewidth=1.5pt,linecolor=blue](6.5,0)(6.5,2)
\psarc[linewidth=1.5pt,linecolor=blue](3,0){.5}{0}{180}
\psarc[linewidth=1.5pt,linecolor=blue](3,2){.5}{-180}{0}
\psline[linewidth=1pt,linecolor=red,linestyle=dashed](4,-0.4)(4,2.4)
\end{pspicture}\;\Bigg)\nonumber\\[16pt]
&\!\!-\!\!&
{2u\over\sin\lambda}\,{1\over s(\xi)s(\xi_\rho)}\;
\begin{pspicture}[shift=-.8](0,0)(6,2)
\rput[bl](0,0){\emptysquarea}
\rput[bl](0,1){\emptysquarea}
\rput[bl](1,0){\emptysquarea}
\rput[bl](1,1){\emptysquarea}
\rput[bl](2,0){\emptysquarea}
\rput[bl](2,1){\emptysquarea}
\rput[bl](3,0){\emptysquarea}
\rput[bl](3,1){\emptysquarea}
\psline[linewidth=1.5pt,linecolor=blue](.5,0)(.5,2)
\psline[linewidth=1.5pt,linecolor=blue](1.5,0)(1.5,2)
\psline[linewidth=1.5pt,linecolor=blue](2.5,0)(2.5,2)
\psline[linewidth=1.5pt,linecolor=blue](5,0)(5,2)
\psline[linewidth=1.5pt,linecolor=blue](6,0)(6,2)
\psarc[linewidth=1.5pt,linecolor=blue](4,0){.5}{90}{180}
\psarc[linewidth=1.5pt,linecolor=blue](4,2){.5}{-180}{-90}
\pspolygon[linewidth=1pt,linecolor=black,fillstyle=solid,fillcolor=lightpurple](4,0)(6,0)(6,2)(4,2)
\rput(5,1){\large $P_N^{(\rho)}$}
\end{pspicture}
\quad+\;\mbox{O}(u^2)\qquad\mbox{}\nonumber
\eea
It follows that
\begin{align}
\vec D(u)&=
I\!-\!{2u\over\sin\lambda}\,\calH+\mbox{O}(u^2)\nonumber \\
&=
I\!-\!{2u\over\sin\lambda}\Big[N\cos\lambda\!-\!(\beta^{-1}\!+\!\frac{\sin\lambda}{2}(\cot\xi\!-\!\cot\xi_\rho))I\!+\!\calH^{(\rho,1)}\Big]+\mbox{O}(u^2)
\label{uExp}
\end{align}
After shifting the zero of energy, the Hamiltonian $\calH^{(\rho,1)}$ acting on ${\cal V}_{\rho,1}^{(N)}$ is
\be
\calH^{(\rho,1)} = -\sum_{j=1}^{N-1}e_j + \frac{1}{s(\xi)s(\xi_\rho)} P^{(\rho)}_N
\label{Hamr1}
\ee
It is similarly shown that the Hamiltonian $\calH^{(\rho,s)}$ acting on ${\cal V}_{\rho,s}^{(N)}$ is
\be
\calH^{(\rho,s)} = -\sum_{j=1}^{N-1}e_j + \frac{1}{s(\xi)s(\xi_\rho)} P^{(\rho)}_N
\label{Hamrs}
\ee

 \subsubsection{Boundary field $\xi$}

\label{appHam}
In the Hamiltonian $\calH^{(\rho,1)}$ (\ref{Hamr1}), the boundary term consists of the generalized TL projector $P^{(\rho)}_N$ multiplied by a coefficient 
$\bdyCoeff(\xi)$ which depends on the boundary field $\xi$
\be
\bdyCoeff(\xi) = \frac{1}{s(\xi)s(\xi_\rho)},\qquad \xi_\rho = \xi + \rho \lambda, \qquad \xi \in (0,\pi)
\label{eq:Ham}
\ee
where we use periodicity to restrict $\xi$ to the interval $(0,\pi)$. Numerically, we find that the conformal properties do not depend on the precise choice for the value of $\xi$ provided $\xi$ is restricted to an appropriate interval. In the sense of the renormalization group (RG), $\xi$ is an irrelevant variable whose precise value does not effect the convergence (in the continuum scaling limit) to the boundary conformal RG fixed point labelled by $r$. However, $\xi$ is a dangerous irrelevant variable in the sense that, if its value is changed too much, the convergence shifts to another boundary conformal RG fixed point labelled by a different value of $r$.

The coefficient $\bdyCoeff(\xi)$ diverges at the endpoints as $\xi\to 0$ or $\xi\to\pi$. If the number of boundary columns is such that $\rho = 0$ mod $p'$, there are no divergences within the interval $\xi \in (0,\pi)$ and the coefficient is either strictly positive or strictly negative throughout this interval. In this case, we specialize the value of $\xi$ to the midpoint value $\xi = \pi/2$. For all other values of $\rho$, there is an additional divergence at a single internal point
\bea
\mbox{$\xi=-\rho\lambda\ $ mod $\pi$}
\eea
This value divides the interval $(0,\pi)$ into two subintervals, one where $\bdyCoeff(\xi)$ is positive containing a local minimum and the other where $\bdyCoeff(\xi)$ is negative containing a local maximum. The values of $\xi$ at these local extrema are
\bea
\xi=
\begin{cases}
\xi_+ = -\frac{\rho\lambda}{2}\ \;\mbox{mod $\pi$}, & \bdyCoeff(\xi_+)>0\\[4pt]
\xi_-  = \frac{\pi-\rho\lambda}{2}\ \;\mbox{mod $\pi$}, &\bdyCoeff(\xi_-)<0
\end{cases}
\eea
In a sense these values are representative of the distinct domains of $\xi$ in which the boundary term acts ferromagnetically and antiferromagnetically respectively. 
These are the basins of attraction for the respective boundary conformal RG fixed points.
Taking these cases into account, we choose to specialize the value of $\xi$ according to 
\be
\xi = 
\begin{cases}
	\min(\xi_+,\xi_-) = [ \frac{\rho p}{p'}] \frac{\pi}{2}, &\quad [\frac{\rho}{p'}] \neq 0\\[4pt]
	\frac{\pi}{2}, &\quad [\frac{\rho}{p'}]= 0
\end{cases}
\label{eq:xiChoice}
\ee
where $[a]$ denotes the fractional part of $a$. The identity $\min(\xi_+,\xi_-) = [ \frac{\rho p}{p'}] \frac{\pi}{2}$ for the specialized value of $\xi$ is proved in Appendix~\ref{equivxi}. 
This specialization of $\xi$ gives conformal data which confirms the conjectured~\cite{PRV} relation between $\rho$ and $r$
\be
\rho = \rho(r) = \Big \lfloor{\frac{rp'}{p}}\Big \rfloor, \qquad r = r(\rho) = \Big \lceil \frac{\rho p}{p'} \Big \rceil
\ee
Increasing $\rho$ in the sequence $r=r(\rho)$ with $p,p'$ fixed, we observe that the specialized value of $\xi$ starts with $\xi_-$ at $\rho=1$ and then switches between 
$\xi_+$ and $\xi_-$ whenever a value of $r$ is repeated. This also gives the correct specialized value $\xi=\max(\xi_+,\xi_-)=\frac{\pi}{2}$ for the cases $ [\frac{\rho}{p'}]= 0$.

\subsection{Free energies}
\label{Free_energy}
In this section, we calculate the bulk and boundary free energies. 
This is achieved by solving a functional equation in the form of an inversion relation satisfied by the ground state eigenvalue $\kappa(u,\xi)=D_0(u,\xi)$ of the double row transfer matrix. 
This method was applied to the bulk free energy by Baxter~\cite{BaxInv82}. The boundary free energies are calculated using the boundary inversion relation methods of \cite{OPB95}. 
The full inversion relation (\ref{fullInversionRelation}) is derived diagrammatically in Appendix~\ref{InversionRelation}. For large $N$, the ground state eigenvalue $D_0(u,\xi)$ factorizes into contributions from the bulk $\kappa_{bulk}(u)$, vacuum boundary $\kappa_0(u)$ and $r$-type seams $\kappa_\rho^{R,L}(u,\xi)$ on the right and left edges. Applying an $r$-type seam only to the right edge gives the factorization
\bea
&D_0(u,\xi)\sim\kappa(u,\xi)={\kappa_{bulk}(u)}^{2N}\kappa_0(u)\kappa_\rho(u,\xi),\quad N\to\infty\qquad&\\[6pt]
&\kappa_{bulk}(u)=\exp(-f_{bulk}(u)),\qquad \kappa_0(u)=\exp(-f_0(u)),\qquad \kappa_{\rho}(u,\xi)=\exp(-f_{\rho}(u,\xi))\qquad&\\[6pt]
&\kappa_{bdy}(u,\rho,\xi)=\kappa_0(u)\kappa_{\rho}(u,\xi)=\exp(-\!f_0(u)\!-\!f_\rho(u,\xi))=\exp(-\!f_{bdy}(u,\rho,\xi))\qquad&
\eea
where $f_{bulk}(u)$ is the bulk free energy per face and $f_{bdy}(u,\rho,\xi)$ is the boundary free energy per double row.
We note that the boundary free energies are independent of $s$. 
For $\rho=1$, there is no $r$-type seam and $\kappa_{\rho}(u,\xi)=1$ and $f_{\rho}(u,\xi)=0$.
Separating $\mbox{O}(N)$ and $\mbox{O}(1)$ terms, it follows from (\ref{fullInversionRelation}) that the full inversion relation factorizes into three separate inversion relations
\begin{align}
        \kappa_{bulk}(u)\kappa_{bulk}(u+\lambda)&= \dfrac{\sin(\lambda+u)\sin(\lambda-u)}{\sin^2\lambda} \label{bulkInversion}\\[6pt]
        \kappa_0(u)\kappa_0(u+\lambda)&=\dfrac{\sin^2\lambda\sin(2\lambda+2u)\sin(2\lambda-2u)}{\sin^22\lambda\sin(\lambda+2u)\sin(\lambda-2u)}
        \label{vacInversion}\\[6pt]
        \kappa_\rho(u,\xi)\kappa_\rho(u+\lambda,\xi)&=\dfrac{\sin(\xi+u)\sin(\xi-u)\sin(\xi+\rho\lambda+u)\sin(\xi+\rho\lambda-u)}{\sin^2\xi\sin^2(\xi+\rho\lambda)}
\label{bdyInversion}
\end{align}
These functional equations are solved for the bulk and boundary free energies in turn subject to certain analyticity assumptions as discussed in Appendix~\ref{analyticity}.  
In some cases, the full solution involves analytic continuation in $\lambda$. Since the analyticity assumptions are checked in many cases but not proved, the analytic solutions of the inversion relations are extensively checked numerically.

\subsubsection{Bulk free energy}
Since we use similar calculations to obtain the boundary free energies, we review in this section the key steps~\cite{BaxInv82} in calculating $\kappa_{bulk}(u)$.
Following Baxter, let us assume that $\kappa_{bulk}(u)$ is analytic and non-zero (ANZ) on the {\it analyticity strip} $|\!\Re u-\frac{\lambda}{2}|\le \min(\lambda,\frac{\pi}{2})$. 
This strip contains the {\it physical strip} $0\le \Re u\le \lambda$. As argued by Baxter, $\kappa(u)$ grows as $\exp(\mp iu)$ as $u\rightarrow\pm i\infty$ in line with the growth of a single face weight. It follows that the second derivative of $\log\kappa_{bulk}(u)$ can be represented on the full analyticity strip by the Fourier/Laplace integral
\be
\frac{d^2}{du^2}\log\kappa_{bulk}(u)=\int_{-\infty}^{\infty}c(t)e^{2ut}\,dt 
\label{laplace:transform:bulk}
\ee
The bulk contribution $\kappa_{bulk}(u)$ satisfies the bulk inversion relation \eqref{bulkInversion} and the crossing symmetry 
\be 
\log\kappa_{bulk}(u)+\log\kappa_{bulk}(u+\lambda)=\log\frac{\sin(\lambda+u)\sin(\lambda-u)}{\sin^2\lambda} ,\quad \log\kappa_{bulk}(u)=\log\kappa_{bulk}(\lambda-u)\label{bulkrelations}
\ee
From the identity
\be 
\frac{d^2}{du^2}\log\sin u=-\int_0^{\infty}\frac{4t\cosh(\pi-2u)t}{\sinh\pi t}\,dt,\qquad 0<\Re u<\pi
\ee
it follows that
\be 
\frac{d^2}{du^2}\log\sin(\lambda+u)\sin(\lambda-u)=-\int_{-\infty}^{\infty}\frac{4t\cosh\left(\pi-2\lambda\right)t}{\sinh\pi t}e^{2ut}\,dt,\qquad |\Re u|<\lambda
\label{identity:bulkInv}
\ee
Substituting \eqref{laplace:transform:bulk} and \eqref{identity:bulkInv} into \eqref{bulkrelations} gives
\be 
c(t)=e^{-2\lambda t}c(-t),\qquad (1+e^{2\lambda t})c(t)=-\frac{4t\cosh\left(\pi-2\lambda\right)t}{\sinh\pi t} 
\ee 
with the solution
\be 
c(t)=-\frac{2te^{-\lambda t}\cosh\left(\pi-2\lambda\right)t}{\sinh\pi t\cosh\lambda t} 
\ee
After integrating once we rewrite the result as
 \begin{align}  
 \frac{d}{du}\log\kappa_{bulk}(u)&=\int_{-\infty}^{\infty}\frac{\cosh(\pi-2\lambda)t\sinh(\lambda-2u)t}{\sinh\pi t\cosh\lambda t}dt+A\nonumber\\
 &=\int_{-\infty}^{\infty}\frac{\cosh(\pi-2\lambda)t}{\sinh\pi t\cosh\lambda t}\big[\sinh ut\cosh(\lambda\!-\!u)t\!-\!\sinh(\lambda\!-\!u)t\cosh ut\big]dt+A
 \end{align}
Integrating again and evaluating the integration constants using the crossing relation and the initial condition at $u=0$ gives 
\begin{align} 
-f_{bulk}(u)=\log\kappa_{bulk}(u)&=\int_{-\infty}^{\infty}\frac{\cosh(\pi-2\lambda)t\sinh ut\sinh(\lambda-u)t}{t\sinh\pi t\cosh\lambda t}\,dt+Au+B\nonumber\\ 
&=\int_{-\infty}^{\infty}\frac{\cosh(\pi-2\lambda)t\sinh ut\sinh(\lambda-u)t}{t\sinh\pi t\cosh\lambda t}\,dt\label{bulkfreeenergy}
\end{align}

\subsubsection{Transfer matrix boundary free energies}
Using similar methods to solve the inversion relations (\ref{vacInversion}) and (\ref{bdyInversion}) subject to appropriate analyticity assumptions gives $\kappa_0(u)$, $\kappa_\rho(u,\xi)$ and the boundary free energies in the physical interval $0\le u\le \lambda$. The details of these calculations are given in Appendix~\ref{laplace:boundary}. 
The results are 
\begin{align} 
-f_{bulk}(u)&=\log\kappa_0(u)=2\int_{-\infty}^{\infty}\frac{\sinh\frac{(3\lambda-\pi)t}{2}\sinh\frac{\lambda t}{2}\sinh ut\sinh(\lambda-u)t}{t\sinh\frac{\pi t}{2}\cosh\lambda t}\,dt,
\qquad 0\le\lambda<\frac{\pi}{2}\\ 
-f_\rho(u,\xi)&=\log{\kappa_\rho}(u,\xi)=2\int_{-\infty}^{\infty}\frac{\cosh(\xi+\XiShift-\pi)t\cosh(\xi-\XiShift) t\sinh ut\sinh(\lambda-u)t}{t\sinh\pi t\cosh\lambda t}\,dt,\ \; \rho\ge 2
\label{finalKappaRho}
\end{align}
where $\XiShift=\big[\frac{\xi+\rho\lambda}{\pi}\big]\pi$ and the first integral diverges at $\lambda=\frac{\pi}{2}$. 
This divergence occurs because the loop fugacity $\beta=2\cos\lambda=0$ vanishes for critical dense polymers ${\cal LM}(1,2)$ causing $\kappa_0(u)$ to vanish. 
The first integral giving the vacuum boundary free energy is analytically continued in $\lambda$ in Appendix~\ref{analyticcontinuation} to give
\begin{align}
-f_0(u)=\log\kappa_0(u)&=\log\frac{\cos u\cos(\lambda-u)}{2\cos\lambda\cos(u-\tfrac{\lambda}{2})}-\int_{-\infty}^{\infty}\frac{\sinh ut\sinh(\lambda-u)t}{t\sinh\pi t\cosh\lambda t}dt
\nonumber\\
&\quad-\int_{-\infty}^{\infty}\frac{\sinh{\lambda t\over2}\sinh({3\lambda\over2}-\pi)t\cosh(\lambda-2u)t}{t\sinh\pi t\cosh\lambda t}dt,\qquad 0\le\lambda\le\pi
\label{finalKappa0}
\end{align}
In accord with the fact that $D_0(u,\xi)=1$, we see that
\bea
\kappa_{bulk}(0)=\kappa_0(0)=\kappa_\rho(0,\xi)=1
\eea

\subsubsection{Hamiltonian limits of bulk and boundary free energies}

In this section we take the Hamiltonian limit of the ground state eigenvalue of $D_0(u,\xi)$. From 
(\ref{HamLimit}) and (\ref{uExp}), we find
\bea
-\half \sin\lambda \lim_{u\to 0} \frac{\partial}{\partial u} \log D_0(u,\xi)=
N\cos\lambda-\beta^{-1}-\half\sin\lambda(\cot\xi\!-\!\cot\xi_\rho)+\calE
\label{calElimit}
\eea
where 
\bea
\calE=N\calE_{bulk}+\calE_{bdy},\qquad \calE_{bdy}=\calE_0+\calE_\rho(\xi)\qquad D_0(u,\xi)\sim \kappa_{bulk}(u)^{2N}\kappa_0(u)\kappa_\rho(u,\xi),\quad N\to\infty
\eea
Keeping just the $\mbox{O}(N)$ terms gives
\bea
-\calE_{bulk}=\cos\lambda+\sin\lambda\,\kappa_{bulk}'(0)=\cos\lambda+\sin\lambda \int_{-\infty}^\infty \frac{\cosh(\pi-2\lambda)t \tanh \lambda t}{\sinh\pi t}\,dt
\eea 
Similarly, setting $\rho=1$, neglecting the $\mbox{O}(N)$ terms and keeping only $\beta^{-1}$ of the $\mbox{O}(1)$ terms in (\ref{calElimit}) gives 
\bea
\calE_0=\half \sin\lambda\int_{-\infty}^{\infty}\frac{\tanh\lambda t \big[1-2\sinh\tfrac{\lambda t}{2}\sinh(\tfrac{3\lambda}{2}-\pi)t\big]}{\sinh\pi t}\,dt+\half
\eea
Finally, keeping the remaining $\mbox{O}(1)$ terms in (\ref{calElimit}) gives 
\bea
-\calE_\rho(\xi)=\sin\lambda \int_{-\infty}^\infty \frac{\cosh(\xi+\XiShift-\pi)t\cosh(\xi-\XiShift) t\sinh\lambda t}{\sinh\pi t\cosh \lambda t}\,dt
-\half\sin\lambda(\cot\xi\!-\!\cot\xi_\rho)
\eea

\begin{figure}[p]
\subfloat[$\rho=1$]{\includegraphics[width = 8cm]{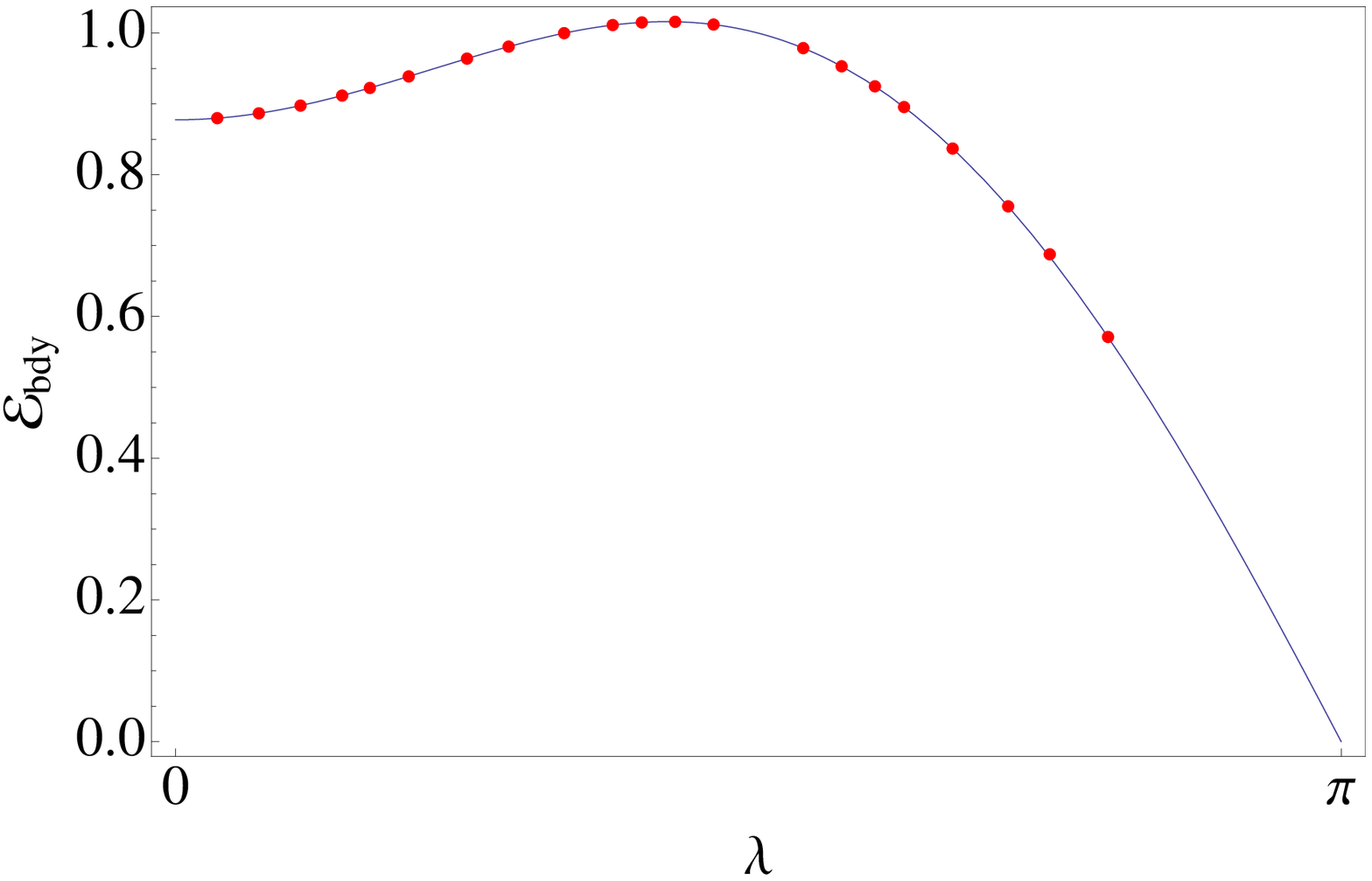}} 
\subfloat[$\rho=2$]{\includegraphics[width = 8cm]{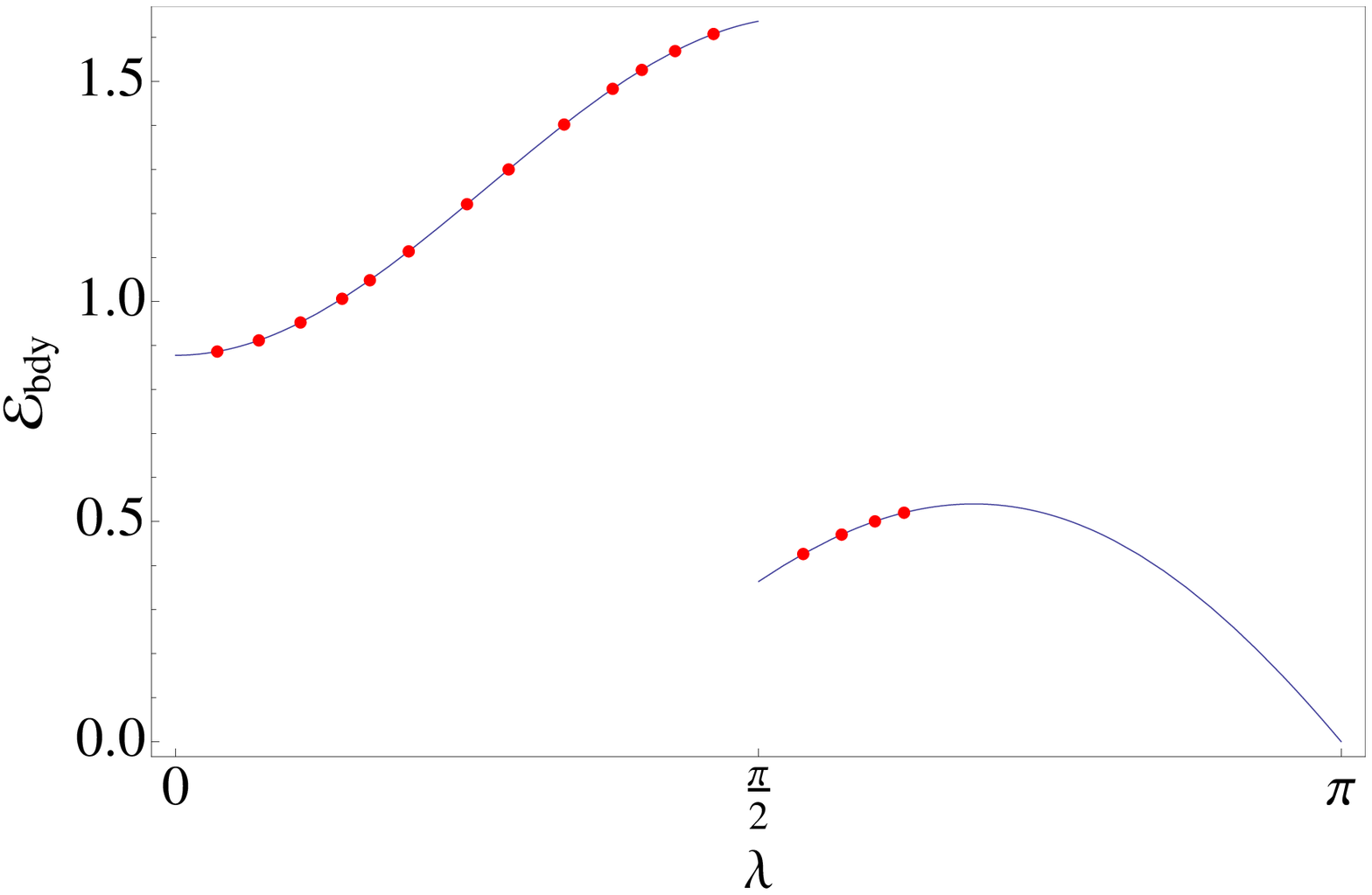}}\\
\subfloat[$\rho=3$]{\includegraphics[width = 8cm]{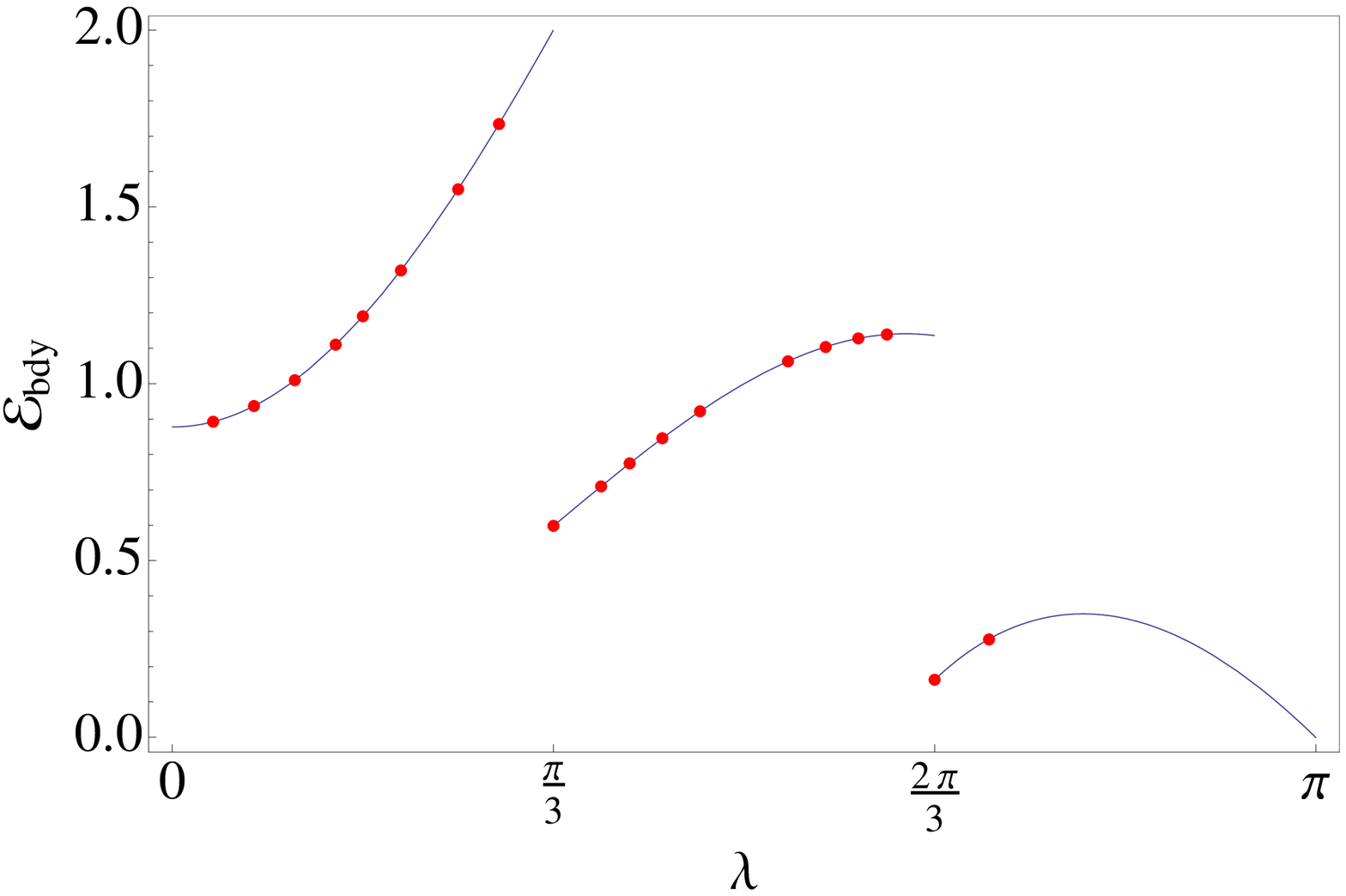}}
\subfloat[$\rho=4$]{\includegraphics[width = 8cm]{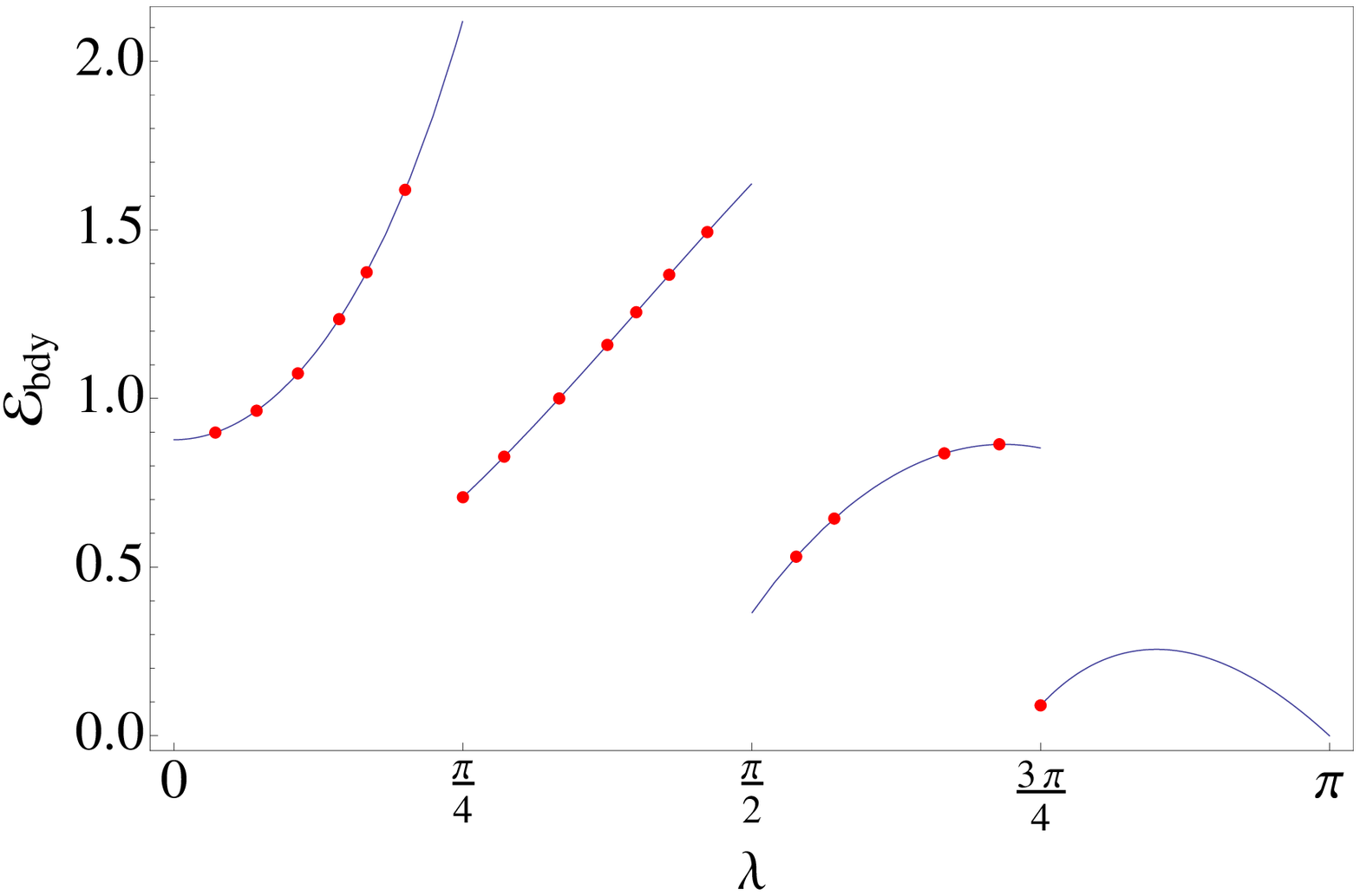}} 
\caption{\label{fig:FBdyPlot}Plots and numerical estimates of the Hamiltonian boundary energies $\calE_{bdy}(\rho,\xi)=\calE_0+\calE_\rho(\xi)$ as a function of $\lambda$ for $\rho = 1,2,3,4$, where $\xi=\frac{\pi}{2}$ if $[\frac{\rho}{p'}]=0$ and $\xi=[\frac{\rho p}{p'}]\frac{\pi}{2}$ otherwise. The jump discontinuities occur at points where $[\frac{\rho}{p'}]=0$ and $\xi = \frac{\pi}{2}$.}
\end{figure}

\begin{figure}[p]
\subfloat{
\includegraphics[width=8.2cm]{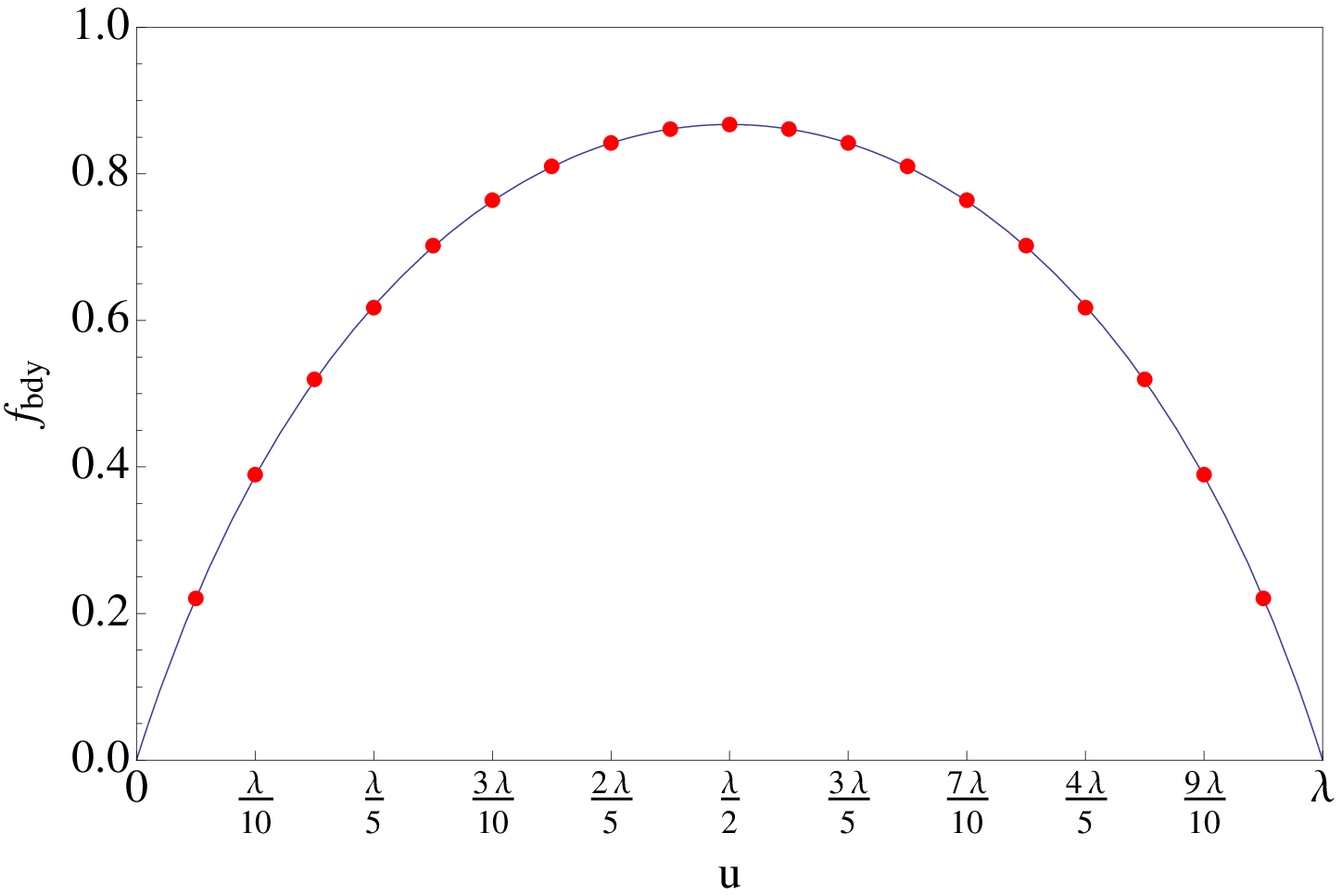}
}
\subfloat{
\includegraphics[width=8cm]{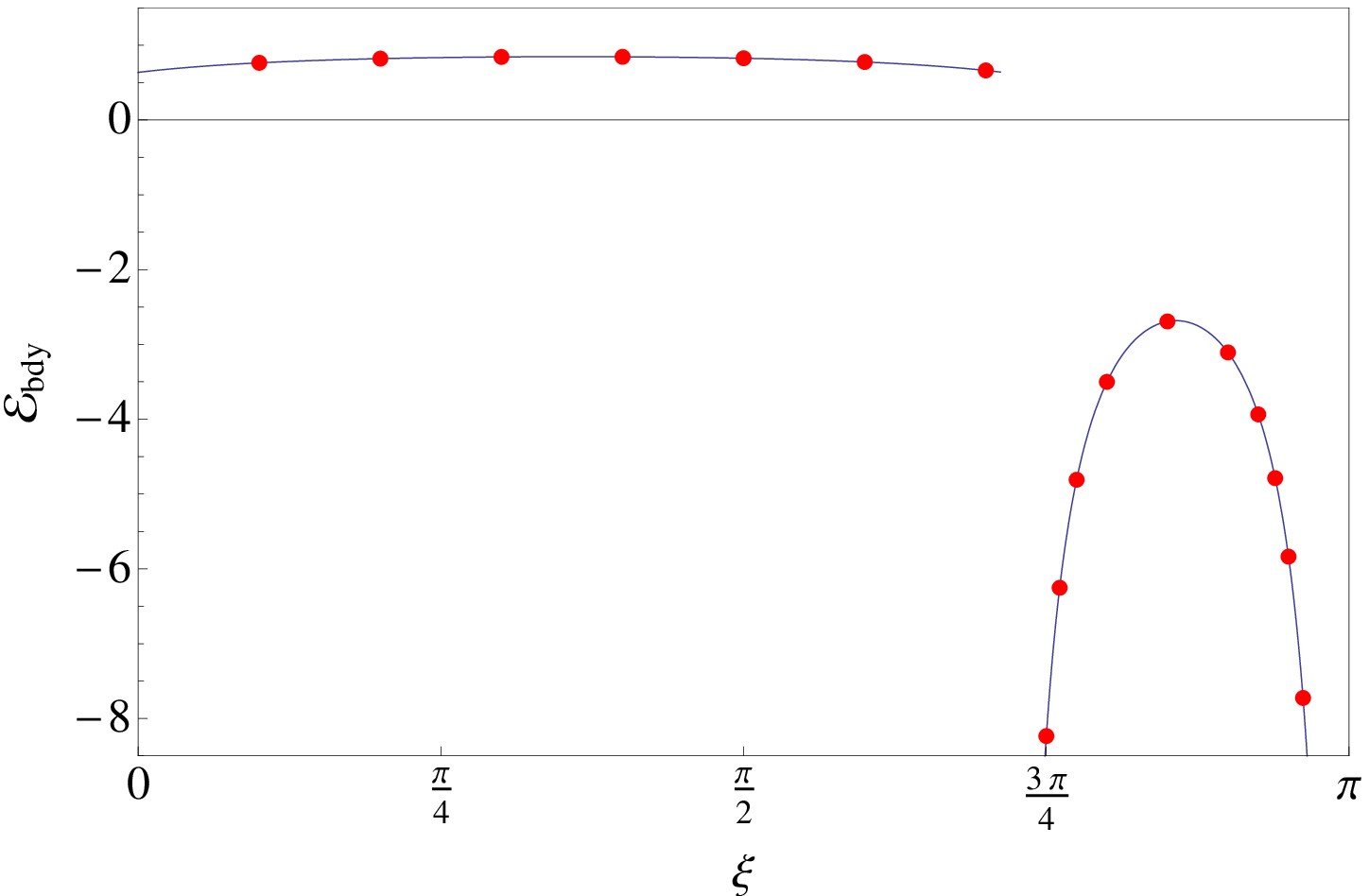}
}
\caption{\label{fig:uxiBdyPlots}Plots and numerical estimates for ${\cal LM}(4,7)$ with $\rho=3$ boundary: (i) the boundary free energy $f_{bdy}(u,\rho,\xi)$ of the double row transfer matrix $\vec D(u,\xi)$ plotted against $u$ in the physical interval with the specialized value of $\xi=[\frac{\rho p}{p'}]\frac{\pi}{2}=\frac{5\pi}{14}$, (ii) the boundary  energy $\calE_{bdy}(\rho,\xi)=\calE_0+\calE_\rho(\xi)$ of the Hamiltonian ${\cal H}^{(\rho,1)}$ plotted against $\xi$.}
\end{figure}

Our analytic formulas for the transfer matrix boundary free energies and Hamiltonian boundary energies have been extensively checked by numerics. Various plots of the Hamiltonian boundary energies 
$\calE_{bdy}(\rho,\xi)=\calE_0+\calE_\rho(\xi)$ are shown in Figures~\ref{fig:FBdyPlot} and \ref{fig:uxiBdyPlots}. In all cases, the agreement between the analytic formulas and numerical estimates is good.


\section{Numerical Finite-Size Spectra}
\label{numericalSection}

In this section, we present our numerical results for the extrapolated eigenvalues of the double row transfer matrix $\vec D(u,\xi)$ (\ref{DTM}) and the associated quantum Hamiltonian ${\cal H}^{(\rho,s)}$ (\ref{Hamrs}) with $(r,s)$ boundary conditions on the right edge of the strip. Numerically, we find all of these eigenvalues are real.

\subsection{Finite-size corrections}
\label{FiniteSize}

Consider the ${\cal LM}(p,p')$ lattice models on a strip with $N$ columns and $N'$ double rows and $(r,s)$ boundary conditions applied on the right edge. 
The lattice partition function is 
\be 
Z_{(1,1)|(r,s)}^{(N,N')}=\mbox{Tr}\,\vec D(u,\xi)^{N'}=\sum_j D_j(u,\xi)^{N'} =\sum_j e^{-N' E_j(u,\xi)},\qquad j=0,1,2,3,\ldots
\ee
where $D_j(u,\xi)$ are the eigenvalues of $\drtm(u,\xi)$ and $E_j$ are their associated energies. In the thermodynamic limit, only the ground state eigenvalue $D_0(u,\xi)$ of the double row transfer matrix in each $(r,s)$ sector contributes to the lattice partition function. 

The conformal data of interest is accessible~\cite{BCN,Aff} through the finite-size corrections to the eigenvalues of the transfer matrix or associated Hamiltonian. 
For the double row transfer matrix eigenvalues, the leading finite-size corrections for large $N$ take the form
\be 
E_j=-\log D_j(u,\xi)=2Nf_{bulk}(u)+f_{bdy}(u,\rho,\xi)+\frac{2\pi\sin\vartheta}{N}\Big(\!-\frac{c}{24}+\Delta_{r,s}+k\Big)+...,\quad k=0,1,2,... \label{fsdtm}
\ee
where $k$ labels the level in the conformal tower. 
The anisotropy angle $\vartheta$~\cite{KimPearce87} and modular nome $q$ are
\bea
\vartheta=\frac{\pi u}{\lambda},\qquad \lambda=\frac{(p'-p)\pi}{p'},\qquad 
q=\exp\!\Big(\!\!-2\pi\,\frac{N'}{N}\sin\vartheta\Big)
\eea
The central charge of the CFT is $c$ while the 
spectrum of conformal weights is given by the possible values of $\Delta_{r,s}$ with excitations or descendants labelled by the non-negative integers $k$.
Similarly, for the associated quantum Hamiltonian $\mathcal{H}^{(\rho,s)}$, the finite-size corrections of the eigenenergies take the form
\be 
E_j=N\calE_{bulk}+\calE_{bdy}(\rho,\xi)+\frac{\pi v_s}{N}\Big(\!-\frac{c}{24}+\Delta_{r,s}+k\Big)+...,\quad k=0,1,2,... 
\label{eq:finite_size_corr_ham}
\ee
where $v_s=\frac{\pi \sin\lambda}{\lambda}$ is the velocity of sound. The Hamiltonian energies $\calE_{bulk}$ and $\calE_{bdy}(\rho,\xi)=\calE_0+\calE_\rho(\xi)$ are determined, up to the shift of the ground state energy, by evaluating the derivative at $u=0$ of $f_{bulk}(u)$ and $f_{bdy}(u,\rho,\xi)$ respectively.

\begin{figure}[p]
\centerline{\includegraphics[width=17cm]{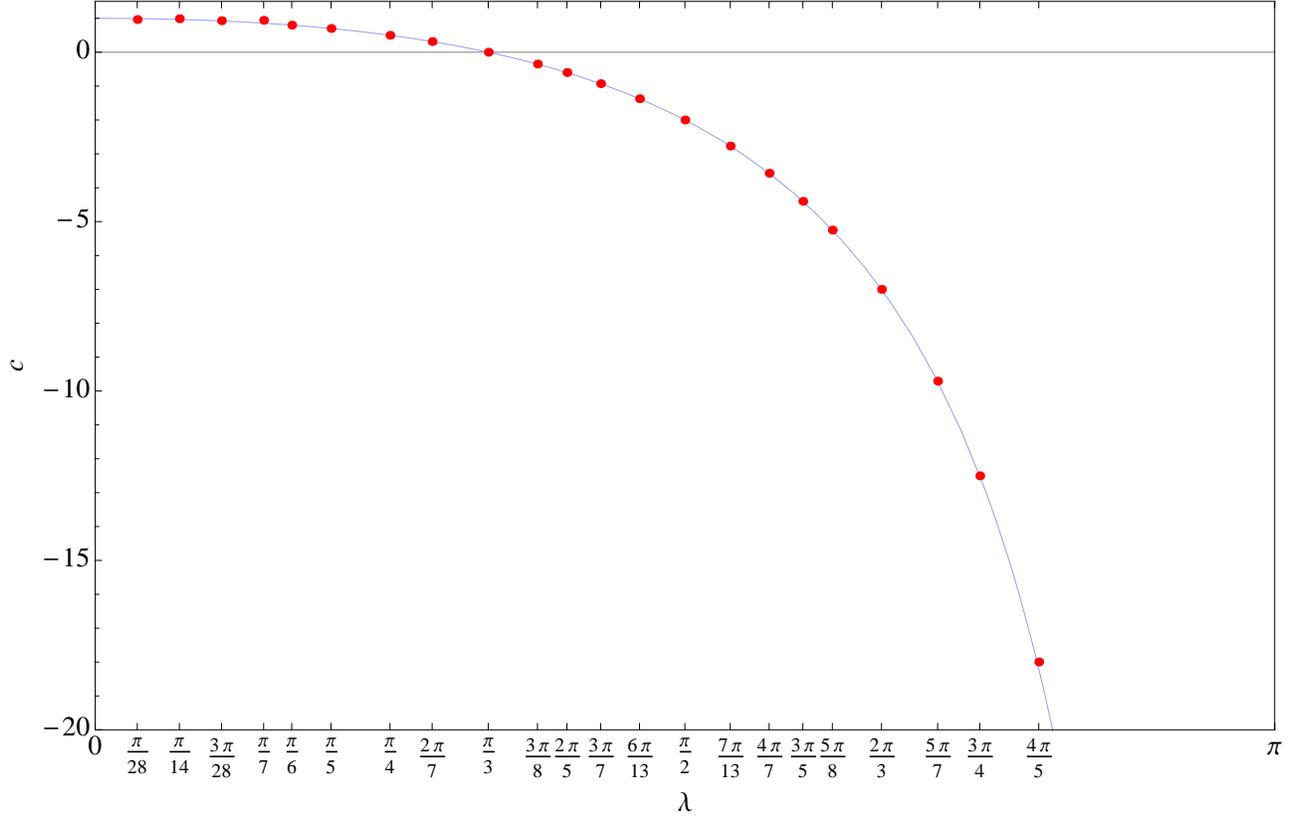}}
\caption{Plot of the central charge $c=(\frac{2\lambda}{\pi}+1)(\frac{3\lambda}{\pi}-1)/(\frac{\lambda}{\pi}-1)$ of the ${\cal LM}(p,p')$ models as a function of $\lambda$. The numerical estimates are obtained from the lowest eigenvalue of the Hamiltonian ${\cal H}^{(1,1)}$ with $(\rho,s)=(r,s)=(1,1)$ boundary conditions.}
\label{cPlot}
\end{figure}
\begin{table}[p]
\center
\begin{tabular}{|c| c | c c| c | c |}
\hline
\multicolumn{6}{|c|}{\rule{0pt}{14pt}Conformal Weights $\Delta_{3,1}$}\\[3pt]
\hline
$(p,p')$ &{$\lambda/\pi$}
	      &\multicolumn{2}{|c|}{Exact} & \multicolumn{1}{|c|}{Estimated}& Error \\
\hline
 $(6,7)$ & 0.143  & $4/3$   & $1.333$ & $1.324$ & $ 0.7\%$\\
  $(5,6)$ & 0.167 & $7/5 $  & $1.4$     & $1.391$ & $ 0.6\%$ \\
$(4,5)$ &  0.2      & $3/2$   & $1.5$     & $1.49996$ & $ 0.003\%$ \\
$(3,4)$ &  0.25  &  $5/3$    & $1.667$ &$1.6671$    &  $ 0.03\%$ \\
$(5,7)$ & 0.286   & $9/5$   & $1.8$     &$1.80004$  & $ 0.002\%$\\
$(2,3)$ & 0.333   & $2$      & $2$        & $2.00005$ & $ 0.002\%$\\
$(3,5)$ &  0.4      & $7/3$   & $2.333$ &$2.3330$  & $ 0.01\%$\\
$(4,7)$ & 0.429   & $5/2$   & $2.5$     &$2.501$  & $ 0.05\%$\\
$(2,5)$ & 0.6       & $4$      & $4$        &$3.9990$  & $ 0.03\%$\\
$(2,7)$ & 0.714   & $11/3$ & $3.667$ &$3.6670$  & $ 0.009\%$\\
\hline
\end{tabular}
\caption{Table comparing exact and numerical estimates of the conformal weight $\Delta^{p,p'}_{3,1}$ for various ${\cal LM}(p,p')$ models. The actual errors are shown in the last column. The numerical estimates are obtained from the lowest eigenvalue of the Hamiltonian ${\cal H}^{(\rho,1)}$ where the values of $\rho=\rho(r)$ and $\xi$ are fixed by (\ref{fixRhoXi}). This data corresponds to the purple curve in Figure \ref{deltar1Plot}.}
\label{deltaTab}
\end{table}
\begin{figure}[p]
\centerline{\includegraphics[width=14.75cm]{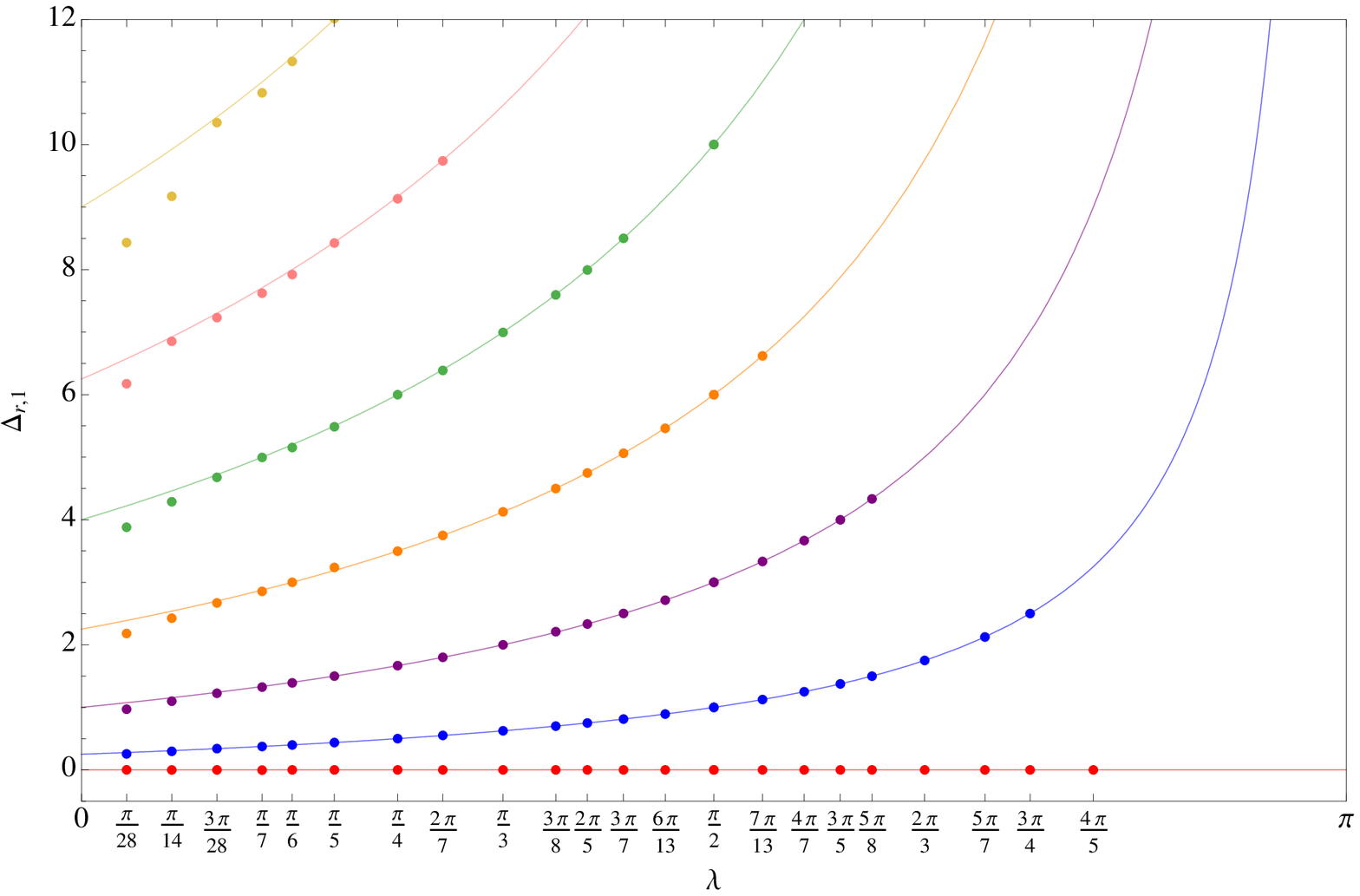}}
\caption{Plots of the conformal weights $\Delta^{p,p'}_{r,1}=(r-1)(\frac{2\lambda}{\pi}+r-1)/4(1-\frac{\lambda}{\pi})$ as a function of $\lambda$ with $r=1,2,\ldots,7$ and $s=1$. The numerical estimates are obtained from the lowest eigenvalue of the Hamiltonian ${\cal H}^{(\rho,1)}$ where the values of $\rho=\rho(r)$ and $\xi$ are fixed by (\ref{fixRhoXi}).}
\label{deltar1Plot}
\end{figure}
\begin{figure}[p]
\centerline{\includegraphics[width=14.75cm]{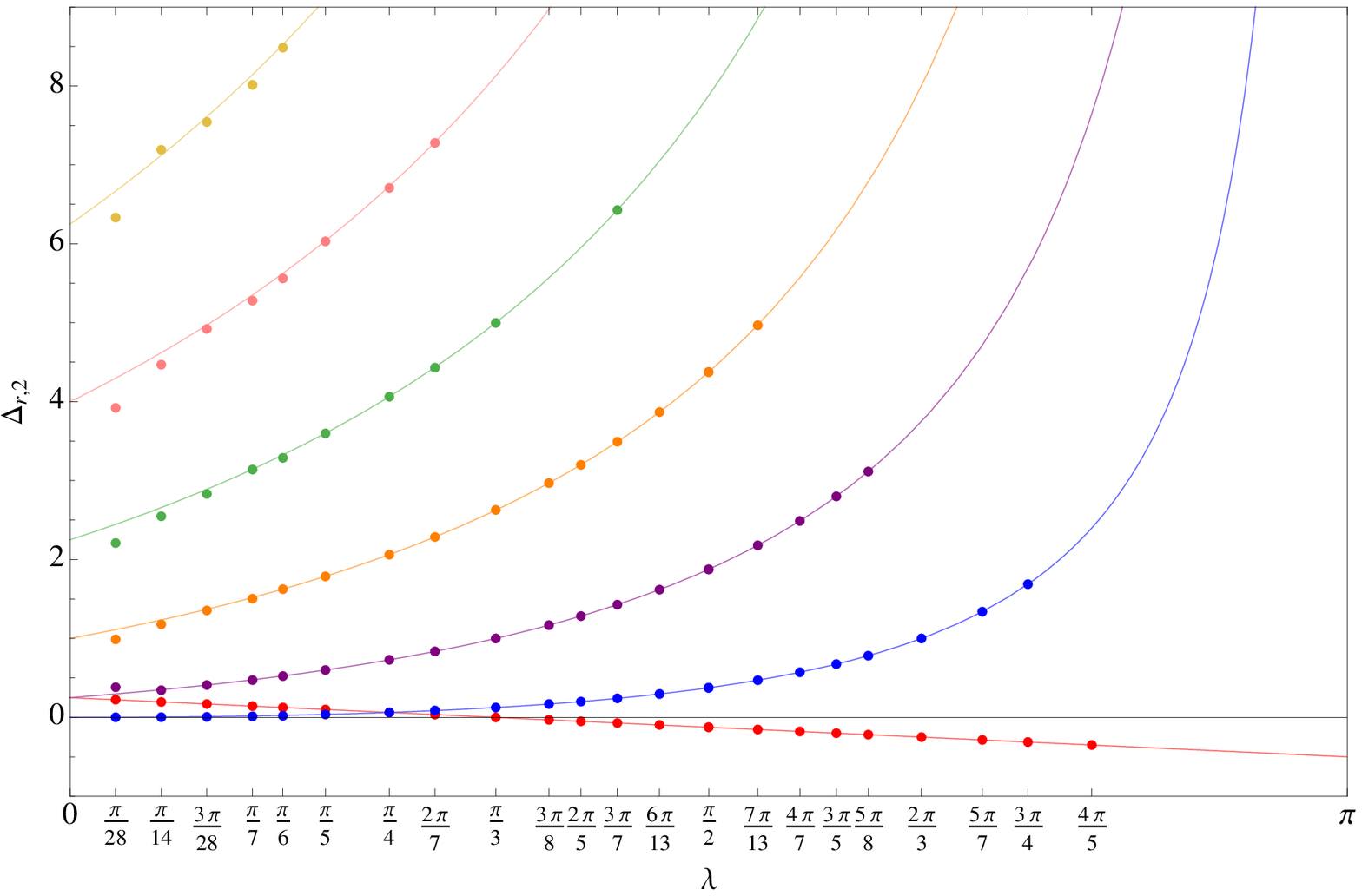}}
\caption{Plots of the conformal weights $\Delta_{r,2}=(\frac{\lambda}{\pi}+r-2)(\frac{3\lambda}{\pi}+r-2)/4(1-\frac{\lambda}{\pi})$ as a function of $\lambda$ with $r=1,2,\ldots,7$ and $s=2$. The numerical estimates are obtained from the lowest eigenvalue of the Hamiltonian ${\cal H}^{(\rho,2)}$ where the values of $\rho=\rho(r)$ and $\xi$ are fixed by (\ref{fixRhoXi}).}
\label{deltar2Plot}
\end{figure}

\subsection{Numerical central charges and conformal weights}
\label{numerical}
In this section, we present details of the numerical calculations for the central charges, conformal weights and conformal partition functions of the ${\cal LM}(p,p)$ models with $(r,s)$ boundary conditions. 
Since it is numerically more efficient, we calculate the conformal spectra using the
Hamiltonians and not the double row transfer matrices. 
Our numerics indicate that the conformal weights $\Delta_{r,s}$ (\ref{confWts}) are indeed obtained by applying boundary conditions on the right edge of the strip implemented by an $r$-type seam of width $\rho-1$ with the boundary field $\xi$ fixed to the specialized value
\be
\rho= \rho(r)= \Big\lfloor \frac{rp'}{p} \Big\rfloor,\qquad \xi=\begin{cases}
	 [ \frac{\rho p}{p'}] \frac{\pi}{2}, &\quad [\frac{\rho}{p'}] \neq 0\\[4pt]
	\frac{\pi}{2}, &\quad [\frac{\rho}{p'}]= 0
\end{cases}\label{fixRhoXi}
\ee
This confirms the conjecture of Pearce, Villani and Rasmussen~\cite{PRV}.

All our numerics were carried out in Mathematica~\cite{Wolfram}. Matrix representatives of the TL generators were obtained as sparse matrices. Due to memory constraints we were limited to bulk sizes out to \mbox{$N\le 32-\rho-s$}. 
For given $p,p',\rho$ and $s$, the numerical Hamiltonian matrices were constructed using the TL generators. The lowest eigenenergies of the Hamiltonians were obtained using Arnoldi methods~\cite{Arnoldi}. From (\ref{eq:finite_size_corr_ham}), estimates of the central charges and conformal weights were obtained using finite length sequences with increasing $N$ to extrapolate the lowest eigenenergy of the Hamiltonian in a given $(r,s)$ sector. The accuracy of the extrapolation to $N=\infty$ is improved by using Vanden Broeck Schwartz sequence 
acceleration~\cite{vandenBroeckSchwartz79}. Since these extrapolated estimates are sensitive to the values of the boundary energies, it is important that the boundary energies are known exactly. 
The numerical estimates of the central charge $c^{p,p'}$ are plotted against the exact analytic curve in Figure \ref{cPlot}. Numerical estimates of the conformal weight $\Delta_{3,1}^{p,p'}$ for various ${\cal LM}(p,p')$ models are tabulated against the exact values in Table~\ref{deltaTab}. This table also gives an indication of the typical errors incurred in our numerical extrapolations. The plots in Figures~\ref{deltar1Plot} and \ref{deltar2Plot} summarize our numerical estimates of the conformal weights $\Delta_{r,s}$ for $s=1,2$ with $r=1,2,\ldots,7$.

\subsection{Numerical conformal partition functions}

The ground state eigenenergy of the Hamiltonian was used, in the various $(r,s)$ sectors, to numerically estimate the central charge and conformal weights. Similarly, the eigenvalues of the first 6--14 finite excitations above the ground state were used to numerically estimate low lying excitations in the conformal towers. Extrapolating sequences based on the finite-size corrections  \eqref{eq:finite_size_corr_ham}, we are able to correctly reproduce the first 3--6 coefficients in the $q$-series of the Kac characters \eqref{KacChars}
\begin{align}
\chi_{1,1}^{p,p'}(q) &=q^{-\frac{c}{24}+\Delta_{1,1}^{p,p'}} (1 + q^2 + q^3 + 2 q^4 + 2 q^5 + 4 q^6 + 4 q^7 + \ldots) \\
\chi_{2,1}^{p,p'}(q) &=q^{-\frac{c}{24}+\Delta_{2,1}^{p,p'}} (1 + q + q^2 + 2 q^3 + 3 q^4 + 4 q^5 + \ldots) \\
\chi_{3,1}^{p,p'}(q) &=q^{-\frac{c}{24}+\Delta_{3,1}^{p,p'}} (1 + q + 2 q^2 + 2 q^3 + 4 q^4 + \ldots) \\
\chi_{4,1}^{p,p'}(q) &=q^{-\frac{c}{24}+\Delta_{4,1}^{p,p'}} (1 + q + 2 q^2 + 3 q^3 + \ldots) \\
\chi_{5,1}^{p,p'}(q) &=q^{-\frac{c}{24}+\Delta_{5,1}^{p,p'}} (1 + q + 2 q^2 + 3 q^3 + \ldots)
\end{align}
where the dependence on $p,p'$ appears only through the conformal weight in the leading fractional power of $q$. 
These and similar results for $s>1$ hold independently of the choice of $p,p'$ and $\rho$ though the precision of the estimates decreases as $\rho$ increases.

\subsection{Finitized characters}

For given fixed $\rho$, the finitized characters are
\bea
\chi^{p,p';(N)}_{r,s}\!(q)
= q^{-{c\over 24}+\Delta_{r,s}^{p,p'}}\,\Big(\gauss{N}{\rule{0pt}{12pt}{N-\rho+s\over 2}}_{\!q}-q^{rs}\gauss{N}{\rule{0pt}{12pt}{N-\rho-s\over 2}}_{\!q}\Big),\qquad
r=r(\rho)=\Big\lceil \frac{\rho p}{p'}\Big\rceil, \qquad \rho=1,2,3,\ldots
\eea
In particular, there can be several finitized characters for given fixed $r$. These finitized characters generalize the finitized characters given in \cite{PRZ}. 
For $q=1$, the finitized characters reduce to the formulas (\ref{countstates}) giving the dimension of the vector space of link states. 
As $N\to\infty$, the finitized characters reproduce the full characters (\ref{KacChars}) since
\bea
 \gauss{N}{n}=\frac{(q)_N}{(q)_{N-n} (q)_n},\qquad \lim_{N\to\infty} \gauss{N}{\rule{0pt}{12pt}\frac{1}{2}N+n}=\frac{1}{(q)_\infty}
\eea
Finally, the finitized characters are fermionic in the sense that are given by $q$-series with nonegative coefficients.

\section{Conclusion}
\label{Conclusion}

In this paper, we have revisited and developed further the study of the logarithmic minimal models ${\cal LM}(p,p')$~\cite{PRZ} with $(r,s)$ Kac boundary conditions. Following~\cite{PRV}, a general construction of these boundary conditions is implemented using diagrammatic fusion~\cite{PRT2014,MDPR14} on $r$- and $s$-type seams of width $\rho-1$ and $s-1$ columns respectively.  
In contrast to the usual fusion construction~\cite{KulReshSkly,BP96,BPO96,BP2001}, based on Wenzl-Jones projectors~\cite{WenzlJones,Wenzl}, this construction is well defined for arbitrary large $\rho$ and $r$.
Applying the $(r,s)$ boundary conditions on the right edge of a strip, our numerics indicate that the known conformal weights $\Delta_{r,s}$ (\ref{confWts}) are indeed obtained
when $\rho=\rho(r)=\lfloor \frac{rp'}{p}\rfloor$ and the boundary field $\xi$ is fixed to the specialized value given in (\ref{fixRhoXi}). 
These results confirm the conjecture of Pearce, Villani and Rasmussen~\cite{PRV}. In principle, the conformal weights $\Delta_{r,s}$ could be calculated analytically using the functional equation methods of Kl\"umper and Pearce~\cite{KlumperPearce1992}. However, so far, the logarithmic $T$- and $Y$-systems have only been derived~\cite{MDPR14} for the vacuum boundary condition $(r,s)=(1,1)$ so this remains an open problem. 

In this paper, the elementary $(r,s)=(1,1)$ {\it vacuum} solution of the BYBE is dressed (\ref{dressedVac}) by $r$- and $s$-type seams to yield integrable $(r,s)$ boundary conditions conjugate to the $(r,s)$ Kac representations and labelled by the integer Kac labels $r,s=1,2,3,\ldots$. These boundary conditions are {\it Neumann} boundary conditions since  loop segments are reflected at the boundary. Introducing elementary boundary triangles of {\it Dirichlet} type~\cite{JSbdy} allows the loop segments to terminate on the boundary. 
An elementary solution of the BYBE is then given by the {\it twist} boundary condition 
\be
\psset{unit=.6cm}
\begin{pspicture}[shift=-0.89](0,0)(1,2)
\pspolygon[fillstyle=solid,fillcolor=lightlightblue](0,1)(1,2)(1,0)(0,1)
\rput(0.65,1){$_u$}
\end{pspicture}
\ =\Gamma(u)\
\begin{pspicture}[shift=-0.89](0,0)(1,2)
\pspolygon[fillstyle=solid,fillcolor=lightlightblue](0,1)(1,2)(1,0)(0,1)
\psarc[linewidth=1.5pt,linecolor=blue](0,1){.7}{-45}{45}
\end{pspicture}
\ +s_0(2u)\
\begin{pspicture}[shift=-0.89](0,0)(1,2)
\pspolygon[fillstyle=solid,fillcolor=lightlightblue](0,1)(1,2)(1,0)(0,1)
\psline[linecolor=blue,linewidth=1.5pt]{-}(0.4,0.6)(1,0.6)
\psline[linecolor=blue,linewidth=1.5pt]{-}(0.4,1.4)(1,1.4)
\end{pspicture}\,,\qquad  \Gamma(u)=\beta_{bdy}\,s_1(\xi-u)[s_1(\xi+u)-s_0(\xi+u)]
\label{twist}
\ee
where $\beta_{bdy}=1$ is the fugacity of the loops that terminate on the boundary. General integrable Robin boundary conditions~\cite{Robin}, which are linear combinations of Neumann and Dirichlet boundary conditions, can be constructed~\cite{PRT14} by dressing the elementary twist boundary condition with $r$- and $s$-type seams. Indeed, critical dense polymers ${\cal LM}(1,2)$ has recently~\cite{PRT14} been solved exactly with Robin boundary conditions yielding conformal weights with half-integer Kac labels
\bea
\Delta_{r,s-\frac{1}{2}}=\tfrac{1}{32}(L^2-4),\qquad L=2s-1-4r,\qquad r,s\in{\Bbb N}
\eea
It would be of interest to repeat the numerical calculations of this paper for the ${\cal LM}(p,p')$ models with Robin boundary conditions to see if the appearance of half-integer Kac labels persists for the general logarithmic minimal models.

\section*{Acknowledgments} Paul Pearce is supported by a Melbourne Research Grant Support Scheme grant. Elena Tartaglia is supported by an Australian Postgraduate Award. We thank J{\o}rgen Rasmussen for discussions and input in the early stages of this project.

\appendix
\section{Properties of the Generalized Temperley-Lieb Projectors}
\label{genTLproj}
In this Appendix, we prove the four properties, given in \eqref{eq:genProjProp1} and \eqref{eq:genProjProp2}, of the generalized Temperley-Lieb projectors \eqref{eq:genproj}
\be
P_N^{(\rho)} = \sum_{k=0}^{\rho-2} (-1)^k U_{\rho-k-2}(\beta/2) e_N^{(k)}, \qquad e_N^{(k)}=\prod_{j=0}^k e_{N+j}
\ee
These properties hold when we impose the diagrammatic fusion rule that the link states cannot form any half-arcs in the $\rho-1$ sites of the boundary. 
These operators are thus restricted to act from the vector space ${\cal V}_{\rho,1}^{(N)}$ to itself or from ${\cal V}_{\rho,s}^{(N)}$ to itself. This action kills any word in the TL algebra not of the form $I$ or $e_j^{(k)}$.

\paragraph{Property 1}
\be
P^{(\rho')}_j P^{(\rho)}_j = U_{\rho'-1}(\beta/2) P^{(\rho)}_j, \qquad N \leq j \leq N+\rho - 1 
\ee
We first rewrite the left side using the definition of the generalized TL projectors and write out the first few terms of the summations to make simplifications easier
\begin{align}
P^{(\rho')}_j P^{(\rho)}_j &=  \Big[\sum_{k=0}^{\rho'-2} (-1)^k U_{\rho'-k-2}(\beta/2) e_N^{(k)} \Big]  \Big[\sum_{n=0}^{\rho-2} (-1)^n U_{\rho-n-2}(\beta/2) e_N^{(n)}\Big]\nn
&= \Big[U_{\rho'-2}(\beta/2)e_N - U_{\rho'-3}(\beta/2)e_Ne_{N+1} + \sum_{k=2}^{\rho'-2} (-1)^k U_{\rho'-k-2}(\beta/2) e_N\ldots e_{N+k} \Big]\nn
& \times \Big[U_{\rho-2}(\beta/2)e_N - U_{\rho-3}(\beta/2)e_Ne_{N+1} + \sum_{n=2}^{\rho-2} (-1)^n U_{\rho-n-2}(\beta/2) e_N\ldots e_{N+n} \Big]\nonumber
\end{align}
Expanding out all 9 terms gives
\begin{align}
P^{(\rho')}_j P^{(\rho)}_j &=
U_{\rho'-2}(\beta/2)U_{\rho-2}(\beta/2)e^2_N - U_{\rho'-2}(\beta/2)U_{\rho-3}(\beta/2)e^2_Ne_{N+1}\nn
&+ \sum_{n=2}^{\rho-2} (-1)^n U_{\rho'-2}(\beta/2)U_{\rho-n-2}(\beta/2) e^2_Ne_{N+1}\ldots e_{N+n}
- U_{\rho'-3}(\beta/2)U_{\rho-2}(\beta/2)e_Ne_{N+1}e_N \nn
& + U_{\rho'-3}(\beta/2)U_{\rho-3}(\beta/2)e_Ne_{N+1}e_Ne_{N+1}
 - \sum_{n=2}^{\rho-2} (-1)^n U_{\rho'-3}(\beta/2)U_{\rho-n-2}(\beta/2) e_Ne_{N+1}e_N\ldots e_{N+n}\nn
 &+\sum_{k=2}^{\rho'-2} (-1)^k U_{\rho'-k-2}(\beta/2) U_{\rho-2}(\beta/2) e_N\ldots e_{N+k}e_N \nn
& - \sum_{k=2}^{\rho'-2} (-1)^k U_{\rho'-k-2}(\beta/2) U_{\rho-3}(\beta/2)e_N\ldots e_{N+k}e_Ne_{N+1}\nn
 &+\sum_{k=2}^{\rho'-2}  \sum_{n=2}^{\rho-2} (-1)^{k+n} U_{\rho'-k-2}(\beta/2) U_{\rho-n-2}(\beta/2) e_N\ldots e_{N+k} e_N\ldots e_{N+n}\nonumber
\end{align}
The last three terms are are killed, by the restricted action on link states, because they would form half-arcs in the boundary when acting on any link state. 
Algebraically, these terms cannot be written in the form $I$ or $e_j^{(k)}$ so they are killed.
For example, term 7  is a scalar multiple of the TL word
\be
e_Ne_{N+1}\ldots e_{N+k}e_N=
\psset{unit=0.7cm}
\begin{pspicture}[shift=-2.25](-0.2,-0.5)(6.75,4.5)
\psarc(0.5,0){0.5}{0}{180}
\psarc(5.5,2.5){0.5}{180}{0}
\psbezier[showpoints=false](0,2.5)(0,1.5)(2,1)(2,0)
\psbezier[showpoints=false](1,2.5)(1,1.5)(3,1)(3,0)
\psbezier[showpoints=false](2,2.5)(2,1.5)(4,1)(4,0)
\psbezier[showpoints=false](4,2.5)(4,1.5)(6,1)(6,0)
\rput(0,2.5){\psarc(0.5,0){0.5}{0}{180}}
\rput(0,3.8){\psarc(0.5,0){0.5}{180}{0}}
\multirput(0,0)(1,0){3}{\psline(4,2.5)(4,3.8)}
\psline(2,2.5)(2,3.8)
\rput(4,1.25){\ldots}
\rput(3,3){\ldots}
\psline[linecolor=red,linestyle=dashed](0.5,0)(0.5,3.8)
\rput(0,-0.25){\scriptsize $N$}
\rput(1,-0.25){\scriptsize $N\!+\!1$}
\rput(2,-0.25){\scriptsize $N\!+\!2$}
\rput(6,-0.25){\scriptsize $N+k$}
\end{pspicture}
=
\begin{pspicture}[shift=-1.75](-0.2,-0.5)(6.75,2.5)
\psarc(0.5,0){0.5}{0}{180}
\psarc(5.5,2.5){0.5}{180}{0}
\psbezier[showpoints=false](2,2.5)(2,1.5)(4,1)(4,0)
\psbezier[showpoints=false](4,2.5)(4,1.5)(6,1)(6,0)
\rput(2,0){\psarc(0.5,0){0.5}{0}{180}}
\rput(0,2.5){\psarc(0.5,0){0.5}{180}{0}}
\rput(4,1.25){\ldots}
\psline[linecolor=red,linestyle=dashed](0.5,0)(0.5,2.5)
\rput(0,-0.25){\scriptsize $N$}
\rput(1,-0.25){\scriptsize $N\!+\!1$}
\rput(2,-0.25){\scriptsize $N\!+\!2$}
\rput(6,-0.25){\scriptsize $N+k$}
\end{pspicture}
\ee
with a half-arc in the boundary, so it is killed. 

We next simplify the first 6 terms using the relations of the TL algebra (\ref{eq:TL3}) with $\beta=U_{1}(\beta/2)$
and the restricted action on link states. 
The remaining terms are regrouped. Combining terms 1 and 4, 2 and 5, 3 and 6 gives
\begin{align}
P^{(\rho')}_j P^{(\rho)}_j &=
\left[U_{\rho'-2}(\beta/2)U_1(\beta/2) - U_{\rho'-3}(\beta/2)\right] U_{\rho-2}(\beta/2)e_N \nn
&- \left[U_{\rho'-2}(\beta/2)U_1(\beta/2) - U_{\rho'-3}(\beta/2)\right]  U_{\rho-3}(\beta/2)e_Ne_{N+1} \nn
&+\sum_{n=2}^{\rho-2} (-1)^k \left[U_{\rho'-2}(\beta/2)U_1(\beta/2) - U_{\rho'-3}(\beta/2)\right] U_{\rho-n-2}(\beta/2)e_N\ldots e_{N+n}
\end{align}
Each term on the right side has the common factor 
\be
U_{\rho'-2}(\beta/2)U_1(\beta/2) - U_{\rho'-3}(\beta/2) = U_{\rho'-1}(\beta/2)
\ee
which simplifies using a standard identity for  Chebyshev polynomials of the second kind
\be
U_{m}(x)U_1(x) = U_{m-1}(x) + U_{m+1}(x)
\ee
Collecting the remaining terms together gives $P_N^{(\rho)}$ as required.

\paragraph{Property 2}
\be
P^{(\rho)}_{N+1} e_N = 0 \label{eq:genProj2}
\ee
Writing the left side using the definition of the generalized TL projector and rearranging using the relations of the TL algebra, we see that none of the words in the summation are of the form $I$ or $e^{(k)}$ so they are all killed
\begin{align}
P^{(\rho)}_{N+1} e_N &=
\Big(\sum_{k=0}^{\rho-2} (-1)^k U_{\rho-k-2}(\beta/2) e_{N+1}\ldots e_{N+k+1}\Big)e_N\nn
 &= \sum_{k=0}^{\rho-2} (-1)^k U_{\rho-k-2}(\beta/2) e_{N+1}e_N e_{N+2}\ldots e_{N+k+1} =0
\end{align}
This result is clear diagrammatically, as each term introduces a closed half-arc in the boundary.

\paragraph{Property 3}
\be
e_N P^{(\rho)}_{N+1} = U_{\rho-1}(\beta/2) e_N - P^{(\rho+1)}_N \label{eq:genProj3}
\ee
Expanding the right side and simplifying, using the relations of the TL algebra, gives the required result
\begin{align}
U_{\rho-1}(\beta/2) e_N - P^{(\rho+1)}_N &= 
U_{\rho-1}(\beta/2) e_N - \sum_{k=0}^{\rho-1}(-1)^k U_{\rho-k-1}(\beta/2) e_N \ldots e_{N+k}\nn
&=U_{\rho-1}(\beta/2) e_N - U_{\rho-1}(\beta/2) e_N - \sum_{k=1}^{\rho-1}(-1)^k U_{\rho-k-1}(\beta/2) e_N \ldots e_{N+k}\nn
&=-\sum_{k=0}^{\rho-2}(-1)^{k+1} U_{\rho-k-2}(\beta/2) e_N \ldots e_{N+k+1}
=e_N P_{N+1}^{(\rho)}
\end{align}

\paragraph{Property 4}
\be
e_N P^{(\rho)}_{N+1} e_N = U_{\rho-2}(\beta/2)e_N \label{eq:genProj4}
\ee
Expanding and simplifying the right side using the TL algebra gives the required result
\begin{align}
e_N P^{(\rho)}_{N+1} e_N &= 
e_N \Big[ \sum_{k=0}^{\rho-2}(-1)^{k} U_{\rho-k-2}(\beta/2) e_{N+1} \ldots e_{N+k+1} \Big] e_N\nn
&= \sum_{k=0}^{\rho-2}(-1)^{k} U_{\rho-k-2}(\beta/2) e_N e_{N+1}e_N e_{N+2} \ldots e_{N+k+1}\nn
&= U_{\rho-2}(\beta/2) e_N +  \sum_{k=1}^{\rho-2}(-1)^{k} U_{\rho-k-2}(\beta/2) e_N e_{N+2} \ldots e_{N+k+1}
=  U_{\rho-2}(\beta/2) e_N
\end{align}
In the last step, the summation is killed since each word introduces a closed half-arc in the boundary.

\section{Specialized Values of $\xi$}
\label{equivxi}
In this Appendix, we show that the two expressions in \eqref{eq:xiChoice} for the specialized value of $\xi$ are in fact equivalent
\be
\min\!\Big(\!-\frac{\rho\lambda}{2}\ \mbox{mod}\ \pi, \frac{\pi-\rho\lambda}{2}\ \mbox{mod}\ \pi \Big) = \Big[ \frac{\rho p}{p'} \Big]\frac{\pi}{2},\qquad \Big[\frac{\rho}{p'}\Big]\neq0
\ee
Taking out the factor of $\pi$, this simplifies to
\begin{align}
\min \Big(\Big[\!-\frac{\rho}{2} +\frac{\rho p}{2p'} \Big], \Big[ \frac{1-\rho}{2} +\frac{\rho p}{2p'} \Big] \Big) = \hf \Big[ \frac{\rho p}{p'} \Big]
\end{align}
Setting $x=\rho p/(2 p')$, using the property that $[x] = [n+x]$ for $n\in\mathbb{Z}$ and adding $\rho$ to both terms inside the minimum, this relation becomes
\begin{align}
\min\!\Big( \Big[ \frac{\rho}{2} + x \Big], \Big[ \frac{\rho+1}{2} + x \Big] \Big) &= \half \left[2 x\right]
\end{align}
For both $\rho$ even and $\rho$ odd, this can be rewritten as
\be
\min( [x], [ x + \half] ) = \half [2 x]
\ee
We prove this simple identity in two cases. For $0<[x]<\half$
\be
\min( [x], [ x + \half] ) = [x] = x = \half (2x) = \half [2x]
\ee
Similarly, for the case $\half<[x]<1$ 
\be
\min( [x], [ x + \half] ) = [x + \half] = \half [2x]
\ee
The last equality follows from the previous case after a change of variables $x = y+\half$ since
\be
[y+1] = \half[1+2y]\quad \Leftrightarrow\quad [y]=\half[2y],\qquad 0<[y]<\half
\ee
We do not need the case $[x]=1/2$ since it can only occur if $\rho p/p'$ is an odd integer. But this cannot occur since $\rho$ is not a multiple of $p'$ and $p,p'$ are coprime.

\section{Derivation of the Inversion Relation}
\label{InversionRelation}
In this Appendix, we derive the inversion relation for the double row transfer matrix $\vec D(u)=\vec D(u,\xi)$ given by (\ref{DTM}). We normalize the double row transfer matrix  with the normalization factor (\ref{normalization}) 
such that it satisfies the crossing symmetry (\ref{cross}) and the initial condition $\vec D(0,\xi)=\vec I$. 
We start by concatenating $\vec D(u)$ with $\vec D(u+\lambda)$
\psset{unit=.9cm}
\setlength{\unitlength}{.9cm}
\begin{align}
&{\cal N}^{(\rho)}(u,\xi){\cal N}^{(\rho)}(u+\lambda,\xi)\vec D(u)\vec D(u+\lambda)\;=\ 
\begin{pspicture}[shift=-1.85](1,0)(8,4)
\facegrid{(2,0)}{(8,4)}
\rput(2.5,.5){\small $u$}
\rput(2.5,1.5){\small $\lambda\!-\!u$}
\rput(4.5,.5){\small $u$}
\rput(4.5,1.5){\small $\lambda\!-\!u$}
\rput(3.5,.5){\small ...}
\rput(3.5,1.5){\small ...}
\rput(5.5,.5){\scriptsize$u\!\!-\!\!\xi_{\rho\!-\!1}$}
\rput(5.5,1.5){\scriptsize$-\!u\!\!-\!\!\xi_{\rho\!-\!2}$}
\rput(6.5,.5){\small ...}
\rput(6.5,1.5){\small ...}
\rput(7.5,.5){\scriptsize $u\!\!-\!\!\xi_{1}$}
\rput(7.5,1.5){\scriptsize $-\!u\!\!-\!\!\xi_{0}$}
\rput(2.5,2.5){\small $\lambda\!+\!u$}
\rput(2.5,3.5){\small $-u$}
\rput(4.5,2.5){\small $\lambda\!+\!u$}
\rput(4.5,3.5){\small $-u$}
\rput(3.5,2.5){\small ...}
\rput(3.5,3.5){\small ...}
\rput(5.5,2.5){\scriptsize$u\!\!-\!\!\xi_{\rho\!-\!2}$}
\rput(5.5,3.5){\scriptsize$-\!u\!\!-\!\!\xi_{\rho\!-\!1}$}
\rput(6.5,2.5){\small ...}
\rput(6.5,3.5){\small ...}
\rput(7.5,2.5){\scriptsize $u\!\!-\!\!\xi_{0}$}
\rput(7.5,3.5){\scriptsize $-\!u\!\!-\!\!\xi_{1}$}
\multirput(2,0)(3,0){2}{\multirput(0,0)(2,0){2}{\multirput(0,0)(0,1){4}{\psarc[linewidth=.5pt,linecolor=red](0,0){.1}{0}{90}}}}
\multiput(1,0)(0,2){2}{\pspolygon[linewidth=0.8pt,linecolor=black,fillstyle=solid,fillcolor=lightlightblue](0,0)(1,1)(0,2)}
\multiput(1.5,0.5)(0,1){4}{\psline[linewidth=1.5pt,linecolor=blue](0,0)(0.5,0)}
\multiput(8,0)(0,2){2}{\pspolygon[linewidth=0.8pt,linecolor=black,fillstyle=solid,fillcolor=lightlightblue](0,1)(1,0)(1,2)}
\multiput(8,0.5)(0,1){4}{\psline[linewidth=1.5pt,linecolor=blue](0,0)(0.5,0)}
\psline[linewidth=1pt,linecolor=red,linestyle=dashed](5,-0.4)(5,4.4)
\multiput(1.65,1)(0,2){2}{\psarc[linewidth=1.5pt,linecolor=blue](0,0){.525}{106}{254}}
\multiput(8.35,1)(0,2){2}{\psarc[linewidth=1.5pt,linecolor=blue](0,0){.525}{-74}{74}}
\end{pspicture}\qquad\qquad
\end{align}
\begin{align}
&=\frac{1}{s(\lambda+2u)s(\lambda-2u)}\;\begin{pspicture}[shift=-1.85](1,0)(12,4)
\multirput(2,0)(7,0){2}{\facegrid{(0,0)}{(3,4)}}
\rput(2.5,.5){\small $u$}
\rput(2.5,1.5){\small $\lambda\!-\!u$}
\rput(4.5,.5){\small $u$}
\rput(4.5,1.5){\small $\lambda\!-\!u$}
\rput(3.5,.5){\small ...}
\rput(3.5,1.5){\small ...}
\rput(9.5,.5){\scriptsize$u\!\!-\!\!\xi_{\rho\!-\!1}$}
\rput(9.5,1.5){\scriptsize$-\!u\!\!-\!\!\xi_{\rho\!-\!2}$}
\rput(10.5,.5){\small ...}
\rput(10.5,1.5){\small ...}
\rput(11.5,.5){\scriptsize $u\!\!-\!\!\xi_{1}$}
\rput(11.5,1.5){\scriptsize $-\!u\!\!-\!\!\xi_{0}$}
\rput(2.5,2.5){\small $\lambda\!+\!u$}
\rput(2.5,3.5){\small $-u$}
\rput(4.5,2.5){\small $\lambda\!+\!u$}
\rput(4.5,3.5){\small $-u$}
\rput(3.5,2.5){\small ...}
\rput(3.5,3.5){\small ...}
\rput(9.5,2.5){\scriptsize$u\!\!-\!\!\xi_{\rho\!-\!2}$}
\rput(9.5,3.5){\scriptsize$-\!u\!\!-\!\!\xi_{\rho\!-\!1}$}
\rput(10.5,2.5){\small ...}
\rput(10.5,3.5){\small ...}
\rput(11.5,2.5){\scriptsize $u\!\!-\!\!\xi_{0}$}
\rput(11.5,3.5){\scriptsize $-\!u\!\!-\!\!\xi_{1}$}
\multirput(2,0)(7,0){2}{\multirput(0,0)(2,0){2}{\multirput(0,0)(0,1){4}{\psarc[linewidth=.5pt,linecolor=red](0,0){.1}{0}{90}}}}
\multiput(1,0)(0,2){2}{\pspolygon[linewidth=0.8pt,linecolor=black,fillstyle=solid,fillcolor=lightlightblue](0,0)(1,1)(0,2)}
\multiput(1.5,0.5)(0,1){4}{\psline[linewidth=1.5pt,linecolor=blue](0,0)(0.5,0)}
\multiput(1.65,1)(0,2){2}{\psarc[linewidth=1.5pt,linecolor=blue](0,0){.525}{106}{254}}
\multiput(12,0)(0,2){2}{\pspolygon[linewidth=0.8pt,linecolor=black,fillstyle=solid,fillcolor=lightlightblue](0,1)(1,0)(1,2)}
\multiput(12,0.5)(0,1){4}{\psline[linewidth=1.5pt,linecolor=blue](0,0)(0.5,0)}
\multiput(12.35,1)(0,2){2}{\psarc[linewidth=1.5pt,linecolor=blue](0,0){.525}{-74}{74}}
\multiput(5,0)(0,3){2}{\psline[linewidth=1.5pt,linecolor=blue](0,0.5)(4,0.5)}
\multiput(5,2)(2,0){2}{\pspolygon[linewidth=0.8pt,linecolor=black,fillstyle=solid,fillcolor=lightlightblue](0,0)(1,1)(2,0)(1,-1)}
\multiput(5,1.5)(2,0){2}{\multiput(0,0)(1.5,0){2}{\multiput(0,0)(0,1){2}{\psline[linewidth=1.5pt,linecolor=blue](0,0)(0.5,0)}}}
\multiput(5,2)(2,0){2}{\psarc[linewidth=.5pt,linecolor=red](0,0){.17}{-45}{45}}
\rput(6,2){\small $2u$}
\rput(8,2){\small $-2u$}
\end{pspicture}\qquad\nonumber\\[16pt]
&=\frac{1}{s(\lambda+2u)s(\lambda-2u)}\;\begin{pspicture}[shift=-1.85](0,0)(9,4)
\multiput(0,0)(0,2){2}{\pspolygon[linewidth=0.8pt,linecolor=black,fillstyle=solid,fillcolor=lightlightblue](0,0)(1,1)(0,2)}
\multiput(0,0)(8,0){2}{\pspolygon[linewidth=0.8pt,linecolor=black,fillstyle=solid,fillcolor=lightlightblue](0,2)(1,1)(2,2)(1,3)}
\multiput(9,0)(0,2){2}{\pspolygon[linewidth=0.8pt,linecolor=black,fillstyle=solid,fillcolor=lightlightblue](0,1)(1,0)(1,2)}
\facegrid{(2,0)}{(8,4)}
\rput(2.5,.5){\small $u$}
\rput(2.5,1.5){\small $\lambda\!+\!u$}
\rput(4.5,.5){\small $u$}
\rput(4.5,1.5){\small $\lambda\!+\!u$}
\rput(3.5,.5){\small ...}
\rput(3.5,1.5){\small ...}
\rput(5.5,.5){\scriptsize$u\!\!-\!\!\xi_{\rho\!-\!1}$}
\rput(5.5,1.5){\scriptsize$u\!\!-\!\!\xi_{\rho\!-\!2}$}
\rput(6.5,.5){\small ...}
\rput(6.5,1.5){\small ...}
\rput(7.5,.5){\scriptsize $u\!\!-\!\!\xi_{1}$}
\rput(7.5,1.5){\scriptsize $u\!\!-\!\!\xi_{0}$}
\rput(2.5,2.5){\small $\lambda\!-\!u$}
\rput(2.5,3.5){\small $-u$}
\rput(4.5,2.5){\small $\lambda\!-\!u$}
\rput(4.5,3.5){\small $-u$}
\rput(3.5,2.5){\small ...}
\rput(3.5,3.5){\small ...}
\rput(5.5,2.5){\scriptsize$-\!u\!\!-\!\!\xi_{\rho\!-\!2}$}
\rput(5.5,3.5){\scriptsize$-\!u\!\!-\!\!\xi_{\rho\!-\!1}$}
\rput(6.5,2.5){\small ...}
\rput(6.5,3.5){\small ...}
\rput(7.5,2.5){\scriptsize $-\!u\!\!-\!\!\xi_{0}$}
\rput(7.5,3.5){\scriptsize $-\!u\!\!-\!\!\xi_{1}$}
\rput(1,2){\small $2u$}
\rput(9,2){\small $-2u$}
\multirput(2,0)(3,0){2}{\multirput(0,0)(2,0){2}{\multirput(0,0)(0,1){4}{\psarc[linewidth=.5pt,linecolor=red](0,0){.1}{0}{90}}}}
\multiput(0.65,1)(0,2){2}{\psarc[linewidth=1.5pt,linecolor=blue](0,0){.525}{106}{254}}
\multiput(9.35,1)(0,2){2}{\psarc[linewidth=1.5pt,linecolor=blue](0,0){.525}{-74}{74}}
\multiput(0,0)(7.5,0){2}{\multiput(0,0)(0,3){2}{\psline[linewidth=1.5pt,linecolor=blue](0.5,0.5)(2,0.5)}}
\multiput(1,1)(6.5,0){2}{\multiput(0,0)(0,1){2}{\psline[linewidth=1.5pt,linecolor=blue](0.5,0.5)(1,0.5)}}
\multiput(0,2)(8,0){2}{\psarc[linewidth=.5pt,linecolor=red](0,0){.17}{-45}{45}}
\psline[linewidth=1pt,linecolor=red,linestyle=dashed](5,-0.4)(5,4.4)
\end{pspicture}
\end{align}
In the second and third equalities we have used the local inversion relation \eqref{Invrel} and YBE \eqref{YBE} to interchange the second and third rows. 

The structure of the inversion identity of \cite{BPO96,MDPR14} is 
\bea
\vec D(u)\vec D(u+\lambda)=f_+(u) I+f_-(u) \vec D^{2,1}(u)\label{InvId}
\eea
where $f_\pm(u)$ are certain scalar functions. 
For large $N$, and for $u$ in the physical interval $0<u<\lambda$, the contribution of the second term on the right side is exponentially small compared to the first term. 
Neglecting the second term gives rise to a scalar inversion relation of the form $D(u)D(u+\lambda)=f_+(u)$ satisfied by all of the eigenvalues $D(u)$ of $\vec D(u)$ for large $N$.
At this point, to project onto the two terms on the right side of the inversion identity (\ref{InvId}), we need to introduce two complementary projectors
\bea
\psset{unit=.6cm}
\frac{1}{\beta}\ 
\begin{pspicture}[shift=-.92](0,0)(2,2)
\pspolygon[linewidth=0.8pt,linecolor=black,fillstyle=solid,fillcolor=lightlightblue](1,0)(2,1)(1,2)(0,1)
\rput(1,1){\small $\lambda$}
\psarc[linewidth=.5pt,linecolor=red](0,1){.17}{-45}{45}
\end{pspicture}
\ =\ \frac{1}{\beta}\ 
\begin{pspicture}[shift=-.92](0,0)(2,2)
\pspolygon[linewidth=0.8pt,linecolor=black,fillstyle=solid,fillcolor=lightlightblue](1,0)(2,1)(1,2)(0,1)
\psarc[linewidth=1.5pt,linecolor=blue](0,1){.72}{-45}{45}
\psarc[linewidth=1.5pt,linecolor=blue](2,1){.72}{135}{-135}
\end{pspicture}
\ ,
\qquad
\frac{1}{\beta}\ 
\begin{pspicture}[shift=-.92](0,0)(2,2)
\pspolygon[linewidth=0.8pt,linecolor=black,fillstyle=solid,fillcolor=lightlightblue](1,0)(2,1)(1,2)(0,1)
\rput(1,1){\small $-\lambda$}
\psarc[linewidth=.5pt,linecolor=red](0,1){.17}{-45}{45}
\end{pspicture}
\ =\ 
\begin{pspicture}[shift=-.92](0,0)(2,2)
\pspolygon[linewidth=0.8pt,linecolor=black,fillstyle=solid,fillcolor=lightlightblue](1,0)(2,1)(1,2)(0,1)
\psarc[linewidth=1.5pt,linecolor=blue](1,2){.72}{-135}{-45}
\psarc[linewidth=1.5pt,linecolor=blue](1,0){.72}{45}{135}
\end{pspicture}
\ -\ \frac{1}{\beta}\ 
\begin{pspicture}[shift=-.92](0,0)(2,2)
\pspolygon[linewidth=0.8pt,linecolor=black,fillstyle=solid,fillcolor=lightlightblue](1,0)(2,1)(1,2)(0,1)
\psarc[linewidth=1.5pt,linecolor=blue](0,1){.72}{-45}{45}
\psarc[linewidth=1.5pt,linecolor=blue](2,1){.72}{135}{-135}
\end{pspicture}
\eea
Fixing the direction of action from left to right, the left projector corresponds to $p_+=\beta^{-1}e_j$ and the right projector to $p_-=I-\beta^{-1}e_j$ in the linear TL algebra. 
The sum of these two projectors is an identity that can be inserted into the previous diagrammatic planar tangle to decompose it into the two terms on the right side of the inversion identity (\ref{InvId}). 
The projector $p_-$ is the Wenzl-Jones projector that implements $2\times 1$ fusion and gives rise to the fused double row transfer matrix $\vec D^{2,1}(u)$. 
By contrast, the projector $p_+$ gives rise to the term proportional to the identity. 

If $I$ denotes the identity tangle acting in the upward direction, then discarding the $p_-$ projector and just keeping the $p_+$ projector in the insertion of the identity gives
\begin{align}
f_+(u)I&=\frac{1}{\beta s(\lambda\!+\!2u)s(\lambda\!-\!2u)}\;\begin{pspicture}[shift=-1.85](0,0)(11,4)
\multiput(0,0)(0,2){2}{\pspolygon[linewidth=0.8pt,linecolor=black,fillstyle=solid,fillcolor=lightlightblue](0,0)(1,1)(0,2)}
\multiput(0.65,1)(0,2){2}{\psarc[linewidth=1.5pt,linecolor=blue](0,0){.525}{106}{254}}
\multiput(0,0)(10,0){2}{\pspolygon[linewidth=0.8pt,linecolor=black,fillstyle=solid,fillcolor=lightlightblue](0,2)(1,1)(2,2)(1,3)}
\multiput(11,0)(0,2){2}{\pspolygon[linewidth=0.8pt,linecolor=black,fillstyle=solid,fillcolor=lightlightblue](0,1)(1,0)(1,2)}
\multiput(11.35,1)(0,2){2}{\psarc[linewidth=1.5pt,linecolor=blue](0,0){.525}{-74}{74}}
\multiput(0,2)(10,0){2}{\psarc[linewidth=.5pt,linecolor=red](0,0){.17}{-45}{45}}
\rput(1,2){\small $2u$}
\rput(11,2){\small $-2u$}
\multiput(2,0)(5,0){2}{\facegrid{(0,0)}{(3,4)}}
\multirput(2,0)(5,0){2}{\multirput(0,0)(2,0){2}{\multirput(0,0)(0,1){4}{\psarc[linewidth=.5pt,linecolor=red](0,0){.1}{0}{90}}}}
\rput(2.5,.5){\small $u$}
\rput(2.5,1.5){\small $\lambda\!+\!u$}
\rput(4.5,.5){\small $u$}
\rput(4.5,1.5){\small $\lambda\!+\!u$}
\rput(3.5,.5){\small ...}
\rput(3.5,1.5){\small ...}
\rput(7.5,.5){\scriptsize$u\!\!-\!\!\xi_{\rho\!-\!1}$}
\rput(7.5,1.5){\scriptsize$u\!\!-\!\!\xi_{\rho\!-\!2}$}
\rput(8.5,.5){\small ...}
\rput(8.5,1.5){\small ...}
\rput(9.5,.5){\scriptsize $u\!\!-\!\!\xi_{1}$}
\rput(9.5,1.5){\scriptsize $u\!\!-\!\!\xi_{0}$}
\rput(2.5,2.5){\small $\lambda\!-\!u$}
\rput(2.5,3.5){\small $-u$}
\rput(4.5,2.5){\small $\lambda\!-\!u$}
\rput(4.5,3.5){\small $-u$}
\rput(3.5,2.5){\small ...}
\rput(3.5,3.5){\small ...}
\rput(7.5,2.5){\scriptsize$-\!u\!\!-\!\!\xi_{\rho\!-\!2}$}
\rput(7.5,3.5){\scriptsize$-\!u\!\!-\!\!\xi_{\rho\!-\!1}$}
\rput(8.5,2.5){\small ...}
\rput(8.5,3.5){\small ...}
\rput(9.5,2.5){\scriptsize $-\!u\!\!-\!\!\xi_{0}$}
\rput(9.5,3.5){\scriptsize $-\!u\!\!-\!\!\xi_{1}$}
\pspolygon[linewidth=0.8pt,linecolor=black,fillstyle=solid,fillcolor=lightlightblue](5,1)(6,0)(7,1)(6,2)
\psarc[linewidth=.5pt,linecolor=red](5,1){.17}{-45}{45}
\rput(6,1){\small $\lambda$}
\multiput(5,2.5)(0,1){2}{\psline[linewidth=1.5pt,linecolor=blue](0,0)(2,0)}
\multiput(5,0)(0,1){2}{\multiput(0,0)(1.5,0){2}{\psline[linewidth=1.5pt,linecolor=blue](0,0.5)(0.5,0.5)}}
\multiput(0,0)(9.5,0){2}{\multiput(0,0)(0,3){2}{\psline[linewidth=1.5pt,linecolor=blue](0.5,0.5)(2,0.5)}}
\multiput(1,1)(8.5,0){2}{\multiput(0,0)(0,1){2}{\psline[linewidth=1.5pt,linecolor=blue](0.5,0.5)(1,0.5)}}
\end{pspicture}\qquad
\end{align}
\begin{align}
&=\frac{\left(s(\lambda+u)s(\lambda-u)\right)^N}{\beta s(\lambda+2u)s(\lambda-2u)}\;\begin{pspicture}[shift=-1.85](0,0)(12,4)
\multiput(0,0)(0,2){2}{\pspolygon[linewidth=0.8pt,linecolor=black,fillstyle=solid,fillcolor=lightlightblue](0,0)(1,1)(0,2)}
\psarc[linewidth=1.5pt,linecolor=blue](0.65,3){.525}{106}{254}
\multiput(0,0)(10,0){1}{\pspolygon[linewidth=0.8pt,linecolor=black,fillstyle=solid,fillcolor=lightlightblue](1,1)(2,2)(3,1)(2,0)}
\multiput(0,0)(10,0){2}{\pspolygon[linewidth=0.8pt,linecolor=black,fillstyle=solid,fillcolor=lightlightblue](0,2)(1,1)(2,2)(1,3)}
\multiput(11,0)(0,2){2}{\pspolygon[linewidth=0.8pt,linecolor=black,fillstyle=solid,fillcolor=lightlightblue](0,1)(1,0)(1,2)}
\multiput(11.35,1)(0,2){2}{\psarc[linewidth=1.5pt,linecolor=blue](0,0){.525}{-74}{74}}
\multiput(0,2)(10,0){2}{\psarc[linewidth=.5pt,linecolor=red](0,0){.17}{-45}{45}}
\rput(1,2){\small $2u$}
\rput(11,2){\small $-2u$}
\facegrid{(3,2)}{(6,4)}
\facegrid{(7,0)}{(10,4)}
\multirput(7,0)(2,0){2}{\multirput(0,0)(0,1){4}{\psarc[linewidth=.5pt,linecolor=red](0,0){.1}{0}{90}}}
\multirput(3,2)(2,0){2}{\multirput(0,0)(0,1){2}{\psarc[linewidth=.5pt,linecolor=red](0,0){.1}{0}{90}}}
\rput(3.5,2.5){\small $\lambda\!-\!u$}
\rput(3.5,3.5){\small $-u$}
\rput(5.5,2.5){\small $\lambda\!-\!u$}
\rput(5.5,3.5){\small $-u$}
\rput(4.5,2.5){\small ...}
\rput(4.5,3.5){\small ...}
\rput(7.5,2.5){\scriptsize$-\!u\!\!-\!\!\xi_{\rho\!-\!2}$}
\rput(7.5,3.5){\scriptsize$-\!u\!\!-\!\!\xi_{\rho\!-\!1}$}
\rput(8.5,2.5){\small ...}
\rput(8.5,3.5){\small ...}
\rput(9.5,2.5){\scriptsize $-\!u\!\!-\!\!\xi_{0}$}
\rput(9.5,3.5){\scriptsize $-\!u\!\!-\!\!\xi_{1}$}
\rput(7.5,.5){\scriptsize$u\!\!-\!\!\xi_{\rho\!-\!1}$}
\rput(7.5,1.5){\scriptsize$u\!\!-\!\!\xi_{\rho\!-\!2}$}
\rput(8.5,.5){\small ...}
\rput(8.5,1.5){\small ...}
\rput(9.5,.5){\scriptsize $u\!\!-\!\!\xi_{1}$}
\rput(9.5,1.5){\scriptsize $u\!\!-\!\!\xi_{0}$}
\multiput(2,2.5)(0,1){2}{\multiput(0,0)(4,0){2}{\psline[linewidth=1.5pt,linecolor=blue](0,0)(1,0)}}
\psarc[linewidth=1.5pt,linecolor=blue](1,1){0.65}{-225}{45}
\psline[linewidth=1.5pt,linecolor=blue](0.5,3.5)(2,3.5)
\psline[linewidth=1.5pt,linecolor=blue](1.5,2.5)(2,2.5)
\psline[linewidth=1.5pt,linecolor=blue](10,0.5)(11.5,0.5)
\psline[linewidth=1.5pt,linecolor=blue](10,1.5)(10.5,1.5)
\psline[linewidth=1.5pt,linecolor=blue](10,2.5)(10.5,2.5)
\psline[linewidth=1.5pt,linecolor=blue](10,3.5)(11.5,3.5)
\psarc[linewidth=1.5pt,linecolor=blue](7,1){0.5}{90}{270}
\psline[linewidth=1.5pt,linecolor=blue](5.5,0)(5.5,2)
\psline[linewidth=1.5pt,linecolor=blue](4.5,0)(4.5,2)
\psarc[linewidth=1.5pt,linecolor=blue](3,1){0.6}{90}{270}
\psarc[linewidth=1.5pt,linecolor=blue](3,2){0.4}{270}{360}
\psarc[linewidth=1.5pt,linecolor=blue](3,0){0.4}{0}{90}
\end{pspicture}
\end{align}
\begin{align}
&=\frac{\left(s(\lambda+u)s(\lambda-u)\right)^N s(2\lambda-2u)}{\beta s(\lambda+2u)s(\lambda-2u)}\;\begin{pspicture}[shift=-1.85](2.5,0)(12,4)
\multiput(11,0)(10,0){1}{\pspolygon[linewidth=0.8pt,linecolor=black,fillstyle=solid,fillcolor=lightlightblue](-1,2)(0,1)(1,2)(0,3)}
\multiput(11,0)(0,2){2}{\pspolygon[linewidth=0.8pt,linecolor=black,fillstyle=solid,fillcolor=lightlightblue](0,1)(1,0)(1,2)}
\multiput(11.35,1)(0,2){2}{\psarc[linewidth=1.5pt,linecolor=blue](0,0){.525}{-74}{74}}
\multiput(10,2)(10,0){1}{\psarc[linewidth=.5pt,linecolor=red](0,0){.17}{-45}{45}}
\rput(11,2){\small $-2u$}
\facegrid{(3,2)}{(6,4)}
\facegrid{(7,0)}{(10,4)}
\multirput(7,0)(2,0){2}{\multirput(0,0)(0,1){4}{\psarc[linewidth=.5pt,linecolor=red](0,0){.1}{0}{90}}}
\multirput(3,2)(2,0){2}{\multirput(0,0)(0,1){2}{\psarc[linewidth=.5pt,linecolor=red](0,0){.1}{0}{90}}}
\rput(3.5,2.5){\small $\lambda\!-\!u$}
\rput(3.5,3.5){\small $-u$}
\rput(5.5,2.5){\small $\lambda\!-\!u$}
\rput(5.5,3.5){\small $-u$}
\rput(4.5,2.5){\small ...}
\rput(4.5,3.5){\small ...}
\rput(7.5,2.5){\scriptsize$-\!u\!\!-\!\!\xi_{\rho\!-\!2}$}
\rput(7.5,3.5){\scriptsize$-\!u\!\!-\!\!\xi_{\rho\!-\!1}$}
\rput(8.5,2.5){\small ...}
\rput(8.5,3.5){\small ...}
\rput(9.5,2.5){\scriptsize $-\!u\!\!-\!\!\xi_{0}$}
\rput(9.5,3.5){\scriptsize $-\!u\!\!-\!\!\xi_{1}$}
\rput(7.5,.5){\scriptsize$u\!\!-\!\!\xi_{\rho\!-\!1}$}
\rput(7.5,1.5){\scriptsize$u\!\!-\!\!\xi_{\rho\!-\!2}$}
\rput(8.5,.5){\small ...}
\rput(8.5,1.5){\small ...}
\rput(9.5,.5){\scriptsize $u\!\!-\!\!\xi_{1}$}
\rput(9.5,1.5){\scriptsize $u\!\!-\!\!\xi_{0}$}
\multiput(6,2.5)(0,1){2}{\psline[linewidth=1.5pt,linecolor=blue](0,0)(1,0)}
\psline[linewidth=1.5pt,linecolor=blue](10,0.5)(11.5,0.5)
\psline[linewidth=1.5pt,linecolor=blue](10,1.5)(10.5,1.5)
\psline[linewidth=1.5pt,linecolor=blue](10,2.5)(10.5,2.5)
\psline[linewidth=1.5pt,linecolor=blue](10,3.5)(11.5,3.5)
\psarc[linewidth=1.5pt,linecolor=blue](7,1){0.5}{90}{270}
\psline[linewidth=1.5pt,linecolor=blue](5.5,0)(5.5,2)
\psline[linewidth=1.5pt,linecolor=blue](4.5,0)(4.5,2)
\psline[linewidth=1.5pt,linecolor=blue](3.5,0)(3.5,2)
\psarc[linewidth=1.5pt,linecolor=blue](3,3){0.6}{90}{-90}
\end{pspicture}\quad
\end{align}
\begin{align}
&={\frac{\left(s(\lambda+u)s(\lambda-u)\right)^{2N}s(2\lambda-2u)\prod\limits_{j=1}^{\rho-1}s(\lambda+u+\xi_j)s(\lambda-u-\xi_j)}{\beta s(\lambda+2u)s(\lambda-2u)}}\nonumber\\[10pt]
&\qquad\qquad\qquad\qquad\qquad\qquad\times\begin{pspicture}[shift=-1.85](1,0)(9.5,4)
\multiput(8.5,0)(7.5,0){1}{\pspolygon[linewidth=0.8pt,linecolor=black,fillstyle=solid,fillcolor=lightlightblue](-1,2)(0,1)(1,2)(0,3)}
\multiput(8.5,0)(0,2){2}{\pspolygon[linewidth=0.8pt,linecolor=black,fillstyle=solid,fillcolor=lightlightblue](0,1)(1,0)(1,2)}
\multiput(8.85,1)(0,2){1}{\psarc[linewidth=1.5pt,linecolor=blue](0,0){.525}{-74}{74}}
\multiput(7.5,2)(8.5,0){1}{\psarc[linewidth=.5pt,linecolor=red](0,0){.17}{-45}{45}}
\rput(8.5,2){\small $-2u$}
\put(4.5,0){\facegrid{(0,0)}{(3,2)}}
\multirput(4.5,0)(2,0){2}{\multirput(0,0)(0,1){2}{\psarc[linewidth=.5pt,linecolor=red](0,0){.1}{0}{90}}}
\rput(5,.5){\scriptsize$u\!\!-\!\!\xi_{\rho\!-\!1}$}
\rput(5,1.5){\scriptsize$u\!\!-\!\!\xi_{\rho\!-\!2}$}
\rput(6,.5){\small ...}
\rput(6,1.5){\small ...}
\rput(7,.5){\scriptsize $u\!\!-\!\!\xi_{1}$}
\rput(7,1.5){\scriptsize $u\!\!-\!\!\xi_{0}$}
\psline[linewidth=1.5pt,linecolor=blue](7.5,0.5)(9,0.5)
\psline[linewidth=1.5pt,linecolor=blue](7.5,1.5)(8,1.5)
\psarc[linewidth=1.5pt,linecolor=blue](4.5,1){0.5}{90}{270}
\psline[linewidth=1.5pt,linecolor=blue](3.5,0)(3.5,4)
\psline[linewidth=1.5pt,linecolor=blue](2.5,0)(2.5,4)
\multiput(5,2)(1,0){2}{\psline[linewidth=1.5pt,linecolor=blue](0,0)(0,2)}
\psline[linewidth=1.5pt,linecolor=blue](1.5,0)(1.5,4)
\psarc[linewidth=1.5pt,linecolor=blue](8.5,3){0.65}{-45}{-135}
\psline[linewidth=1.5pt,linecolor=blue](7,2)(7,4)
\end{pspicture}\qquad\qquad
\end{align}
\begin{align}
&\quad={\frac{[s(\lambda\!+\!u)s(\lambda\!-\!u)]^{2N}s(2\lambda\!-\!2u)s(2\lambda\!+\!2u)\prod\limits_{j=1}^{\rho-1}s(\lambda\!+\!u\!+\!\xi_j)s(\lambda\!-\!u\!-\!\xi_j)s(\lambda\!+\!u\!-\!\xi_j)s(\lambda\!-\!u\!+\!\xi_j)}{ s(\lambda\!+\!2u)s(\lambda\!-\!2u)}}\,I
\end{align}
In propagating the projector $p_+$ around the closed loop, we used the drop-down property and the planar identities
\bea
\psset{unit=.65cm}
\begin{pspicture}[shift=-1.85](0,0)(3,4)
\multiput(0,0)(0,2){2}{\pspolygon[linewidth=0.8pt,linecolor=black,fillstyle=solid,fillcolor=lightlightblue](0,0)(1,1)(0,2)}
\multiput(1,0)(6.5,0){1}{\pspolygon[linewidth=0.8pt,linecolor=black,fillstyle=solid,fillcolor=lightlightblue](-1,2)(0,1)(1,2)(0,3)}
\multiput(0,2)(6.5,0){1}{\psarc[linewidth=.5pt,linecolor=red](0,0){.17}{-45}{45}}
\rput(1,2){\small $2u$}
\psarc[linewidth=1.5pt,linecolor=blue](1,1){0.65}{135}{45}
\psarc[linewidth=1.5pt,linecolor=blue](1,3){0.65}{90}{225}
\psline[linewidth=1.5pt,linecolor=blue](1,3.65)(2.5,3.65)
\psline[linewidth=1.5pt,linecolor=blue](1.5,2.5)(2.5,2.5)
\end{pspicture}\hspace{-8pt}
= s(2\lambda-2u)
\begin{pspicture}[shift=-0.9](0,0)(1.3,2)
\psarc[linewidth=1.5pt,linecolor=blue](1,1){0.6}{90}{-90}
\end{pspicture},
\qquad\quad
\begin{pspicture}[shift=-1.85](6,0)(8.5,4)
\multiput(7.5,0)(6.5,0){1}{\pspolygon[linewidth=0.8pt,linecolor=black,fillstyle=solid,fillcolor=lightlightblue](-1,2)(0,1)(1,2)(0,3)}
\multiput(7.5,0)(0,2){2}{\pspolygon[linewidth=0.8pt,linecolor=black,fillstyle=solid,fillcolor=lightlightblue](0,1)(1,0)(1,2)}
\multiput(6.5,2)(6.5,0){1}{\psarc[linewidth=.5pt,linecolor=red](0,0){.17}{-45}{45}}
\rput(7.5,2){\small $-2u$}
\psarc[linewidth=1.5pt,linecolor=blue](7.5,3){0.65}{-45}{225}
\psarc[linewidth=1.5pt,linecolor=blue](7.5,1){0.65}{-90}{45}
\psline[linewidth=1.5pt,linecolor=blue](7.5,0.35)(6,0.35)
\psline[linewidth=1.5pt,linecolor=blue](7,1.5)(6,1.5)
\end{pspicture}
= s(2\lambda+2u) 
\begin{pspicture}[shift=-0.9](0,0)(1.5,2)
\psarc[linewidth=1.5pt,linecolor=blue](0.5,1){0.6}{-90}{90}
\end{pspicture}
\eea

After simplification, we obtain the full matrix inversion relation
\begin{align}
\vec D(u)\vec D(u+\lambda)&=\left(\dfrac{\sin(\lambda\!+\!u)\sin(\lambda\!-\!u)}{\sin^2\lambda}\right)^{2N}\!\!\times\dfrac{\sin^2\!\lambda\sin(2\lambda\!+\!2u)\sin(2\lambda\!-\!2u)}{\sin^22\lambda\sin(\lambda\!+\!2u)\sin(\lambda\!-\!2u)}\nonumber\\ 
&\qquad\quad\times\dfrac{\sin(\xi\!+\!u)\sin(\xi\!-\!u)\sin(\xi\!+\!\rho\lambda\!+\!u)\sin(\xi\!+\!\rho\lambda\!-\!u)}{\sin^2\xi\sin^2(\xi\!+\!\rho\lambda)}\;\vec I
\end{align}
\normalsize
The double row transfer matrices $\left\{ \vec D(v),v\in\mathbb{C}\right\}$ constitute a one-parameter of commuting matrices which we assume are simultaneously diagonalizable. If $\kappa(u,\xi)=D_0(u,\xi)$ denotes the ground state eigenvalue of $\vec D(u)$ in the $(r,s)$ sector, then it must satisfy the scalar inversion relation
\begin{align}
\kappa(u,\xi)\kappa(u+\lambda,\xi)&=\left(\frac{\sin(\lambda\!+\!u)\sin(\lambda\!-\!u)}{\sin^2\lambda}\right)^{2N}\!\!\times\frac{\sin^2\!\lambda\sin(2\lambda\!+\!2u)\sin(2\lambda\!-\!2u)}{\sin^22\lambda\sin(\lambda\!+\!2u)\sin(\lambda\!-\!2u)}\nonumber\\ 
&\times\frac{\sin(\xi\!+\!u)\sin(\xi\!-\!u)\sin(\xi\!+\!\rho\lambda\!+\!u)\sin(\xi\!+\!\rho\lambda\!-\!u)}{\sin^2\xi\sin^2(\xi\!+\!\rho\lambda)}
\label{fullInversionRelation}
\end{align}

\section{Analyticity of $\kappa(u,\xi)$}
\label{analyticity}
To obtain the bulk and boundary free energies, we must solve the inversion relation (\ref{fullInversionRelation}) for $\log\kappa(u,\xi)=\log D_0(u,\xi)$ where $D_0(u,\xi)$ is the ground state eigenvalue of the double row transfer matrix $\vec D(u,\xi)$ in the $(r,s)$ sector. 
In fact, all of the eigenvalues $D(u,\xi)$ of the double row transfer matrix $\vec D(u,\xi)$ satisfy the inversion relation \eqref{fullInversionRelation}, with $D(0,\xi)=1$, but we will only be interested in finite excitations above the ground state $D_0(u,\xi)$ in the various $(r,s)$ sectors. 
Following Baxter~\cite{BaxInv82}, we need to represent the second derivative of $\log\kappa(u,\xi)$ by the Fourier/Laplace integral
\be
F(u)=\frac{d^2}{du^2}\log\kappa(u,\xi)=\int_{-\infty}^{\infty}c(t)e^{2ut}\,dt,\qquad u\in{\Bbb C}
\label{FourierLaplace}
\ee
For this integral to exist, we need $F(u)$ to be analytic in a suitable vertical strip in the complex $u$ plane with vanishing asymptotics $\lim_{u\to\pm i\infty}F(u)=0$. 
The asymptotic requirement is satisfied since, following the arguments of Baxter, $\kappa_{bulk}(u)$ grows as $\exp(\mp iu)$ as $u\rightarrow\pm i\infty$ in line with the growth of a single face weight. Similarly arguments apply to $\kappa_0(u)$ and $\kappa_\rho(u,\xi)$. The analyticity in the thermodynamic limit $N\to\infty$ can be inferred from the analyticity for finite $N$. For finite $N$, the entries of the double row transfer matrices are Laurent polynomials in $e^{iu}$ (hence analytic) and periodic in $u$ with period $\pi$. These statements hold equally for the eigenvalues $D(u,\xi)$ since we assume the family of double row transfer matrices $\vec D(u,\xi)$ can be simultaneously diagonalized using eigenvectors which are independent of $u$.
For finite $N$, we deduce that $\log D(u,\xi)$ is singular at any complex zero of the Laurent polynomial $D(u,\xi)$ but is otherwise analytic. 
The analyticity of each eigenvalue is thus determined by the pattern of zeros of $D(u,\xi)$ in the complex $u$ plane. Using periodicity, we only need to consider the {\em periodicity strip}
\bea
\mbox{Periodicity Strip:}\qquad\quad \big|\!\Re \big(u-\frac{\lambda}{2}\big)\big|<\frac{\pi}{2}
\eea
Using numerical data for finite $N$, we can therefore extrapolate the patterns of zeros in the periodicity strip to deduce the analyticity as $N$ becomes large.
These patterns of zeros must be compatible with the crossing symmetry $D(u,\xi)=D(\lambda-u,\xi)$, that is they must be invariant under reflection through the point $u=\frac{\lambda}{2}$. 
In addition, since the eigenvalues $D(u,\xi)$ are real, the patterns of zeros must also be invariant under complex conjugation. For this reason, we only need to consider the patterns of zeros in the upper-half plane.

For finite excitations of the ${\cal LM}(p,p')$ models in all $(r,s)$ sectors, we observe that zeros called long 2-strings, consisting of zeros at $x_L+iy_j$ and $x_R+iy_j$, accumulate along the edges of the {\em analyticity strip}\/ for large $N$
\bea
\mbox{Analyticity Strip:}\qquad x_L=\mbox{max}\big(\!\!-\frac{\lambda}{2},-\frac{\pi}{2}+\frac{\lambda}{2}\big)<\Re u<\mbox{min}\big(\frac{3\lambda}{2},\frac{\pi}{2}+\frac{\lambda}{2}\big)\label{anaStrip}=x_R
\eea
As $N\to\infty$, the zeros of the long 2-strings become dense along the edges of the analyticity strip forming a barrier to analytic continuation. 
For $\lambda<\frac{\pi}{2}$, the analyticity strip $|\!\Re u-\frac{\lambda}{2}|<\lambda$ is of width $2\lambda$. For $\lambda>\frac{\pi}{2}$, the analyticity strip $|\!\Re u-\frac{\lambda}{2}|<\tfrac{\pi}{2}$ is of width $\pi$ corresponding to the full periodicity strip. In both cases, the analyticity strip is  centered about $\Re u=\frac{\lambda}{2}$. 
Although we use the term 2-string we note that, for $\lambda\ge \frac{\pi}{2}$, $x_L+iy_j\equiv x_R+iy_j$ due to periodicity.

The analyticity strip contains the \textit{physical strip}
\bea
\mbox{Physical Strip:}\qquad\quad 0\le\Re u\le\lambda
\eea
For real $u$ in this strip, the weights $s(u)$ and $s(\lambda-u)$ of the face operator (\ref{eq:face1}) are physical in the sense that they are nonnegative.
The inversion relation \eqref{fullInversionRelation}, or equivalently the three inversion relations (\ref{bulkInversion}),  (\ref{vacInversion}) and (\ref{bdyInversion}), can be straightforwardly solved using Fourier/Laplace transforms if $D_0(u,\xi)$ is free of zeros in the analyticity strip. 
However for finite excitations, in addition to the long 2-strings, we typically find a finite number of zeros inside the analyticity strip in the form of 1-strings ($u_j=\frac{\lambda}{2}+iy_j$) or short 2-strings (of the form $u_j=\pm x+ iy_j$).
These $\mbox{O}(1)$ zeros, which only enter the inversion relation for the boundary free energies, need to be removed to obtain an ANZ function before the solution of the boundary inversion relation can proceed. Strictly speaking, in solving the inversion relation, it suffices if $D_0(u,\xi)$ is ANZ in an open strip containing the physical strip. The solution can then be extended by analytic continuation to any larger strip in which $D_0(u,\xi)$ is analytic. 

Let us first consider the case where $D(u,\xi)$ has a single 1-string  in the analyticity strip of the form $u_j=\frac{\lambda}{2}+iy_j$ with its complex conjugate $\bar{u}_j=\frac{\lambda}{2}-iy_j$. 
The inversion relation is not sufficient to fix the $y$ coordinate of a 1-string. To remove a 1-string and its complex conjugate, we need a function $f(u)$ satisfying
\bea
f(u_j)=f(\bar{u}_j)=0,\qquad f(u)f(u+\lambda)=1,\qquad f(u)=f(\lambda-u)
\eea
Two solutions to these equations are
\bea
 f(u)=\pm\tan\Big(\frac{\pi}{2\lambda}(u-u_j)\Big)\tan\Big(\frac{\pi}{2\lambda}(u-\bar{u}_j)\Big)
\eea
It follows that $D(u,\xi)/f(u)$ is ANZ in the analyticity strip (including at $u=u_j$) and satisfies the inversion relation and crossing symmetry. In this way, all such 1-string zeros can be removed from $D(u,\xi)$. Furthermore, for finite excitations, it is known~\cite{KlumperPearce1992} that the $y_j$ coordinates of the 1-strings in the upper-half plane (not on the real axis) scale as $\log N$ so that $y_j\to\infty$ as $N\to\infty$. But in this limit
\bea
\lim_{y_j\to\pm\infty} f(u)= \lim_{y_j\to\pm\infty}\tan\Big(\frac{\pi}{2\lambda}(u-u_j)\Big)\tan\Big(\frac{\pi}{2\lambda}(u-\bar{u}_j)\Big)=1
\eea
In this limit these 1-strings are infinitely far from the real axis so they do not effect the free energies.
It follows that any 1-strings not on the real axis can be removed without effecting the solution to the inversion relation in the thermodynamic limit. The overall sign of the product of $f(u)$ functions can be chosen to ensure a positive value at $u=0$. We observe that there are no 1-strings on the real axis.
 
We expect the same arguments to hold for short 2-strings. For example, a function $f(u)$ which acts to remove short 2-strings was found in \cite{BLZ}. 
Again, physically, these zeros are not expected to contribute to boundary free energies as $y_j\to\pm i\infty$ in the thermodynamic limit. 
Actually, we observe that short 2-strings are absent in the case of the ground state eigenvalue $D_0(u,\xi)$ in all $(r,s)$ sectors so, for our purposes, their removal is not required.

Lastly, we consider short 2-strings on the real axis which appear in pairs by crossing symmetry. These $\mbox{O}(1)$ zeros relate again to the boundary free energies and their locations are usually $\xi$-dependent. The only $\xi$-independent short 2-string on the real axis occurs for $\kappa_0(u)$ with zeros at $u=\frac{\pi}{2}$ and $u=\lambda-\frac{\pi}{2}$. In this case and all other cases we find that the zeros of $D_0(u,\xi)$ on the real axis are such that the real zeros of $D_0(u,\xi)D_0(u+\lambda,\xi)$ are compatible with the zeros of the given right sides of the boundary inversion relations. 

\section{Solution of the Inversion Relation for $\kappa_0(u)$ and $\kappa_\rho(u,\xi)$}
\label{laplace:boundary}
In this appendix we derive the vacuum and $r$-type boundary free energies $f_0(u)$ and $f_\rho(u,\xi)$. More explicitly, we solve the boundary inversion relations (\ref{vacInversion}) and (\ref{bdyInversion}) for $\kappa_0(u)$ and $\kappa_\rho(u,\xi)$ using Fourier/Laplace transforms. As in the calculation of $\kappa_{bulk}(u)$, we need to assume that $\log\kappa_0(u)$ and $\log\kappa_\rho(u,\xi)$ are analytic and nonzero (ANZ) in suitable vertical strips in the complex $u$ plane. This assumption is justified for finite excitations, after the removal of 1- and 2-strings, as discussed in Appendix~\ref{analyticity}.

The vacuum contribution $\kappa_0(u)$ satisfies crossing and the inversion relation \eqref{vacInversion}
\be 
\log\kappa_0(u)+\log\kappa_0(u+\lambda)=\log\frac{\sin^2\lambda\sin(2\lambda+2u)\sin(2\lambda-2u)}{\sin^22\lambda\sin(\lambda+2u)\sin(\lambda-2u)}, \qquad |\!\Re u|<\epsilon
\label{logvacInversion}
\ee
for some small $\epsilon$. In this strip 
\be 
\frac{d^2}{du^2}\log\frac{\sin(2\lambda+2u)\sin(2\lambda-2u)}{\sin(\lambda+2u)\sin(\lambda-2u)}=8\int_{-\infty}^{\infty}\frac{t\sinh\frac{\pi-3\lambda}{2}t\sinh\frac{\lambda}{2}t}{\sinh\frac{\pi}{2}t}\,e^{2ut}\,dt 
\ee
It follows that the coefficient $c(t)$ of the Fourier/Laplace transform (\ref{FourierLaplace}) satisfies
\be 
c(t)=e^{-2\lambda t}c(-t),\quad (1+e^{2\lambda t})c(t)=\frac{8t\sinh\frac{\pi-3\lambda}{2}t\sinh\frac{\lambda}{2}t}{\sinh\frac{\pi}{2}t},\quad
c(t)=\frac{4te^{-\lambda t}\sinh\frac{\pi-3\lambda}{2}t\sinh\frac{\lambda}{2}t}{\sinh\frac{\pi}{2}t\cosh\lambda t} 
\ee 
Integrating twice and evaluating the integration constants gives 
\begin{align} 
\log\kappa_0(u)&=\int_{-\infty}^{\infty}\frac{\sinh\frac{\pi-3\lambda}{2}t\sinh\frac{\lambda}{2}t}{t\sinh\frac{\pi}{2}t\cosh\lambda t}e^{-(\lambda-2u)t}dt+Au+B\nonumber\\ 
&=-2\int_{-\infty}^{\infty}\frac{\sinh\frac{\pi-3\lambda}{2}t\sinh\frac{\lambda}{2}t\sinh ut\sinh(\lambda-u)t}{t\sinh\frac{\pi}{2} t\cosh\lambda t}\,dt,
\qquad 0\le\lambda<\frac{\pi}{2}
\label{kappa0}
\end{align}
It is to be stressed that this integral diverges at $\lambda=\frac{\pi}{2}$. We analytically continue this solution to the full interval $0\le\lambda\le\pi$ in Appendix~\ref{analyticcontinuation}.

The $r$-type boundary contribution $\kappa_\rho(u,\xi)$ satisfies crossing and the inversion relation \eqref{bdyInversion}
\be 
\log\kappa_\rho(u,\xi)+\log\kappa_\rho(u+\lambda,\xi)=\log\frac{\sin(\xi+u)\sin(\xi-u)\sin(\xi+\rho\lambda+u)\sin(\xi+\rho\lambda-u)}{\sin^2\xi\sin^2(\xi+\rho\lambda)},
\quad |\!\Re u|<\epsilon\label{logrhoInv}
\ee
where in this strip
\begin{align} 
\frac{d^2}{du^2}\log\sin(\xi\!+\!u)\sin(\xi\!-\!u)\sin(\xi\!+\!\rho\lambda\!+\!u)\sin(\xi\!+\!\rho\lambda\!-\!u)
=-8\int_{-\infty}^{\infty}\frac{t\cosh(\pi\!-\!2\xi\!-\!\rho\lambda)t\cosh\rho\lambda t}{\sinh\pi t}\,e^{2ut}dt 
\end{align}
It follows that
\be 
c(t)=e^{-2\lambda t}c(-t),\qquad (1+e^{2\lambda t})c(t)=-\frac{8t\cosh\left(\pi-2\xi-\rho\lambda\right)t\cosh\rho\lambda t}{\sinh\pi t}
\ee 
with the solution
\be 
c(t)=-\frac{4te^{-\lambda t}\cosh\left(\pi-2\xi-\rho\lambda\right)t\cosh\rho\lambda t}{\sinh\pi t\cosh\lambda t} 
\ee 
Integrating twice and evaluating the integration constants gives 
\be 
\log{\kappa_\rho(u,\xi)}=2\int_{-\infty}^{\infty}\frac{\cosh(2\xi+\rho\lambda-\pi)t\cosh\rho\lambda t\sinh ut\sinh(\lambda-u)t}{t\sinh\pi t\cosh\lambda t}\,dt,\qquad \rho\ge 2
\label{kappaDvgt}
\ee

For given $\lambda$, the integral in (\ref{kappaDvgt}) is not convergent for all values of $\xi$ and $\rho$ of interest. To fix this problem, we observe that the inversion relation is invariant under the replacement $\rho\lambda\mapsto\rho\lambda-k\pi$ on the right side of (\ref{logrhoInv}) for any $k\in{\Bbb Z}$. We therefore look for a $k\in{\Bbb Z}$ such that, after the substitution, the integral is convergent for all values of $\xi$ and $\rho$ of interest with $u$ in the physical strip $0\le \Re(u)\le \lambda$. 
Notice that we cannot have $x=(\xi+\rho\lambda)/\pi\in{\Bbb Z}$ since the denominator in (\ref{logrhoInv}) would vanish. 
We find that the integral is convergent if 
\be 
|2\xi+\rho\lambda-(k+1)\pi|+|\rho\lambda-k\pi|<\pi,\qquad \xi\in(0,\pi)
\ee
This is equivalent to 
\be 
(k-x)(k+1-x)<0,\qquad x=\frac{\xi+\rho\lambda}{\pi} 
\ee
so that we require $k<x<k+1$. The unique value of $k$ which satisfies this inequality is $k=\big\lfloor\frac{\xi+\rho\lambda}{\pi}\big\rfloor$. 
Defining $\XiShift=\big[\frac{\xi+\rho\lambda}{\pi}\big]\pi$, the convergent form of the solution to the inversion relation is
\be 
\log{\kappa_\rho}(u,\xi)=2\int_{-\infty}^{\infty}\frac{\cosh(\xi+\XiShift-\pi)t\cosh(\xi-\XiShift) t\sinh ut\sinh(\lambda-u)t}{t\sinh\pi t\cosh\lambda t}\,dt,\qquad \rho\ge 2
\ee

\section{Analytic Continuation of $\kappa_0(u)$}
\label{analyticcontinuation}
In this appendix, we analytically continue the integral expression (\ref{kappa0}) for $\kappa_0(u)$ to the full interval $0\le \lambda\le\pi$. 
The divergence at $\lambda=\pi/2$ arises because of the factors of $\beta=2\cos\lambda=\sin 2\lambda/\sin\lambda$ in the inversion relation \eqref{logvacInversion}. 
To analytically continue the solution of the inversion relation, we write
\be 
\log\beta\kappa_0=\log2+\log\cos\lambda-2\int_{-\infty}^{\infty}\frac{\sinh\frac{\pi-3\lambda}{2}t\sinh\frac{\lambda}{2}t\sinh ut\sinh(\lambda-u)t}{t\sinh\frac{\pi}{2} t\cosh\lambda t}\,dt \ee
and use the identity
\be 
\log\cos\lambda=-\int_{-\infty}^{\infty}\frac{\sinh^2\lambda t}{t\sinh\pi t}\,dt,\qquad -{\pi\over2}<\lambda<{\pi\over2} 
\ee
The divergent terms at $\lambda=\frac{\pi}{2}$ then cancel on the right side giving 
\begin{align}
\log\beta\kappa_0&=\log2-4\!\int_{-\infty}^{\infty}\frac{\sinh{\lambda t\over2}\big[(\sinh{\lambda t\over2}\cosh^2{\lambda t\over2}\cosh\lambda t+\sinh{(\pi-3\lambda)t\over 2}\sinh ut\sinh(\lambda-u)t\cosh{\pi t\over2}\big]}{t\sinh\pi t\cosh\lambda t}\,dt\qquad\nonumber\\ 
&=\log2-\!\int_{-\infty}^{\infty}\frac{\sinh{\lambda t\over2}}{2t\sinh\pi t\cosh\lambda t}\Big[\sinh(\tfrac{5\lambda}{2}-2u)t+\sinh(\tfrac{5\lambda}{2}-2u-\pi) t+\sinh(\tfrac{\lambda}{2}+2u)t\nonumber \\
&\qquad\qquad+\sinh(\tfrac{\lambda}{2}+2u-\pi)t +\sinh\tfrac{3\lambda t}{2}-\sinh(\tfrac{5\lambda}{2}-\pi)t-\sinh\tfrac{\lambda t}{2}+\sinh(\pi-\tfrac{\lambda}{2})t\Big] dt
\label{twoInt}\\
&=-\!\int_{-\infty}^{\infty}\frac{\sinh{\lambda t\over2}[\sinh(\!{5\lambda\over2}\!-\!2u)t\!+\!\sinh(\!{\lambda\over2}\!+\!2u)t]}{2t\sinh\pi t\cosh\lambda t}\,dt
-\!\int_{-\infty}^{\infty}\frac{\sinh{\lambda t\over2}\sinh({3\lambda\over2}\!-\!\pi)t\cosh(\lambda\!-\!2u)t}{t\sinh\pi t\cosh\lambda t}\,dt\nonumber
\end{align}
In the last step, the $\log2$ term cancels with the four last integrals.

The previous transformation analytically continues the vacuum boundary free energy to the interval $\lambda-\frac{\pi}{2}<u<\frac{\pi}{2}$. However, we need to perform another analytic continuation to reach $u=0$ for $\lambda>\frac{\pi}{2}$. 
This is related to the appearance of a (real) short 2-string with zeros at $u=\frac{\pi}{2}$, $u=\lambda-\frac{\pi}{2}$ which causes divergences of $\log\kappa_0(u)$. 
The divergent terms at these two points appear in the first integral in the last line of (\ref{twoInt}).
To carry out the additional analytic continuation, we use the formula
\begin{align}
\log\cos u&=-\int_{-\infty}^{\infty}\frac{\sinh^2u t}{t\sinh\pi t}\,dt,\qquad -{\pi\over2}<u<{\pi\over2} 
\end{align}
Straightforward calculation then gives the two results
\begin{align}
-\!\log\cos(\lambda\!-\!u)-\!\int_{-\infty}^{\infty}\frac{\sinh{\lambda t\over2}\sinh\left({5\lambda\over2}\!-\!2u\right)\!t}{2t\sinh\pi t\cosh\lambda t}\,dt
&=\!\!\int_{-\infty}^{\infty}\frac{\cosh(2u\!-\!2\lambda)t\!+\!\cosh(2u\!-\!\lambda)t\!-\!2\cosh\lambda t}{4t\sinh\pi t\cosh\lambda t}\,dt\quad\\
-\!\log\cos u-\!\int_{-\infty}^{\infty}\frac{\sinh{\lambda t\over2}\sinh\left({\lambda\over2}\!+\!2u\right)\!t}{2t\sinh\pi t\cosh\lambda t}\,dt
&=\!\!\int_{-\infty}^{\infty}\frac{\cosh2ut\!+\!\cosh(2u\!-\!\lambda)t\!-\!2\cosh\lambda t}{4t\sinh\pi t\cosh\lambda t}\,dt
\end{align}
Removing these additional divergences thus leads to a convergent integral on a full period
\begin{align} 
\log\frac{\beta\kappa_0(u)}{\cos u\cos(\lambda-u)}=&\int_{-\infty}^{\infty}\frac{\cosh(2u-2\lambda)t+\cosh2ut+2\cosh(2u-\lambda)t-4\cosh\lambda t}{4t\sinh\pi t\cosh\lambda t}\,dt\nonumber \\ 
&-\!\int_{-\infty}^{\infty}\frac{\sinh{\lambda t\over2}\sinh({3\lambda\over2}\!-\!\pi)t\cosh(\lambda\!-\!2u)t}{t\sinh\pi t\cosh\lambda t}\,dt,
\qquad -\frac{\pi}{2}<u-\frac{\lambda}{2}<\frac{\pi}{2}
\end{align}
After simplifying of the integrands, it follows that
\begin{align}
\log\frac{\beta\cos(u-\tfrac{\lambda}{2})\,\kappa_0(u)}{\cos u\cos(\lambda-u)}&=-\int_{-\infty}^{\infty}\frac{\sinh ut\sinh(\lambda-u)t}{t\sinh\pi t\cosh\lambda t}\,dt
-\int_{-\infty}^{\infty}\frac{\sinh{\lambda t\over2}\sinh({3\lambda\over2}-\pi)t\cosh(\lambda-2u)t}{t\sinh\pi t\cosh\lambda t}\,dt
\end{align}

\end{document}